\pdfoutput=1
\documentclass[cernpreprint, UKenglish, texmf, orcidlogo]{atlasdoc}
\usepackage{atlaspackage}
\usepackage{atlasbiblatex}

\usepackage{atlasphysics}
\graphicspath{{logos/}{figures/}}

\usepackage{siunitx}
\usepackage{tikz}
\usepackage{todonotes}
\usepackage{pifont}
\usepackage{makecell}
\usepackage{setspace}
\usepackage{multirow}
\usepackage{adjustbox}
\usepackage{standalone}
\usepackage{scalerel}
\usepackage{rotating}
\usepackage{tabularht}
\usepackage{pgfplots,tikz}
\usetikzlibrary{calc,decorations.markings,arrows}
\usepackage[scaled=1]{helvet}
\usepackage{sansmath}
\usepackage{amsmath,latexsym}
\usepackage{wasysym}
\usetikzlibrary{shapes.misc,positioning}
\tikzset{cross/.style={cross out, draw=black, minimum size=2*(#1-\pgflinewidth), inner sep=0pt, outer sep=0pt},cross/.default={3pt}}
\definecolor{myyellow}{rgb}{0.96,0.742,0.29}
\definecolor{myblue}{rgb}{0.1,0.32,0.738}
\pgfplotsset{compat=newest}
\usepackage{ANA-HIGG-2022-17-PAPER-defs}

\providecommand{\checkmark}{\ding{51}}

\AtlasTitle{Interpretations of the ATLAS measurements of Higgs
boson production and decay rates and differential cross-sections
in $pp$ collisions at $\sqrt{s}=13$~TeV}

\AtlasAbstract{%
Measurements of the Higgs boson production times decay rates and differential
cross-sections have recently been performed by the ATLAS experiment
in several decay channels using up to $139$\hspace{.5ex}fb${}^{-1}$ of
proton--proton collision data at $\sqrt{s}=13$\hspace{.5ex}TeV recorded
at the Large Hadron Collider.
This paper presents multiple interpretations of these Higgs boson
measurements.
Measurements of production-mode cross-sections,
simplified template cross-sections and fiducial
differential cross-sections in different decay channels are reparameterised
in terms of the impact of Standard Model effective field theory
operators, and constraints are reported on the corresponding Wilson
coefficients.
Production and decay rate measurements are interpreted in
UV-complete extensions
of the Standard Model, namely the two-Higgs-doublet model (2HDM) near the
alignment limit and the Minimal Supersymmetric Standard Model (MSSM)
for various MSSM benchmark scenarios.
The constraints on the 2HDM parameters
$(\cos(\beta-\alpha), \textrm{tan}\beta)$
and the MSSM parameters $(m_A, \textrm{tan}\beta)$ are complementary to
those obtained from direct searches for additional Higgs bosons.
}

\author{The ATLAS Collaboration}

\AtlasRefCode{HIGG-2022-17}

\PreprintIdNumber{CERN-EP-2024-017}

\AtlasJournalRef{JHEP 11 (2024) 097}
\AtlasDOI{10.1007/JHEP11(2024)097}

\hypersetup{pdftitle={ATLAS document},pdfauthor={The ATLAS Collaboration}}

\begin{document}

\maketitle

\setcounter{tocdepth}{3}
\tableofcontents

\newpage

%

\section{Introduction}

Since a particle ($H$) with attributes consistent with those of the Higgs
boson~\cite{Englert:1964et, Higgs:1964ia,Higgs:1964pj,Guralnik:1964eu,
Higgs:1966ev,Kibble:1967sv} of the Standard Model (SM) of particle physics
was discovered~\cite{HIGG-2012-27, CMS-HIG-12-028} by the ATLAS and
CMS experiments, its properties have been measured with proton--proton
($pp$) collision data produced by the Large Hadron Collider (LHC) at CERN.
The ATLAS and CMS Collaborations recently presented their first results on the
combination of Higgs boson production and decay rate measurements with
up to 139\,$\ifb$ of 13\,\TeV~$pp$ collision
data~\cite{HIGG-2021-23, CMS-HIG-22-001}, collected during the full
LHC Run~2 in the years 2015 through 2018.
Two interpretations of the ATLAS measurements are presented here.
The first one is based on an effective field theory (EFT) framework
of the Standard Model.
The second interpretation relies on specific theories that provide
ultraviolet (UV) completions of the SM, specifically several benchmark scenarios
within the two-Higgs doublet
model (2HDM)~\cite{Lee:1973iz,Gunion:2002zf,Branco:2011iw,Heinemeyer:2013tqa}
or within a minimal supersymmetric extension of the Standard Model
(MSSM)~\cite{Fayet:1976et,Fayet:1977yc,Farrar:1978xj,Fayet:1979sa,
Dimopoulos:1981zb}.

In addition to the production and decay rates mentioned above,
the fiducial differential cross-sections for Higgs boson production
as a function of selected event kinematics
have been measured by the ATLAS
Collaboration with the same data sample in the
\hfourl~($\ell=e,\mu$)~\cite{HIGG-2018-29}
and \Hyy~\cite{HIGG-2019-13} decay channels, among
others~\cite{HIGG-2018-49,HIGG-2020-25}.
The fiducial differential cross-sections measured as a function of the
total transverse momentum\footnote{\AtlasCoordFootnote}
of the Higgs boson decay products in the two decay channels,
$\pT^{4\ell}$ and $\pT^{\gamma\gamma}$, are used in this
paper to constrain anomalous Higgs boson couplings to gluons and top quarks,
and an anomalous Higgs-gluon-top-antitop quark coupling, within the
same EFT framework.
These couplings are also probed using the EFT
interpretation of Higgs boson production and decay rate measurements;
the sensitivities obtained with the two approaches are compared.

Effective field theories provide a model-independent approach to describe
new physics at high mass scales,
systematically improvable with higher-order perturbative calculations,
to parametrise the effects of candidate theories beyond the Standard
Model (BSM) that reduce to the Standard Model at low energies.  In the SM
Effective Field Theory (SMEFT), the effects of BSM dynamics at energy
scales $\Lambda$ that are large in comparison to the
vacuum-expectation-value (vev) $v$ of the SM Higgs field
($\Lambda \gg v$) can be parameterised at
low energies, $E \ll \Lambda$, in terms of higher-dimensional
operators that are built up from the Standard Model fields and respect its
symmetries.  Measurements of (fiducial) cross-sections can be used
to place constraints on the Wilson coefficients associated with the
SMEFT operators, corresponding to limits on BSM physics at some fixed scale
$\Lambda$.

The methodology employed here for the EFT interpretation of
the combined measurements of Higgs boson production and decay rates
is similar to the one used by the individual
analyses.
The interpretation is applied to the combined measurements
of the Higgs boson production cross-sections times branching ratios in the
following decay channels:
$\hzz(\to 4\ell)$, \hyy, \hbb, \htt, $\hww(\to e\nu\mu\nu)$, \hzg, and
\hmm~\cite{HIGG-2018-28, HIGG-2020-16, HIGG-2018-51,HIGG-2018-52,
HIGG-2019-04, HIGG-2021-08, HIGG-2020-23, HIGG-2019-09, HIGG-2021-20,
HIGG-2018-42, HIGG-2019-14}.

A similar approach is used for the interpretation of the fiducial
\pT-differential cross-sections in the \hfourl\ and \Hyy\ channels.
For this study the \pT\ distribution was chosen as it was previously
measured in both channels
and is sensitive to anomalous Higgs boson couplings to top-quarks and gluons.
The method resembles closely that described in Ref.~\cite{HIGG-2019-13}
for the joint interpretation of five differential cross-sections in the
\hyy\ decay channel in terms of
a few dimension-6 SMEFT operators, but only the cross-section
measurements as a function of $\pT^{4\ell}$ and $\pT^{\gamma\gamma}$
are used, and a smaller number of operators is probed.
On the other hand, the joint interpretation with
a second decay channel with similar statistical precision
increases the sensitivity to the probed BSM operators.

In the interpretations presented in the second part of this paper,
performed in the context of UV-complete BSM theories,
two additional measurements of
production and decay rates, based on a partial Run~2 data sample collected
in the years 2015 and 2016, are considered:
\hww\ in the $WH$ and $ZH$ production modes~\cite{HIGG-2016-07},
and $H\to \text{multileptons}$ in the $t\bar{t}H$ production
mode~\cite{HIGG-2017-02}.
In the latter, the selection requirement of two prompt leptons
primarily selects \hww\ and \htt\ decays, with a small additional
contribution of \hzz\ decays.

In a previous publication~\cite{HIGG-2018-57}, measurements were
interpreted within the hMSSM model~\cite{Djouadi:2013uqa}.
In this paper, in addition to the hMSSM model,
seven more recent benchmark MSSM
scenarios~\cite{Bahl:2018zmf,Bahl:2019ago} are tested.
In addition, an interpretation
within
the 2HDM near the alignment limit, in which the light
neutral Higgs boson of the 2HDM is SM-like\footnote{The terms SM-like
and analogous ones are used throughout this paper to denote particles with
similar properties to those of the named particle.}, is also reported.
These constraints are compared with those inferred from the limits on the
SMEFT Wilson coefficients after matching the 2HDM parameters to the
SMEFT operators.

The paper is structured as follows. Section~\ref{sec:input} gives an
overview of the Higgs boson production-mode cross-sections and decay rate
measurements and of the fiducial differential cross-section measurements that
are used to derive the results presented.
Section~\ref{sec:eft}, devoted to the EFT interpretation of the
measurements, starts with a description
of the general approach of deriving limits on Wilson coefficients
(Section~\ref{sec:eft-intro}). This is followed by the details of the specific
choices of Wilson coefficients probed
in the interpretations of the combined measurements of Higgs boson
production-mode cross-sections and decay rates (Section~\ref{sec:eft-results})
or of the measurements of the fiducial differential cross-sections as a
function of $\pTH$
(Section~\ref{sec:eft-results-diff}), together with the corresponding
constraints.
The interpretations in UV-complete extensions of the SM are
detailed in Section~\ref{sec:bsm}.
The approach taken to interpret the Higgs boson production-mode
cross-sections and decay rate measurements in terms of constraints on
the parameters of the 2HDM or the MSSM, and the derived results,
are described in Sections~\ref{sec:2hdm} and~\ref{sec:mssm},
respectively.
Finally, Section~\ref{sec:conclusions}
presents the conclusions.


%
%

\section{Measurements of Higgs boson cross-sections}
\label{sec:input}

\subsection{Data samples}

The results presented in this article are based on up to 139~\ifb of
$pp$ collision data
collected by the ATLAS detector~\cite{PERF-2007-01, ATLAS-TDR-19,
PIX-2018-001} in the years 2015 to 2018, with the LHC operating at a
centre-of-mass energy of 13\,\TeV.
The decay channels, targeted production modes and integrated luminosity
of the data sample used in each input analysis
are summarised in Table~\ref{tab:lumi}, where
the simplified template cross-section measurements labelled `STXS' are identical to those used in
Ref.~\cite{HIGG-2021-23}.
The analyses used for this interpretation were published while the
uncertainty in the combined 2015--2018 integrated luminosity was
1.7\%. It was derived from the calibration of the luminosity scale using
$x$--$y$ beam-separation scans, following a methodology similar to that
detailed in Ref.~\cite{DAPR-2013-01},
and using the LUCID-2~detector~\cite{LUCID2}
for the baseline luminosity measurements.
Subsequently an improved estimate of the integrated luminosity
was made~\cite{DAPR-2021-01}, with a slightly higher central value
$(+0.8\%)$ and half the uncertainty (0.8\%).
The results described in the following sections assume
the older, less precise luminosity calibration; they would not be significantly
affected if recomputed using the latest one.

\begin{table}[!htbp]
\caption{The decay channels (1st column), targeted production
modes (2nd column) and integrated luminosity $\mathcal{L}$
(3rd column) of the data sample used
for each analysis included in the combined measurements
of Higgs boson production and decay rates,
couplings and simplified template cross-sections (labelled as
`STXS' in the `Binning' column), or in the measurements of
fiducial $\pT$-differential cross-sections (labelled as
`differential' in the `Binning' column).
The references for the input analyses (5th column) and
information about which input analyses are considered for the
SMEFT-based interpretation (6th column) and the
interpretations in terms of UV-complete models (7th column)
are also provided.
The STXS-0$^{*}$ binning only distinguishes inclusive production processes, whereas the STXS-1.2 and differential
binnings partition production processes in kinematic regions at a varying level of granularity.
The full definitions of these binning schemes for the
measurements and the corresponding signal yield
parameterisations are detailed in Section~\ref{sec:sigyieldparam}.
In all the analyses, the tiny expected contribution from associated
$b \bar{b} H$ production is applied as a correction to ggF as
described in the text.}
\vspace{3pt}
\begin{center}

\centering
\resizebox{\linewidth}{!}{
\begin{tabular}{llrcccc}
\toprule
\multicolumn{2}{c}{Analysis}                                                                  & $\mathcal{L}\phantom{xx}$          & \multirow{2}{*}{Reference}       & \multirow{2}{*}{Binning} & \multirow{2}{*}{SMEFT} & 2HDM and \\
Decay channel         & Production mode                                                       & \([\textrm{fb}^{-1}]\)    & {}                               &                          &            & (h)MSSM       \\
\midrule
\multirow{2}{*}{\hyy} & \multirow{2}{*}{(ggF, VBF, $WH$, $ZH$, \ttH, $tH$)}                   & \multirow{2}{*}{$139\phantom{.1}$}   & \cite{HIGG-2020-16}              & STXS-1.2                 & \checkmark & \checkmark \\
&                                                                       &                          & \cite{HIGG-2019-13}              & differential             & \checkmark (subset) &            \\
\\
\multirow{2}{*}{\hzz} & \multirow{2}{*}{($ZZ^*\to 4\ell$: ggF, VBF, $WH+ZH$, $\ttH + tH$)}   & \multirow{2}{*}{$139\phantom{.1}$}   & \cite{HIGG-2018-28}               & STXS-1.2                 & \checkmark & \checkmark \\
&                                                                       &                          & \cite{HIGG-2018-29}              & differential             & \checkmark (subset) &            \\
& ($ZZ^*\to \ell\ell\nu\bar\nu/\ell\ell q\bar{q}$: \tttH\ multileptons) & $36.1$                   & \cite{HIGG-2017-02}              & STXS-0$^*$               &            & \checkmark \\
\\
\htt                  & (ggF, VBF, $WH+ZH$, $\ttH+tH$)                                        & $139\phantom{.1}$                    & \cite{HIGG-2019-09}              & STXS-1.2                 & \checkmark & \checkmark \\
& (\tttH\ multileptons)                                                 & $36.1$                   & \cite{HIGG-2017-02}              & STXS-0$^*$               &            & \checkmark \\
\\
\hww                  & (\tggF, \tVBF)                                                        & $139\phantom{.1}$                    & \cite{HIGG-2021-20}              & STXS-1.2                 & \checkmark & \checkmark \\
& ($WH$, $ZH$)                                                          & $36.1$                   & \cite{HIGG-2017-14}              & STXS-0$^*$               &            & \checkmark \\
& (\tttH\ multileptons)                                                 & $36.1$                   & \cite{HIGG-2017-02}              & STXS-0$^*$               &            & \checkmark \\
\\
\hbb                  & ($WH$, $ZH$)                                                          & $139\phantom{.1}$                    & \cite{HIGG-2018-51,HIGG-2018-52} & STXS-1.2                 & \checkmark & \checkmark \\
& (\tVBF)                                                               & $126\phantom{.1}$                    & \cite{HIGG-2019-04}              & STXS-1.2                 & \checkmark & \checkmark \\
& ($\tttH+tH$)                                                          & $139\phantom{.1}$                    & \cite{HIGG-2020-23}              & STXS-1.2                 & \checkmark & \checkmark \\
& (boosted Higgs bosons: inclusive production)                          & $139\phantom{.1}$                    & \cite{HIGG-2021-08}              & STXS-1.2                 & \checkmark & \checkmark \\
\\
\hzg                  & (inclusive production)                                                & $139\phantom{.1}$                    & \cite{HIGG-2018-42}              & STXS-0$^*$               & \checkmark & \checkmark \\
\hmm                  & ($\tggF + \tttH + tH$, $\tVBF + WH + ZH$)                             & $139\phantom{.1}$                    & \cite{HIGG-2019-14}              & STXS-0$^*$               & \checkmark & \checkmark \\
\bottomrule
\end{tabular}
}


\end{center}
\label{tab:lumi}
\end{table}

\subsection{Simulation of the Standard Model signal}
\label{sec:mc}

Monte Carlo (MC) simulated event samples of SM Higgs bosons produced
in $pp$ collisions and decaying into the final states
considered in this study were generated and normalised
to SM predictions calculated for a Higgs boson mass
$m_H=125.09$~\GeV~\cite{HIGG-2014-14}.
For each Higgs boson decay mode, the branching ratio that was used
corresponds to the theoretical calculation at the highest available
order of accuracy in perturbation theory~\cite{deFlorian:2016spz}.

Higgs boson production via gluon--gluon fusion was simulated using the
\powhegbox~v2~\cite{Nason:2004rx,Frixione:2007vw,Alioli:2010xd,Alioli:2008tz}
NNLOPS implementation~\cite{Hamilton:2013fea,Hamilton:2015nsa}.  The
event generator uses HNNLO~\cite{Catani:2007vq} to reweight the
inclusive Higgs boson rapidity distribution produced by the
next-to-leading-order (NLO) generation of $pp\to H+\textrm{parton}$ events,
with the scale of each parton emission determined using the MiNLO
procedure~\cite{Hamilton:2012rf}. The PDF4LHC15~\cite{Butterworth:2015oua}
next-to-next-to-leading-order (NNLO)
parton distribution functions (PDFs)
were used for the central prediction and uncertainty,
except for the $H\to b\bar{b}$ and $H\to Z\gamma$
measurements, which used the \NNPDF[3.0]~\cite{Ball:2014uwa} NNLO PDFs.
The sample is normalised such that it reproduces the total
cross-section predicted by a next-to-next-to-next-to-leading-order
(N$^3$LO) quantum chromodynamics (QCD) calculation with NLO electroweak (EW) corrections
applied~\cite{deFlorian:2016spz,Anastasiou:2015ema,Anastasiou:2016cez,Actis:2008ug,Anastasiou:2008tj}.
The NNLOPS generator reproduces the Higgs boson \pT
distribution predicted by the NNLO
plus next-to-next-to-leading logarithm (NNLL) calculation of
\hres~\cite{Grazzini:2013mca}, which
includes the effects of top- and bottom-quark masses and uses
dynamical renormalisation and factorisation scales.

The \tVBF\ and $q\bar{q} \to VH$ ($gg\to ZH$)
production processes were simulated at NLO (LO)
accuracy in QCD using the \powhegbox~v2~\cite{Nason:2009ai} generator
where the simulation of $q\bar{q} \to VH$
relies on improved NLO calculations~\cite{Hamilton:2012np}.
The PDF4LHC15 NLO set of PDFs was used,
except for the $H\to b\bar{b}$ and $H\to Z\gamma$
measurements, which used the \NNPDF[3.0] NLO PDFs.
In the vector-boson fusion (VBF) $H\to b\bar{b}$ analysis,
NLO EW corrections were calculated using \MGNLO~3.0
and applied as a function of the generated Higgs
boson transverse momentum.
The
\tVBF\ sample is normalised to an approximate-NNLO QCD cross-section
with NLO EW corrections applied~\cite{deFlorian:2016spz,
Ciccolini:2007jr,Ciccolini:2007ec,Bolzoni:2010xr}.  The
\tVH\ samples are normalised to cross-sections calculated at NNLO in
QCD with NLO EW corrections~\cite{Brein:2003wg,Denner:2011id}
and additional NLO QCD corrections~\cite{Altenkamp:2012sx} for the
$gg\to ZH$ subprocess~\cite{deFlorian:2016spz}.

Higgs boson production in association with a top--antitop quark pair (\ttH)
was simulated at NLO accuracy in QCD using the \powhegbox~v2\ generator,
except for the $H\to\mu\mu$ analysis and the measurements based on 36.1~\ifb
of data, which used the \amc~2.6~\cite{MadGraph} generator.
The PDF4LHC15 NLO set of PDFs was used in the \Hyy\ and \hfourl\ analyses,
while the \NNPDF[3.0] (N)NLO set of PDFs was used in the other ones.
In all cases, the sample is normalised to a calculation with NLO
QCD and EW corrections~\cite{deFlorian:2016spz,Beenakker:2002nc,
Dawson:2003zu,Yu:2014cka, Frixione:2014qaa, Frixione:2015zaa}.
Higgs boson production in association with a single top
quark in the $t$-channel (\tHq) or with an additional $W$ boson (\WtH)
were simulated at NLO accuracy using \amc~2.6 and either the
\NNPDF[3.0(N)NLO] PDF set
or the CT10 NLO PDF set.
The two \tH\ samples are normalised to NLO
QCD calculations~\cite{deFlorian:2016spz,Demartin:2015uha}.

All parton-level events were input to \PYTHIA[8.2]~\cite{Sjostrand:2007gs}
to model the Higgs boson decay, parton showering, hadronisation, and
multiple parton interactions.  The generators were interfaced to
\PYTHIA[8.2]\ using the AZNLO~\cite{STDM-2012-23} or
A14~\cite{ATL-PHYS-PUB-2014-021} sets of tuned parameters (tunes).

Particle-level events were passed through a
\geant~\cite{Agostinelli:2002hh}
simulation of the ATLAS
detector~\cite{SOFT-2010-01} and reconstructed using the same analysis
software as used for the data. Event pile-up was included in the
simulation by overlaying inelastic $pp$ collisions, such that the
distribution of the average number of interactions per bunch crossing
reproduces that observed in the data.
The inelastic $pp$ collisions were simulated
with \PYTHIA[8.1]\ using the \MSTW[LO]~\cite{Martin:2009iq} set of PDFs with
the A2 tune~\cite{ATL-PHYS-PUB-2012-003}, or using
the \NNPDF[2.3LO]~\cite{Ball:2012cx} set of PDFs with the
A3 tune~\cite{ATL-PHYS-PUB-2016-017}.

\subsection{Signal yield parameterisation}
\label{sec:sigyieldparam}

In all analyses listed in Table~\ref{tab:lumi},
the likelihood function for each signal region $k$, with one or more bins $r$,
is modelled as

\begin{eqnarray}
\label{eq:Lbasic}
L(\vec{N_k} | \vec{\mu}, \vec{\theta}) &=& \prod_{r} \text{Poisson}\left( N_{k,r} | s_k(\vec{\mu},\vec{\theta}) \cdot f_s^{k,r}(\vec{\theta})~+~ b_{k,r}(\vec{\theta}) \right),
\end{eqnarray}

\noindent where $N_{k,r}$ is the observed event count of bin $r$ in region $k$,
$s_k$ is the expected signal count in region $k$,
$f_s^{k,r}$ is the expected fraction of the signal in region $k$ that is
contained in bin $r$,
and $b_{k,r}$ represents the expected event count from background processes.
The ensemble of \textit{parameters of interest} $\vec{\mu}$ describes the
Higgs boson signal normalisation,
while $\vec{\theta}$ represents the set of \textit{nuisance parameters}
taking into account the systematic uncertainties that originate from
theoretical and experimental sources, as well as additional degrees of freedom
without prior constraints such as background yields or normalisations in
some of the input channels.
The full likelihood function is then the product of the likelihood
functions for each signal region $k$ and of Gaussian or log-normal
probability density functions that constrain the nuisance parameters.

Depending on the level of detail implemented in each analysis, the signal
yield parameters $\vec{\mu}$ can be indexed by Higgs boson production process
($i$), decay mode ($X$), and fiducial phase space region defined at the
particle level ($k^{\prime}$).
Analysis region ${k}$, defined at the reconstruction level, is typically
chosen to match the particle-level region ${k^\prime}$ as closely as possible,
to reduce the extrapolation uncertainty.
As reconstruction-level selections do not generally correspond exactly to
particle-level regions, multiple particle-level regions will
contribute to the signal yield $s_k$.

For the measurements listed in Table~\ref{tab:lumi},
two distinct signal parameterisation strategies are followed
as defined in Eq.~\eqref{eq:Lsk} and Eq.~\eqref{eq:Lsk_diff}.
For analyses labelled as `STXS-0$^{*}$' or `STXS-1.2',
the signal yield for region $k$ is modelled
as a scale factor $\mu^{i,k^\prime,X}_k$ applied to the SM Higgs boson
production cross-section times branching ratio,
for each Higgs boson production process $i$ and decay $X$,
in a fiducial region $k^\prime$ defined at the particle level.
Alternatively, for analyses labelled as `differential', the signal yield is
modelled as a cross-section $\sigma_{\textrm{fid.}}^{k^\prime,X}$ describing the
sum of all production processes, separately for each Higgs boson decay mode
$X$ and fiducial region $k^\prime$ defined at the particle level.
The corresponding parameterisations of the signal yield $s_k$ in terms
of the parameters of interest $\vec{\mu_k} = \{\mu^{i,k^\prime,X}_k\}$ or $\{\mu^{k^\prime,X}_k\}$
and of the nuisance parameters $\vec\theta$ are:

\begin{eqnarray}
\label{eq:Lsk}
s_k^\textrm{STXS}\left(\vec{\mu_k},\vec{\theta}\right) & = & \mathcal{L} \times  \sum\limits_{i,k^\prime,X}   \mu^{i,k^\prime,X}_k \times (\sigma \times \BR)_\text{SM,(N(N))NLO}^{i,k^\prime,X}(\vec{\theta}) \times \epsilon^{i,k^\prime,X}_{\textrm{STXS},k}(\vec{\theta}), \\
\label{eq:Lsk_diff}
s_{k}^\textrm{diff.}\left( \vec{\mu_k},\vec{\theta}\right) & = &  \mathcal{L} \times \sum\limits_{k^\prime,X} \mu^{k^\prime,X}_k \times \sigma_{\textrm{SM,fid.}}^{k^\prime,X}(\vec{\theta}) \times \epsilon^{k^\prime,X}_{\textrm{diff.},k}(\vec{\theta}),
\end{eqnarray}

\noindent where $\mathcal{L}$ is the integrated luminosity,
$(\sigma \times \BR)_\text{SM,(N(N))NLO}^{i,k^\prime,X}$ is the
calculation, at the highest available order, of the SM Higgs boson
cross-section for the production process $i$ in particle-level region
$k^\prime$ multiplied by the SM Higgs boson branching ratio to the
final state $X$, and $\sigma_{\textrm{SM,fid.}}$ is the SM Higgs boson production
cross-section for the fiducial region $k^\prime$, summed over all production modes.
The factors $\epsilon^{i,k^\prime,X}_{\textrm{STXS},k}$ and
$\epsilon^{k^\prime,X}_{\textrm{diff.},k}$ represent the products of acceptance
times efficiency of the reconstruction-level region $k$ for the
particle-level fiducial phase space region $k^\prime$
and Higgs boson decay $X$ (in production mode $i$ for the STXS
interpretation).

For each interpretation based on a particular model
(SMEFT, 2HDM, or MSSM) with a vector of model
parameters $\vec{\alpha}$, the original signal parameters $\vec{\mu}$ are replaced with expressions that parameterise
the model predictions, i.e., $\vec{\mu} \rightarrow \vec{\mu}(\vec{\alpha})$,
so that the likelihood of Eq.~\eqref{eq:Lbasic} is directly
expressed in terms of the parameters $\vec{\alpha}$.
Then, constraints on these parameters can be directly inferred from
the modified likelihood expression. The model-specific
re-parameterisations of the signal parameters are detailed in
Sections~\ref{sec:eft} and~\ref{sec:bsm}.

The acceptance factors $\epsilon_{\textrm{STXS},k}^{i, k^\prime,X}$ and
$\epsilon_{\textrm{diff.},k}^{k^\prime,X}$, as well as the signal bin
fractions $f_s^{k,r}$,
are derived assuming SM Higgs boson kinematics. For
interpretations of the measurements in physics models that
significantly alter  kinematic distributions, additional correction
factors may be needed to account for changes in the acceptance and
signal shape as a function of BSM model parameters. These are discussed
when applicable in Sections~\ref{sec:eft} and~\ref{sec:bsm}.

\subsubsection{Fiducial differential cross-section measurements}

For the interpretation of fiducial differential cross-section measurements
in the \hgg\ and \hZZllll\ analyses, the differential distributions
in the transverse momentum of the Higgs boson decay products are considered.
The  \hZZllll\ analysis defines nine analysis regions with boundaries in
$\pT^{4\ell}$ at
\[\{0, 10, 20, 30, 45, 60, 80, 120, 200, 300\}~\GeV,\]
\noindent whereas the \hgg\ analysis defines a finer granularity with 20
analysis regions with boundaries in $\pT^{\gamma\gamma}$ at
\[\{0, 5, 10, 15, 20, 25, 30, 35, 45, 60, 80, 100, 120, 140, 170,
200, 250, 300, 450, 650, 13000\}~\GeV,\]
\noindent resulting in a total of 29
fiducial cross-section measurements that
are interpreted in Section~\ref{sec:eft-results-diff}.
Further details of these analyses, including the full definitions of the
fiducial regions, can be found in Refs.~\cite{HIGG-2019-13,HIGG-2018-29}.

\subsubsection{Production and decay rate and simplified template
cross-section measurements}

For the interpretations based on the SMEFT framework presented in
Section~\ref{sec:eft}, measurements of `simplified template
cross-sections' of various Higgs boson production processes
in the regions of phase space defined within the STXS
framework~\cite{deFlorian:2016spz,LesHouches,
Berger:2019wnu, Amoroso:2020lgh} are used.
The inclusive production process classes defined at `Stage-0'
in the STXS scheme are
\bbH\ production, \ttH\ and \tH\ processes,
\qqtoHqq\ processes (with contributions from both \VBF\ production and
quark-initiated \VH\  production with a hadronic decay of the gauge boson),
$WH$ and $ZH$ production (including \ggtoZH, followed by any leptonic decay of
the vector boson \VHlep)
and finally the \ggtoH\ process, consisting of \ggF\ and \ggtoZH,
$Z\to q\bar{q}$ production.
Since the acceptances for \ggtoH\ and \bbH\ production are similar for
all input analyses, the \bbH\ production mode is modelled as a
$1\%$ increase of the \ggtoH\ yield in each region of production phase
space~\cite{deFlorian:2016spz}.
Theory uncertainties for the \ggtoH, \qqtoHqq, and \ttH\ processes are
defined as in Refs.~\cite{HIGG-2018-28,HIGG-2020-16}, while those of the
\VHlep\ process follow the scheme described in
Ref.~\cite{ATL-PHYS-PUB-2018-035}.

The analyses listed in Table~\ref{tab:lumi} as using the `STXS-0$^*$' binning
provide measurements of the scale factors $\mu^{i,X}$
for the product of the inclusive cross-section of one or more production
modes $i$, with $i$~=~ggF, VBF, $WH$, $ZH$, $ttH$,
$tH$\footnote{The classification of production modes used here differs
in one respect from the Stage-0 STXS definition: in the
STXS definition $qq \to VH(qq)$  production and \VBF\ production are both
labelled as a common \qqtoHqq\  production mode, whereas these production
modes are labelled separately here,
based on the corresponding definition/settings of the MC event generators.},
and branching ratio of the Higgs boson decay mode $X$.
Analyses of the most sensitive production modes performed with the
full Run~2 data sample, labelled `STXS-1.2' in Table~\ref{tab:lumi},
report measurements of signal yield scale factors in the more fine-grained
`Stage-1.2' definition of Higgs boson production cross-sections.
At Stage-1.2, the Stage-0 regions are partitioned in particle-level
kinematic volumes, for instance in a few coarse intervals of Higgs
boson transverse momentum ($\pT^H$), which are helpful
for distinguishing momentum-dependent and momentum-independent SMEFT
operators.
The ensemble of parameters
$\mu^{i,k^\prime,X}_\mathrm{STXS-1.2}$ describes
deviations in differential distributions, with the level of detail
controlled by the number of particle-level regions that are defined.
The precision with which the more fine-grained set of scale factors
$\mu^{i,k^\prime,X}_\mathrm{STXS-1.2}$ can be measured depends on the design
of the analysis and the amount of available data.
As individual analyses provide only limited sensitivity to some of the
Stage-1.2 categories with the current data sample, some of these categories
are merged compared to the Stage-1.2 definitions given in
Ref.~\cite{deFlorian:2016spz}.
Details of the merged category definitions in each analysis are provided
in Refs.~\cite{HIGG-2020-16,HIGG-2018-28,HIGG-2019-09,HIGG-2021-20,
HIGG-2018-51,HIGG-2018-52,HIGG-2019-04,HIGG-2020-23,HIGG-2021-08}.
The grouping of the production modes for STXS-0$^*$ and STXS-1.2 regions
is also shown in the second column of Table~\ref{tab:lumi}.

After merging, a total
of 78 signal yield parameters are extracted from the STXS
measurements listed in Table~\ref{tab:lumi}.
The kinematic definitions of these 78 STXS signal yield parameters, along
with their measured signal strengths (i.e., their ratios to the
corresponding SM predictions) and uncertainties, are shown in
Figures~\ref{fig:stxs_muplot_WW_ZZ}--\ref{fig:stxs_muplot_bb_tautau_mumu}.
While the data underlying the measurements in these figures is the same as
that used for the interpretation of the combined Higgs boson measurements
shown in Ref.~\cite{HIGG-2021-23}, the granularity of the STXS model used
here is substantially larger, providing additional constraining power for
the parameters of the SMEFT interpretation of the data.
The $\chi^2$ for the compatibility between the observed values and
the SM hypothesis is 50 for 78 degrees of freedom, corresponding to a
$p$-value  of 99.4\%.

For the MSSM and tree-level 2HDM interpretations of Section~\ref{sec:bsm},
no predictions are available for all production modes for the full
STXS Stage-1.2 phase space partitioning.
In this case, the interpretation is performed based on the STXS-0$^*$
production mode scale factors measured by each input analysis.

\begin{figure}[p]
\centering
\includegraphics[width=0.8\textwidth,page=1]{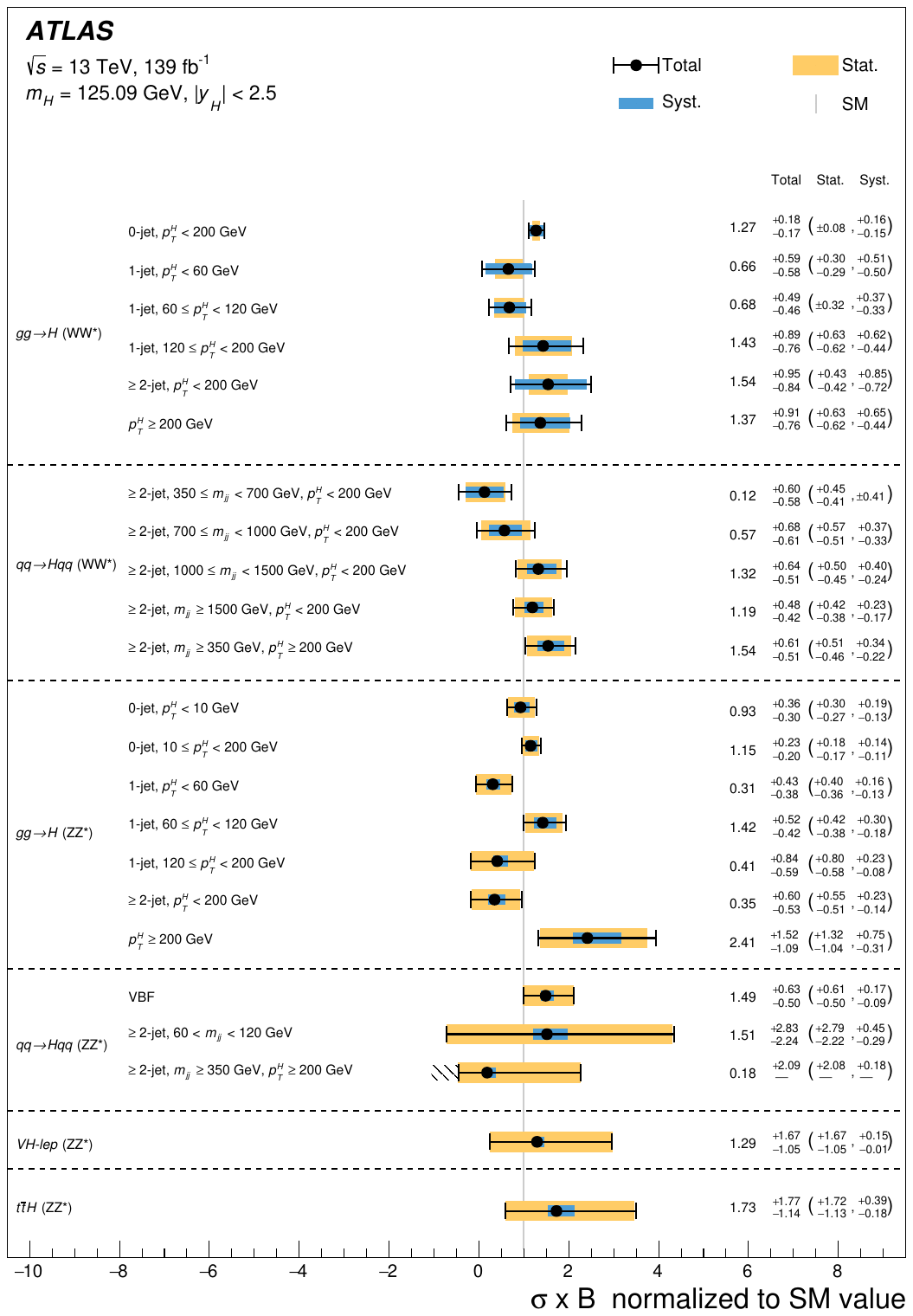}
\caption{
Measured signal strength for each of the $H\to WW^*$ and $H\to ZZ^*$ STXS
categories included in the combination.
In the $H\to ZZ^*$ analysis, $VH$ production events with the vector
boson $V$ decaying into leptons are assigned to the $VH$-lep category.
In one category the negative uncertainty is truncated, as indicated
by the hatched region, which corresponds to cross-section values for which
the total yield prediction in an analysis region is negative.
In that case, the negative uncertainty and its breakdown in terms of
statistical and systematic components are not reported,
and replaced with a long dash (--).
}
\label{fig:stxs_muplot_WW_ZZ}
\end{figure}

\begin{figure}[p]
\centering
\includegraphics[width=0.75\textwidth,page=1]{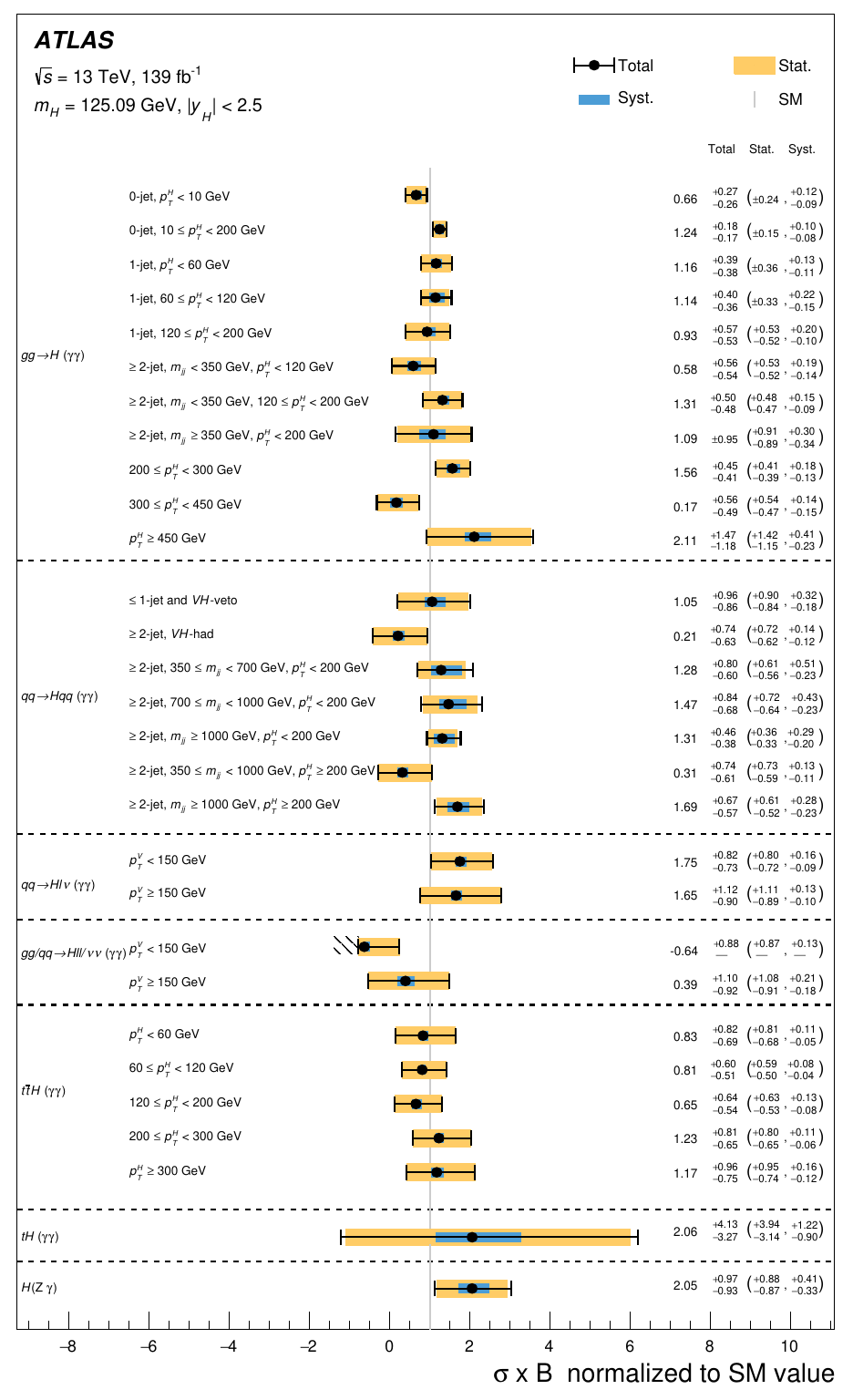}
\caption{
Measured signal strength for each of the $H\to\gamma\gamma$
and $H\to Z\gamma$ STXS categories included in the combination.
In the $H\to\gamma\gamma$ analysis, events with two or more
particle-level jets are assigned to the $VH$-veto ($VH$-had)
categories if the leading dijet invariant mass $m_{jj}$ is in the
range $m_{jj}<60$~\GeV\ or $120\leq m_{jj}<350$~\GeV\ ($60\leq m_{jj}<120$~\GeV).
In one category the uncertainty is truncated, as indicated by the
hatched region, which corresponds to cross-section values for which the
total yield prediction in an analysis region is negative.
In that case, the negative uncertainty and its breakdown in terms of
statistical and systematic components are not reported,
and replaced with a long dash (--).
}
\end{figure}

\begin{figure}[p]
\centering
\includegraphics[width=0.8\textwidth,page=1]{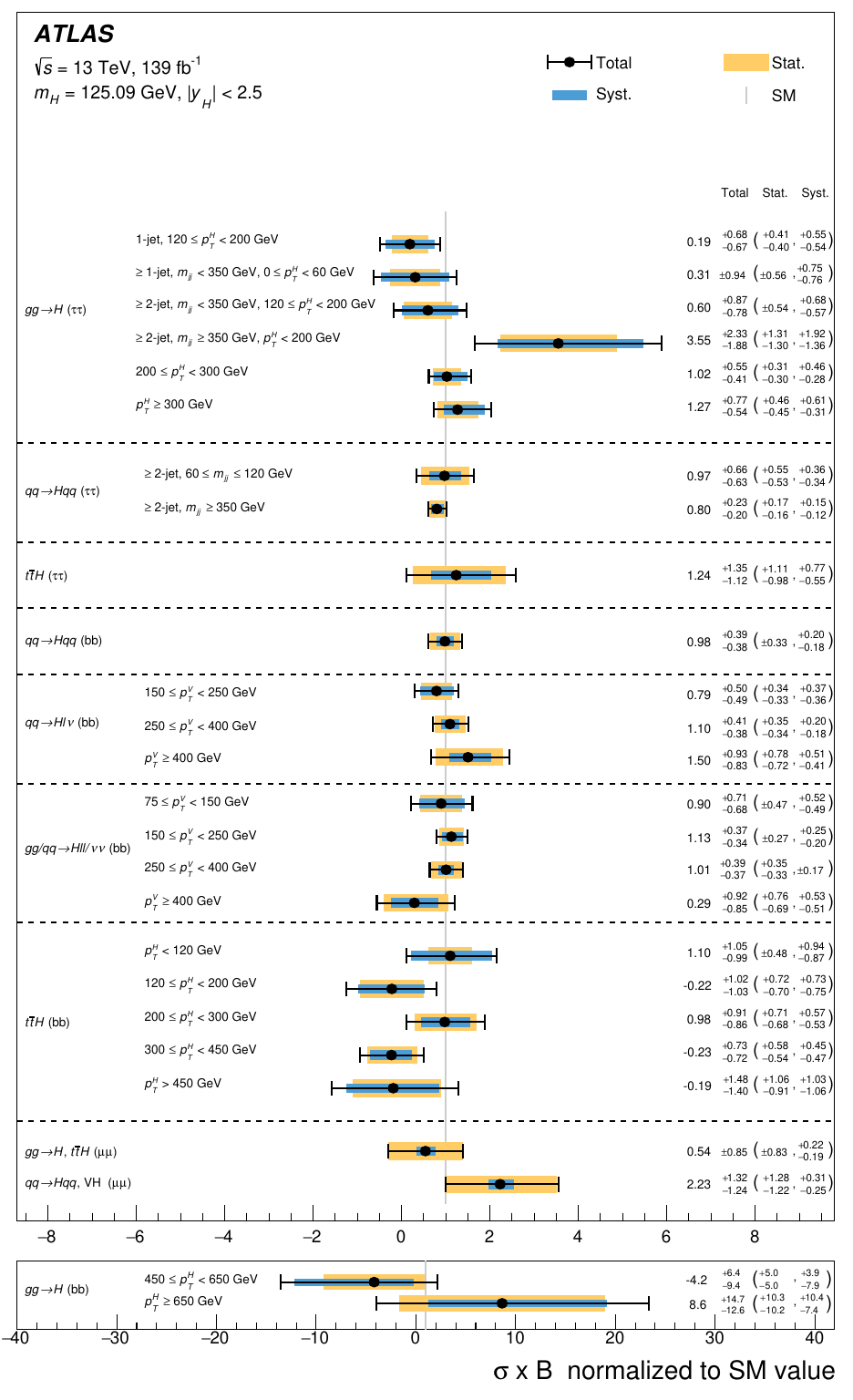}
\caption{
Measured signal strength for each of the $H\to \tau\tau$,
$b\bar{b}$, and $\mu\mu$
STXS categories implemented.
}
\label{fig:stxs_muplot_bb_tautau_mumu}
\end{figure}


\FloatBarrier

\section{Interpretations based on SM Effective Field Theory}
\label{sec:eft}
%

%
\subsection{Methodology of effective field theory interpretations}
\label{sec:eft-intro}

The SM effective field theory provides an
elegant language to encode the modifications of the Higgs boson properties
induced by a wide class of BSM theories.
Within the mathematical
language of the SMEFT, the effects of BSM dynamics at high energies
$\Lambda$, well above the electroweak scale
$v = 246$\,\GeV ($(\Lambda/v)^2 \gg 1$), can
be parameterised at low energies, $E \ll \Lambda$, in terms of
higher-dimensional operators built up from the SM fields
and respecting its symmetries such as gauge invariance.
This yields an effective Lagrangian:
\begin{equation}
\mathcal{L}_\text{SMEFT} = \mathcal{L}_\text{SM} + \sum\limits_i^{N_{d=6}} \frac{c_i}{\Lambda^2} \mathcal{O}_i^{(6)} + \sum\limits_j^{N_{d=8}} \frac{b_j}{\Lambda^4} \mathcal{O}_j^{(8)} + \dots,
\label{eq:smeft}
\end{equation}
\noindent where $\mathcal{L}_\text{SM}$ is the SM Lagrangian,
${\mathcal{O}_i^{(6)}}$ and ${\mathcal{O}_j^{(8)}}$ represent a
complete set of operators of mass-dimensions $d = 6$ and $d =8 $,
and $c_i$, $b_j$ are the corresponding dimensionless Wilson coefficients.
Operators with $d = 5$ and $d = 7$ violate lepton and/or baryon
number conservation and are
not considered in this study.
The effective theory expansion in Eq.~\eqref{eq:smeft} is robust, fully
general, and can be systematically matched to explicit
UV-complete BSM scenarios.

The goal of the analysis is to constrain the
$d=6$ Wilson coefficients that correspond to operators that either
directly or indirectly impact Higgs boson couplings to SM
particles~\cite{Brivio:2017vri,ATL-PHYS-PUB-2019-042}.
Contributions of operators of mass-dimension
$d=8$, which are suppressed by $1/\Lambda^2$ relative to
the leading effects from $d=6$ operators and whose impact
on Higgs boson production and decay in the kinematic regions
of interest are not fully calculated, are not considered.
The `Warsaw' basis~\cite{Warsaw} is used, which
forms a complete set of all $\mathcal{O}_i^{(6)}$ operators in
Eq.~\eqref{eq:smeft} allowed for by the SM gauge symmetries. This basis is
widely used in EFT constraints in various fields of particle physics.

The `top' flavour symmetry scheme for EFT operators is assumed,
following the recommendation of Ref.~\cite{Aguilar-Saavedra:2018ksv}.
In this scheme, quarks of the first two generations and quarks of the 3rd
generation are described by independent fields, a $\text{U(}2)^3$ symmetry is
imposed on the light quarks, and the quark sector of the EFT is assumed to
have no mixing; all lepton generations are modelled independently.
Table~\ref{tab:CoeffOperators} lists the operators considered in the
analysis of STXS data, and their corresponding Wilson coefficients $c_i$.
Only CP-even $d=6$ operators for which the $\Lambda^{-2}$-suppressed
contribution to any of the STXS measurements shown in
Figures~\ref{fig:stxs_muplot_WW_ZZ}--\ref{fig:stxs_muplot_bb_tautau_mumu}
exceeds 0.1\% relative to the SM prediction at $c_i/\Lambda^2=1/\TeV^2$
are listed.
The EFT analysis of the differential data only considers a subset of the
coefficients listed, $\cHG, \ctH$ and $\ctG$, due to the more restricted
sensitivity of the differential analyses.

A value of $\Lambda = 1\,\TeV$ is assumed, coefficients for alternative values of $\Lambda=X$ can be
obtained through a scaling of the results presented in this
article by a factor $(X/1\,\TeV)^2$.
All Wilson coefficients are assumed to be real.

\begin{table}[btp]
\centering
\caption{Wilson coefficients $\vec{c}$ and corresponding $d=6$
SMEFT operators $\mathcal{O}_j$ used in this analysis. The operator and coefficient notation
follows the convention of Ref.~\cite{SMEFTsim}.
}
\label{tab:CoeffOperators}
\vspace{5pt}

%
%
%
%
%
%
%
%
%

\begin{tabularht}{16cm}{c c}
\toprule
Wilson coefficient & Operator \\
\hline\interrowfill
\cH        &  \scalebox{0.8}{\defH} \\ \interrowfill
\cHbox &  \scalebox{0.8}{\defHbox} \\ \interrowfill
\cG &  \scalebox{0.8}{\defG} \\ \interrowfill
\cW &  \scalebox{0.8}{\defW} \\ \interrowfill
\cHDD &  \scalebox{0.8}{\defHDD} \\ \interrowfill
\cHG &  \scalebox{0.8}{\defHG} \\ \interrowfill
\cHB &  \scalebox{0.8}{\defHB} \\ \interrowfill
\cHW &  \scalebox{0.8}{\defHW} \\ \interrowfill
\cHWB &  \scalebox{0.8}{\defHWB} \\ \interrowfill
\hline \interrowfill
\cHloneone &  \scalebox{0.8}{\defHlOneOne} \\ \interrowfill
\cHlonetwo &  \scalebox{0.8}{\defHlOneTwo} \\ \interrowfill
\cHlonethree &  \scalebox{0.8}{\defHlOneThree}\\ \interrowfill
\cHlthreeone &  \scalebox{0.8}{\defHlThreeOne} \\ \interrowfill
\cHlthreetwo &   \scalebox{0.8}{\defHlThreeTwo} \\ \interrowfill
\cHlthreethree &  \scalebox{0.8}{\defHlThreeThree} \\ \interrowfill
\cHeone &  \scalebox{0.8}{\defHeOne} \\ \interrowfill
\cHetwo &  \scalebox{0.8}{\defHeTwo} \\ \interrowfill
\cHethree &  \scalebox{0.8}{\defHeThree} \\ \interrowfill
\cHjone &  \scalebox{0.8}{\defHjOne} \\ \interrowfill
\cHjthree &  \scalebox{0.8}{\defHjThree} \\ \interrowfill
\cHu &  \scalebox{0.8}{\defHu} \\ \interrowfill
\cHd &  \scalebox{0.8}{\defHd} \\ \interrowfill
\cHQone &  \scalebox{0.8}{\defHQOne} \\ \interrowfill
\cHQthree &  \scalebox{0.8}{\defHQThree} \\ \interrowfill
\cHt &  \scalebox{0.8}{\defHt} \\
\cHb &  \scalebox{0.8}{\defHb} \\
\bottomrule
\end{tabularht}
\begin{tabularht}{16cm}{c c}
\toprule
Wilson coefficient & Operator \\
\hline  \interrowfill
\cQjoo & \scalebox{0.8}{\defQjoo} \\\interrowfill
\cQjoe & \scalebox{0.8}{\defQjoe}  \\\interrowfill
\cQjto & \scalebox{0.8}{\defQjto}  \\\interrowfill
\cQjte & \scalebox{0.8}{\defQjte}  \\\interrowfill
\cjjto &  \scalebox{0.8}{\defjjto} \\\interrowfill %
\ctuo & \scalebox{0.8}{\deftuo}  \\\interrowfill
\ctue & \scalebox{0.8}{\deftue} \\ \interrowfill %
\ctdo & \scalebox{0.8}{\deftdo} \\ \interrowfill %
\ctde & \scalebox{0.8}{\deftde} \\ \interrowfill %
\cQuo & \scalebox{0.8}{\defQuo}  \\\interrowfill
\cQue & \scalebox{0.8}{\defQue}  \\\interrowfill
\cQdo & \scalebox{0.8}{\defQdo}  \\\interrowfill
\cQde & \scalebox{0.8}{\defQde}  \\\interrowfill
\ctjo & \scalebox{0.8}{\deftjo}  \\\interrowfill
\ctje & \scalebox{0.8}{\deftje}  \\\interrowfill
\hline \interrowfill
\ceHtwo &  \scalebox{0.8}{\defeHTwo} \\ \interrowfill
\ceHthree &  \scalebox{0.8}{\defeHThree} \\ \interrowfill
\cuH &  \scalebox{0.8}{\defuH} \\ \interrowfill
\ctH &  \scalebox{0.8}{\deftH} \\ \interrowfill
\cbH &  \scalebox{0.8}{\defbH} \\ \interrowfill
\hline \interrowfill
\ctG &  \scalebox{0.8}{\deftG} \\ \interrowfill
\ctW &  \scalebox{0.8}{\deftW} \\ \interrowfill
\ctB &  \scalebox{0.8}{\deftB} \\   \interrowfill
\hline \interrowfill
\cllonetwoone &  \scalebox{0.8}{\defllOneTwoOne} \\
\bottomrule
\end{tabularht}


\end{table}

\subsubsection{Simulation of the impact of SMEFT operators}

The impact of the $d=6$ SMEFT operators listed in
Table~\ref{tab:CoeffOperators} are computed with two
universal  \textsc{FeynRules} output (UFO) models of
\madgraph~\cite{MadGraph} normalised to theoretical
calculations of production cross-sections and decay
branching ratios as follows.

Calculations for Higgs boson production modes with tree-level diagrams
are performed with  \textsc{SMEFTsim}~\cite{SMEFTsim} assuming the `top' flavour symmetry scheme, following the recommendation of
Ref.~\cite{Aguilar-Saavedra:2018ksv} and providing as input
$G_F = 1.1663787\times 10^{-5}~\GeV^{-2}$,
$m_Z = 91.1876~\GeV$,
and $m_W = 80.387~\GeV$,
where $G_F$ is the Fermi constant and $m_Z$ and $m_W$ are the
$Z$ and $W$ boson masses respectively.
Cross-sections are calculated at NLO accuracy in QCD for \ggF,
\ggZH{} and $H\to gg$ with SMEFTatNLO~\cite{SMEFTatNLO} and at NLO
EW accuracy for SMEFT-SM interference terms in
\hgg{}~\cite{Dedes:2018seb} and \hzg~\cite{Dedes:2019bew}.
Lowest-order calculations in QCD for $m_H=125.09$~\GeV\
are used for all other production and decay modes.
SMEFT modifications to the background processes in the included
analyses are not considered, but their impact on the SMEFT measurements
of the signal is expected to be small as most background distributions
are constrained by control regions.

In the simulation, the following requirements are applied at
particle level: the minimal
jet transverse momentum is $\pT > 20$~\GeV and an angular
separation $\Delta R>0.05$ between two jets or
two leptons is required, to avoid divergences
in the matrix element calculation.
For all events,
\pythia{} was used for the simulation of parton
showering, where the Higgs boson decay is based on the width
calculated by \madgraph\ for the EFT parameters in
question.
In \ggH\ events a matching was performed with the MLM
algorithm~\cite{Mangano:2006rw,Mrenna:2003if}
to remove phase space overlap between the jets
from the matrix element and the shower.

For analyses based on the STXS methodology, the  \textsc{Rivet} program~\cite{Buckley:2010ar} is used
to analyse the simulated events, where the computation of high-level kinematic quantities
and classification of the events according to their STXS region is performed by the
\texttt{HiggsTemplateCrossSections} in accordance with definitions in~\cite{deFlorian:2016spz}.\footnote{The  \textsc{Rivet} algorithm has been modified to classify events in which a Higgs boson and two leptons
arise from the same production vertex as VH production events.
This change ensures the proper classification of events with leptons
from off-shell $V$ decays, since \madgraph\ only saves on-shell
intermediate particles. Contributions from off-shell
$V$ bosons are small in the SM, but can be enhanced by SMEFT operators.}

The cross-section predictions for a specific process, calculated
as described above, are estimated as the sum of three terms:
\begin{equation}
\label{eq:EFT_crossSection}
\sigma_\text{SMEFT} = \sigma_\text{SM} + \sigma_\text{int} + \sigma_\text{BSM},
\end{equation}
\noindent where $\sigma_\text{SM}$ is the SM cross-section,
$\sigma_\text{int}$ describes the interference between the SMEFT
operators (BSM processes) and SM operators, and $\sigma_\text{BSM}$ is
the cross-section involving exclusively SMEFT operators.  When
considering only $d=6$ SMEFT operators, it follows from
Eq.~\eqref{eq:smeft} that $\sigma_\text{int}$ consists of terms
involving a single $d=6$ SMEFT operator, suppressing each term by a
factor $\Lambda^{-2}$, and that $\sigma_\text{BSM}$ contains terms
involving products of two $d=6$ SMEFT operators, suppressing each term by
a factor $\Lambda^{-4}$. For this reason, the impact of the
$\sigma_\text{BSM}$ term is generally expected to be small, though its
impact may still be non-negligible in certain regions of phase space,
e.g., when energy scales are of order $\Lambda$.

Additional terms suppressed by a factor $\Lambda^{-4}$ occur in the
general SMEFT expansion of Eq.~\eqref{eq:smeft},
i.e., the lowest-order cross-section terms generated by
$d=8$ operators that involve interference between SM and these
operators.
As only a subset of Higgs boson processes have been calculated with $d=8$
operators~\cite{Corbett:2021cil,Corbett:2023yhk,Martin:2023fad},
the list of terms suppressed by a factor $\Lambda^{-4}$ considered
is incomplete and the effect of the missing $d=8$ terms relative to
$\sigma_\text{BSM}$ is not known. However, their effect may be of
a magnitude comparable to that of products of two $d=6$ SMEFT operators,
as the suppression factor is at the same order.
For this reason, SMEFT interpretations are
presented both with and without the $\sigma_\text{BSM}$ contribution
to give a general indication of the sensitivity of the analysis to
$\Lambda^{-4}$ terms.

To reduce the perturbative QCD uncertainty in the extraction of the SMEFT Wilson
coefficients, the parameterisation of the predicted cross-section
of Eq.~\eqref{eq:EFT_crossSection}
is computed as a relative correction to the SM prediction computed
at the highest available order for each process:
\begin{equation}\label{eq:relCorrectionToSM_EFT}
\sigma_\text{SMEFT}^i = \sigma^{i\text{,((N)N)NLO}}_\text{SM}\times\left( 1 + \frac{\sigma^{i\text{,(N)LO}}_\text{int}}{\sigma^{i\text{,(N)LO}}_\text{SM}} + \frac{\sigma^{i\text{,(N)LO}}_\text{BSM}}{\sigma^{i\text{,(N)LO}}_\text{SM}} \right).
\end{equation}
\noindent This calculation strategy assumes that
higher-order terms have the same relative effect on $\sigma_\textrm{int}$ and $\sigma_\textrm{BSM}$ as on $\sigma_\textrm{SM}$~\cite{Degrande:2016dqg} and that the SMEFT operator predominantly modifies
the production process $i$ under consideration.

The predictions are further modified by the impact of SMEFT
operators on Higgs boson decay branching ratios. Since the Higgs boson is
a narrow, scalar particle, and only on-shell production is considered
in this analysis, its production cross-section and decay width
factorise. The impact of SMEFT operators on production and decay
therefore also factorises and can be derived independently.  Thus, the
cross-section for a given Higgs boson production process $i$ in particle-level
region $k^\prime$ and for a given decay mode $H\to X$ is
\begin{equation}
(\sigmaTimesBR)_\mathrm{SMEFT}^{i,k^\prime,H\to X}  =  \sigma^{i,k^\prime}_\mathrm{SMEFT} \times \BR^{H\to X}_\mathrm{SMEFT} = \left(\sigma^{i,k^\prime}_\text{SM} + \sigma^{i,k^\prime}_\text{int} + \sigma^{i,k^\prime}_\text{BSM}\right) \times
\left( \frac{\Gamma^{H\to X}_\text{SM} + \Gamma^{H\to X}_\text{int} + \Gamma^{H\to X}_\text{BSM}}{\Gamma^{H}_\text{SM} + \Gamma^{H}_\text{int} + \Gamma^{H}_\text{BSM}} \right).
\nonumber
\end{equation}

\noindent The factorised SMEFT prediction is calculated with ratios as in
Eq.~\eqref{eq:relCorrectionToSM_EFT} to utilise the SM prediction at the
highest available order:
\begin{equation}
(\sigmaTimesBR)_\mathrm{SMEFT}^{i,k^\prime,H\to X}  = (\sigmaTimesBR)^{i,k^\prime,H\to X}_\text{SM,(N(N))NLO} \left( 1 + \frac{\sigma^{i,k^\prime}_\text{int,(N)LO}}{\sigma^{i,k^\prime}_\text{SM,(N)LO}} + \frac{\sigma^{i,k^\prime}_\text{BSM,(N)LO}}{\sigma^{i,k^\prime}_\text{SM,(N)LO}} \right)
\left( \frac{1 + \frac{\Gamma^{H\to X}_\text{int}}{\Gamma^{H\to X}_\text{SM}} + \frac{\Gamma^{H\to X}_\text{BSM}}{\Gamma^{H\to X}_\text{SM}}}{1 + \frac{\Gamma^{H}_\text{int}}{\Gamma^{H}_\text{SM}} + \frac{\Gamma^{H}_\text{BSM}}{\Gamma^{H}_\text{SM}}} \right),
\label{eq:sigmaTimesBR_eft}
\end{equation}

\noindent where the ratios $\sigma_\text{int}/\sigma_\text{SM}$ and
$\Gamma_\text{int}/\Gamma_\text{SM}$ have a linear dependence on SMEFT
operators and are suppressed by a factor $\Lambda^{-2}$, and the ratios
$\sigma_\text{BSM}/\sigma_\text{SM}$ and
$\Gamma_\text{BSM}/\Gamma_\text{SM}$ have a quadratic dependence on
SMEFT operators and are suppressed by a factor $\Lambda^{-4}$.
In the analysis, these ratios are parameterised as
\begin{eqnarray}\label{eq:EFTparam_production1}
\frac{\sigma^{i,k^\prime}_{\text{int}}}{\sigma^{i,k^\prime}_\text{SM}} = \sum\limits_j A^{\sigma_{i,k^\prime}}_{j} c_j  ~~~~~& &  \frac{\sigma^{i,k^\prime}_{\text{BSM}}}{\sigma^{i,k^\prime}_\text{SM}} = \sum\limits_{j,l\ge j} B^{\sigma_{i,k^\prime}}_{jl} c_j c_l \\
\frac{\Gamma^{H\to X}_{\text{int}}}{\Gamma^{H\to X}_\text{SM}} = \sum\limits_j A^{\Gamma^{H\to X}}_{j} c_j & &  \frac{\Gamma^{H\to X}_{\text{BSM}}}{\Gamma^{H\to X}_\text{SM}} = \sum\limits_{j,l\ge j} B^{\Gamma^{H\to X}}_{jl} c_j c_l \\
\frac{\Gamma^{H}_{\text{int}}}{\Gamma^{H}_\text{SM}} = \sum\limits_j A^{\Gamma^{H}}_{j} c_j  & &  \frac{\Gamma^{H}_{\text{BSM}}}{\Gamma^{H}_\text{SM}} = \sum\limits_{j,l\ge j} B^{\Gamma^{H}}_{jl} c_j c_l,
\label{eq:EFTparam_production3}
\end{eqnarray}
\noindent with
\begin{eqnarray}
\label{eq:EFTtotalwidth}
A_j^{\Gamma^H}   =  \frac{\sum\limits_X{\Gamma_\mathrm{SM}^{H\to X}A_j^{\Gamma^{H\to X}}}}{\sum\limits_X{\Gamma_\mathrm{SM}^{H\to X}}} ~~~~~ & &
B_{jl}^{\Gamma^H}  =  \frac{\sum\limits_X{\Gamma_\mathrm{SM}^{H\to X}B_{jl}^{\Gamma^{H\to X}}}}{\sum\limits_X{\Gamma_\mathrm{SM}^{H\to X}}}.
\end{eqnarray}
\noindent In Eq.~\eqref{eq:EFTtotalwidth},
the sum includes all possible final states $X$ with up to four
elementary particles produced by one of the following Higgs boson
decays: $H\to f\bar{f}$ ($f=c,b,\mu,\tau$), $H\to\gamma\gamma$,
$H\to Z\gamma$, and $H\to VV^*$ ($V=W,Z$).
All $A_j^{\sigma_{i,k^\prime}}$, $A_j^{\Gamma^{H\to X}}$,
$B_{jl}^{\sigma_{i,k^\prime}}$ and $B_{jl}^{\Gamma^{H\to X}}$
coefficients are constant factors obtained
from simulation that express the sensitivity of the process to the
operators $\mathcal{O}_{j}$  and $\mathcal{O}_{l}$
that correspond to the Wilson coefficients
$c_{j}$ and $c_l$, where the indices $j,l$ run over all non-negligible operators.
The coefficients $A_j$ and $B_{jl}$ are proportional
to $\Lambda^{-2}$ and $\Lambda^{-4}$ respectively.
The values of all SM Higgs boson branching ratios that enter
Eq.~\eqref{eq:EFTtotalwidth} are taken from Ref.~\cite{deFlorian:2016spz}.

For the interpretation of the differential analyses, where the measured
parameters $\mu^{k^\prime,X}$ do not label individual
production processes, the relative fractions of the different processes
are predicted according to the SMEFT calculations as in
Eq.~\eqref{eq:relCorrectionToSM_EFT}.

Using Eqs.~\eqref{eq:sigmaTimesBR_eft}--\eqref{eq:EFTparam_production3},
two statistical models are constructed for the interpretation of the
data: a linearised variant that only considers terms suppressed
by up to a factor $\Lambda^{-2}$, and a variant including quadratic terms
with suppression factor $\Lambda^{-4}$.

\subsubsection{Cross-section calculation with linear terms}

In a scenario where $\Lambda^{-4}$-suppressed contributions are
ignored, the predicted deviation of the cross-section, partial width
and total width from their SM values can each be explicitly linearised
as a function of the Wilson coefficients $\vec{c}$. Ignoring all
$\Lambda^{-4}$-suppressed BSM terms in Eq.~\eqref{eq:sigmaTimesBR_eft},
and using the parameterisation of
Eqs.~\eqref{eq:EFTparam_production1}--\eqref{eq:EFTparam_production3}, the
expression for the cross-section times branching ratio reduces to
\begin{eqnarray}
(\sigmaTimesBR)_\mathrm{SMEFT}^{i,k^\prime,H\to X}  &=& (\sigmaTimesBR)^{i,k^\prime,H\to X}_\text{SM,((N)N)NLO} \times \left( 1 + \frac{\sigma^{i,k^\prime}_\text{int,(N)LO}}{\sigma^{i,k^\prime}_\text{SM,(N)LO}}  \right) \times
\left( \frac{1 + \frac{\Gamma^{H\to X}_\text{int}}{\Gamma^{H\to X}_\text{SM}}}{1 + \frac{\Gamma^{H}_\text{int}}{\Gamma^{H}_\text{SM}}} \right) \nonumber \\
&=& (\sigmaTimesBR)^{i,k^\prime,H\to X}_\text{SM,((N)N)NLO} \times \left( 1 +  \sum\limits_j A^{\sigma_{i,k^\prime}}_{j} c_j   \right) \times
\left( \frac{1 + \sum\limits_j A^{\Gamma^{H\to X}}_{j} c_j }{1 + \sum\limits_j A^{\Gamma^{H}}_{j} c_j} \right), \nonumber \\
&=& (\sigmaTimesBR)^{i,k^\prime,H\to X}_\text{SM,((N)N)NLO} \times \left( \frac{ 1 +  \sum\limits_j \left( A^{\sigma_{i,k^\prime}}_{j} + A^{\Gamma^{H\to X}}_{j} \right) c_j + O\left(\Lambda^{-4}\right) }{1 + \sum\limits_j A^{\Gamma^{H}}_{j} c_j + O\left(\Lambda^{-4}\right) } \right), \label{eq:eftlinmodel}
\end{eqnarray}

\noindent where all higher order terms in the expansion are
suppressed by power $\Lambda^{-4}$ or beyond.

A subsequent Taylor expansion of the width ratio expression of
Eq.~\eqref{eq:eftlinmodel} and truncation of terms beyond $\Lambda^{-2}$
would result in a completely linearised expression for
$\sigmaTimesBR$.
However, as the parametric dependence of a Wilson
coefficient in a branching fraction $f$ is effectively of the form
$c/(1 + f \cdot c)$, the linearity in $c$ of this expression assumed
for small values of $c$ does not hold for large values of $c$.
This effect is particularly pronounced for operators
affecting Higgs boson decays with a large branching fraction, e.g., \cbH,
and operators that have a large measured uncertainty that allows large values of $c$ in the error propagation.
For this reason, a full Taylor expansion of the ratio is not performed.

\subsubsection{Cross-section calculation including quadratic terms}

The SMEFT prediction including the available terms proportional to
$\Lambda^{-4}$ is:
\begin{eqnarray}
(\sigmaTimesBR)_\mathrm{SMEFT}^{i,k^\prime,H\to X}  &=& (\sigmaTimesBR)^{i,k^\prime,H\to X}_\text{SM,((N)N)NLO} \scriptstyle \left( 1~+~\sum\limits_j A^{\sigma_{i,k^\prime}}_{j} c_j~+~\sum\limits_{j,l\ge j} B^{\sigma_{i,k^\prime}}_{jl} c_j c_l \right) \left( \frac{1~+~\sum\limits_j A^{\Gamma^{H\to X}}_{j} c_j ~+~ \sum\limits_{j,l\ge j} B^{\Gamma^{H\to X}}_{jl} c_j c_l}{1 ~+~  \sum\limits_j A^{\Gamma^{H}}_{j} c_j ~+~ \sum\limits_{j,l\ge j} B^{\Gamma^{H}}_{jl} c_j c_l} \right), \nonumber\\
& = & (\sigmaTimesBR)^{i,k^\prime,H\to X}_\text{SM,((N)N)NLO}  \cdot \nonumber \\
&& \hspace{-0.7cm} \scriptstyle \Biggl( \frac{ 1 ~+~  \sum\limits_j \left( A^{\sigma_{i,k^\prime}}_{j} ~+~ A^{\Gamma^{H\to X}}_{j} \right) c_j ~+~ \sum_{j,l} \Big(A_j^{\sigma_{i,k^\prime}} A_l^{\Gamma_{H \to X}}\Big) c_jc_l ~+~ \sum_{j,l\geq j} \Big( B_{jl}^{\sigma_{i,k^\prime}} ~+~ B_{jl}^{\Gamma_{H \to X}} \Big)c_j c_l ~+~ O\left(\Lambda^{-6}\right)}{1 ~+~ \sum\limits_j \left(A^{\Gamma^{H}}_{j}\right) c_j ~+~ \sum\limits_{j,l\geq j}\left(B_{jl}^{\Gamma_H}\right) c_jc_l ~+~ O\left(\Lambda^{-6}\right)} \Biggl)  \label{eq:eftquadmodel}
\end{eqnarray}

\noindent where both the numerator and denominator are a second-order Taylor
expansion resulting in a linearised expression for terms of
order $\Lambda^{-2}$ and a quadratic  expression for terms of
order $\Lambda^{-4}$.
Similarly to Eq.~\eqref{eq:eftlinmodel}, and for the same reason, the width
ratio expression in Eq.~\eqref{eq:eftquadmodel} is not subjected to a
further Taylor expansion.
While the set of operators proportional to $\Lambda^{-4}$ considered
in Eq.~\eqref{eq:eftquadmodel} is not complete due to missing
$d=8$ operators, as noted earlier a comparison of results obtained
with the linear statistical model and the model including quadratic terms is indicative of
the sensitivity of the measurements to terms suppressed by $\Lambda^{-4}$.

Figure~\ref{fig:aij_impact_diff} illustrates the impact of the
linear terms $A_j$
for the three SMEFT operators considered in the
interpretation of the fiducial differential cross-section
measurements, whereas
Figure~\ref{fig:bij_impact}
shows the impact of linear terms and the impact including quadratic terms on production
($A^{\sigma_{i,k^\prime}}_j$) and decay $\left(\frac{1+A^{\Gamma^{H\to
X}}_j}{1+A^{\Gamma^H}_j}-1\right)$
rates.

\begin{figure}[hbt]
\centering
\includegraphics[width=1.0\textwidth]{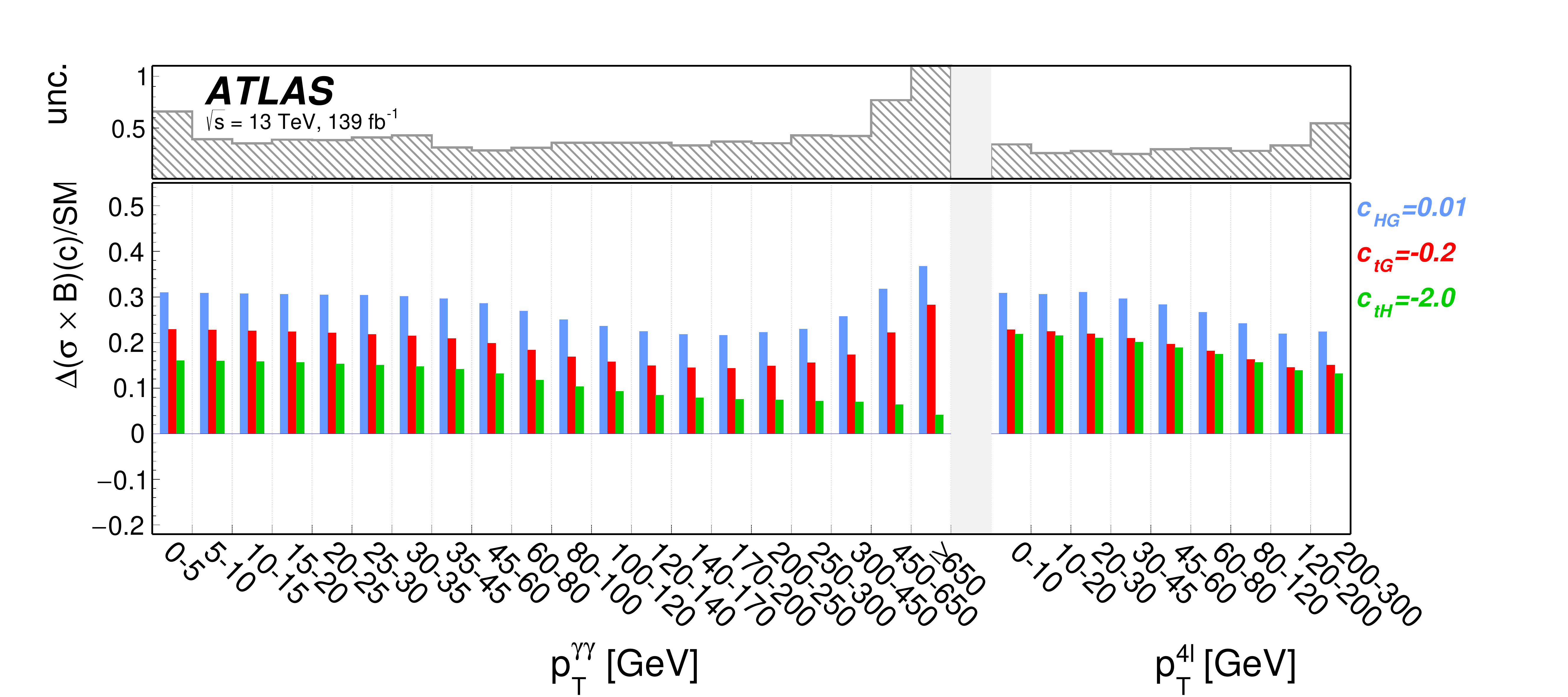}

\caption{
Expected impact of the SMEFT operators on the fiducial
differential distributions of the transverse momentum of the Higgs
boson decay products in \hgg~and \hzz~decays, relative to
the SM cross-section, assuming the linearised SMEFT model.
The values of the Wilson coefficients, specified in the legend, are
chosen to show the distribution
of the operator impact in the same range as the typical uncertainty
of the measurement.
To judge the experimental sensitivity to constrain the operators from
the data in the listed fiducial regions, the total uncertainty
in the measurement in each region is shown in the top panel.
For presentational clarity, the uncertainty of low
precision regions is clipped off in the plot.
The impact of these three operators in the STXS analysis is shown
in matching colours in the bottom panel of
Figure~\ref{fig:bij_impact} for coefficient variations of
identical magnitude.
}
\label{fig:aij_impact_diff}
\end{figure}

\begin{figure}[!hbt]
\centering
\includegraphics[width=1.0\textwidth]{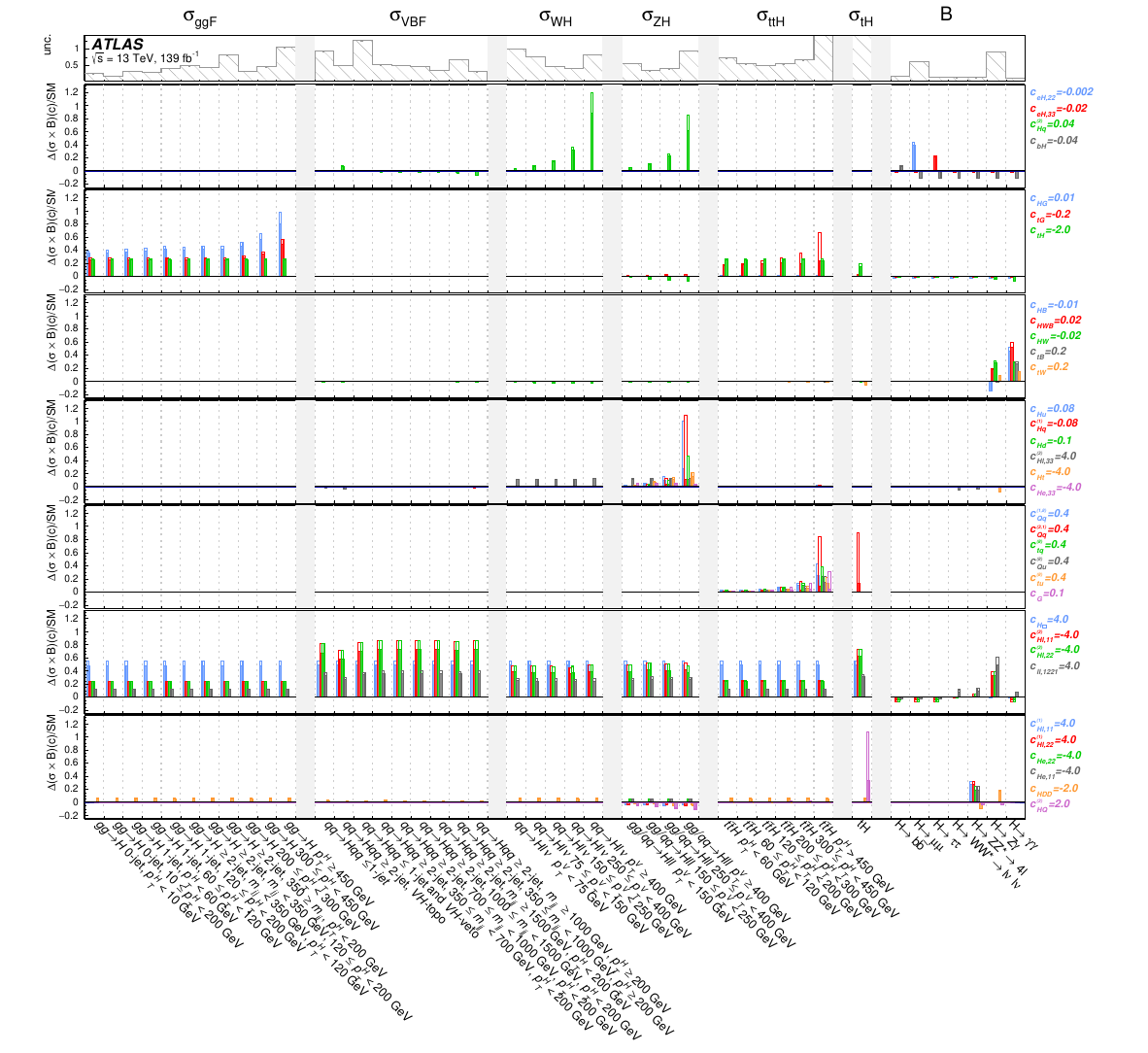}
\caption{Expected impact of the most relevant SMEFT operators on the
observables $O$ (cross-sections in STXS-1.2 regions or decay branching
ratios), relative to the SM predictions,
for the linearised SMEFT model (shaded histogram) and the
SMEFT model including quadratic terms (open histogram).
The values of the Wilson coefficients, specified in the legend, are
chosen to show the distribution
of the operator impact in the same range as the typical uncertainty
of the measurement.
To judge the experimental sensitivity to constrain the operators
from the data, the total
uncertainty in the measurement of each observable is
shown in the top panel.
For presentational clarity, the uncertainty of low
precision measurements is clipped off in the plot.
}
\label{fig:bij_impact}
\end{figure}

The relative importance of the quadratic term increases linearly with the
considered variation of the corresponding Wilson coefficient.
Figure~\ref{fig:bij_impact} shows that the quadratic terms have the
highest impact in the high-$\pT$ regions of $VH$ production
for the coefficients \cHq[3], \cHq[1], \cHd, and \cHu,
on high-$\pT$ (inclusive) $t\bar{t}H$ production for
the coefficients \ctG, \cQjoe, \cQjto, \ctje, \cQue, \ctue,
(\cG) and in the $tH$ channel for
the coefficients $\cHQthree$ and \cQjto.

\subsubsection{Assumptions on SM Higgs boson kinematics}

With the insertion of the SMEFT cross-section predictions of
Eqs.~\eqref{eq:eftlinmodel} or~\eqref{eq:eftquadmodel} in
Eqs.~\eqref{eq:Lsk} and~\eqref{eq:Lsk_diff} describing the expected
signal yield $s_k$ for each analysis region $k$, i.e.,
\begin{align*}
\mu^{i,k^\prime,X} \to  \frac{(\sigmaTimesBR)_{\text{SM}+\Lambda^{-2,-4}}^{i,k^\prime,H\to X}}{(\sigmaTimesBR)^{i,k^\prime,H \to X}_\text{SM,((N(N))NLO}} & \textrm{~in Eq.~\eqref{eq:Lsk}}, \\
\mu^{k^\prime,X} \to  \frac{\sigma_{\text{SM}+\Lambda^{-2,-4},\text{fid.}}^{k^\prime,H\to X}}{\sigma^{k^\prime,H \to X}_\text{SM,fid.}} & \textrm{~in Eq.~\eqref{eq:Lsk_diff}}, \\
\end{align*}
\noindent the effect of SMEFT cross-section modifications on the
inclusive Higgs boson signal yield for each particle-level region
$k^\prime$ is fully taken into account for every reconstruction-level
region $k$.
However, the SMEFT operators may also affect other terms of the
likelihood, notably the efficiency times acceptance factors
$\epsilon_k^{i,X,k^\prime}$ occurring in Eqs.~\eqref{eq:Lsk}
and~\eqref{eq:Lsk_diff}, as well as the signal yield distribution
inside each region $k$ represented by the factor $f_s^{r,k}$ occurring
in Eq.~\eqref{eq:Lbasic}.

In both the linear SMEFT model and the model including quadratic terms, it is assumed
that the theoretical systematic uncertainties assigned to the acceptance factors $\epsilon_k^{i,X,k^\prime}$,
which are fully taken into account in the SM cross-section prediction, are large compared to the effect of the possible acceptance
changes induced by SMEFT operators in Higgs boson production through the full validity range of the SMEFT model,
and that this additional SMEFT related theory uncertainty can therefore be neglected.

This assumption is motivated by the similarity of reconstruction-level
analysis regions to STXS and differential fiducial regions, which are
designed to be relatively insensitive to acceptance changes induced by
SMEFT operators. In addition, the rather fine binning of the
Stage-1.2 STXS regions reduces further the possibility of significant
variations of the acceptance factors inside each region.
The effect of SMEFT operators on other observables
used in the definition of the analysis regions, e.g., through
multivariate discriminants, is assumed to be negligible.

Conversely, the effect of SMEFT operators on Higgs boson decays can
strongly affect the acceptance factors $\epsilon_k^{i,X}$, since
decays are not limited to a restricted fiducial phase space in the
STXS framework.
For 2-body decays, such acceptance effects are generally
small~\cite{HIGG-2020-16}.
However, the four-body $\hfourl$ and $\hww \to \ell \nu \ell \nu$
decays are significantly impacted by acceptance effects, hence for
these decays the impact factors $A_j^{H \to X}$ and $B_{jl}^{H \to X}$
are recalculated including an approximate implementation of the
reconstruction-level requirements at the particle level.
The impact of acceptance effects is evaluated for all operators in
Table~\ref{tab:CoeffOperators} that can affect $\Gamma^{H \to ZZ^*}$ or
$\Gamma^{H \to WW^*}$, i.e., \cHbox{}, \cHDD{}, \cHW{}, \cHB{},
\cHWB{}, \cHloneone{}, \cHlonetwo{}, \cHlthreeone{}, \cHlthreetwo{},
\cHeone{}, \cHetwo{} and \cllonetwoone{}, and is observed to be
non-negligible for four of these,  \cHW{}, \cHB{}, \cHWB{},
and \cHl3{}. For these, the corresponding recalculated factors $A_j^{H \to
X}, B_{jl}^{H \to X} (X=WW^*,ZZ^*)$ with approximate acceptance modelling
are used in the STXS SMEFT analysis.
Figure~\ref{fig:acc_wwzz} illustrates the effect of the \cHW\ operator
on the acceptance in the \hww~and \hzz~decay modes.
The acceptance correction is not used in the differential analysis of
\hzz, as none of the three operators analyzed for that measurement are
affected. For all other Higgs boson decays, the acceptance effects from
SMEFT operators are neglected.

\begin{figure}[!hbt]
\centering
\subfloat[]{\includegraphics[width=0.50\textwidth]{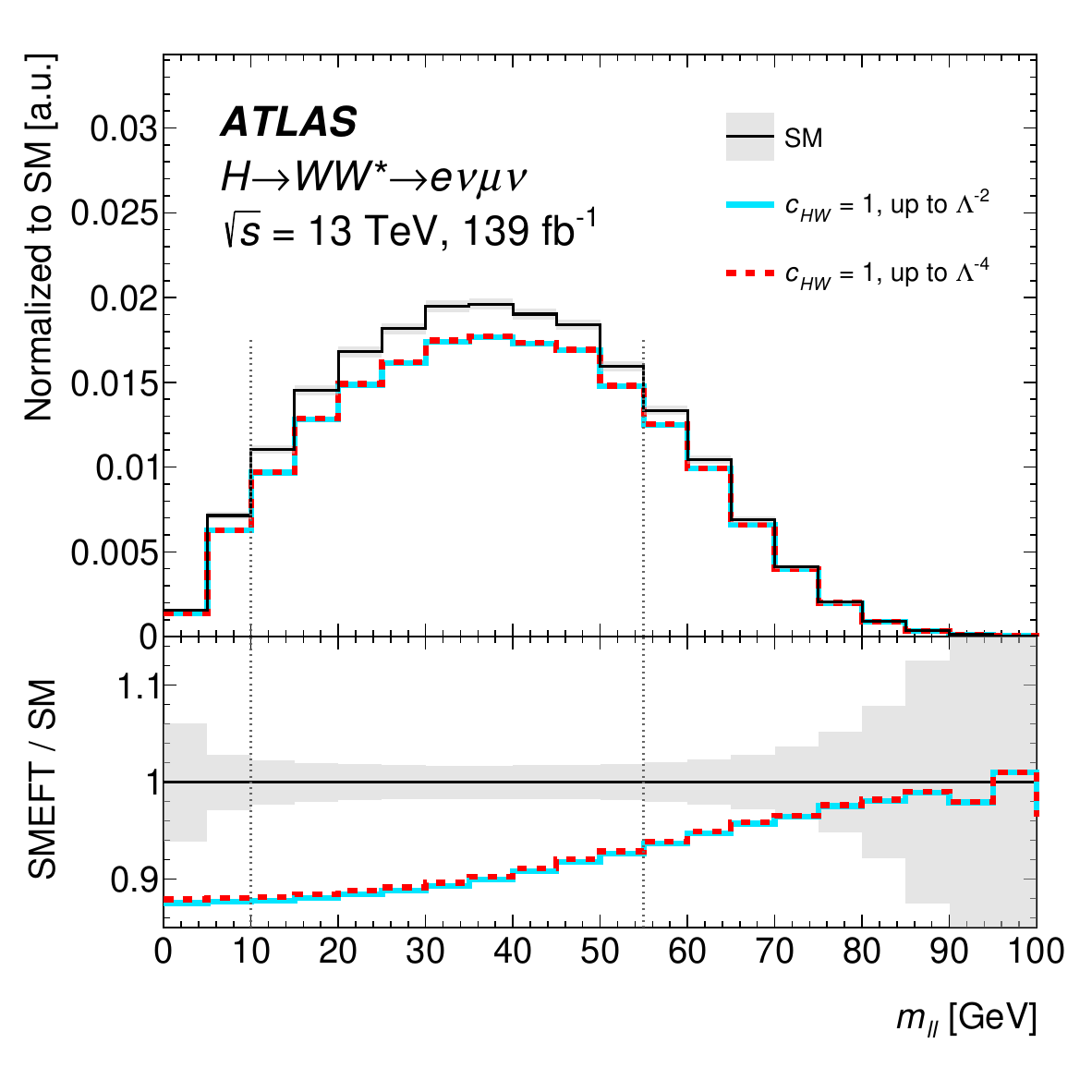}}
\subfloat[]{\includegraphics[width=0.50\textwidth]{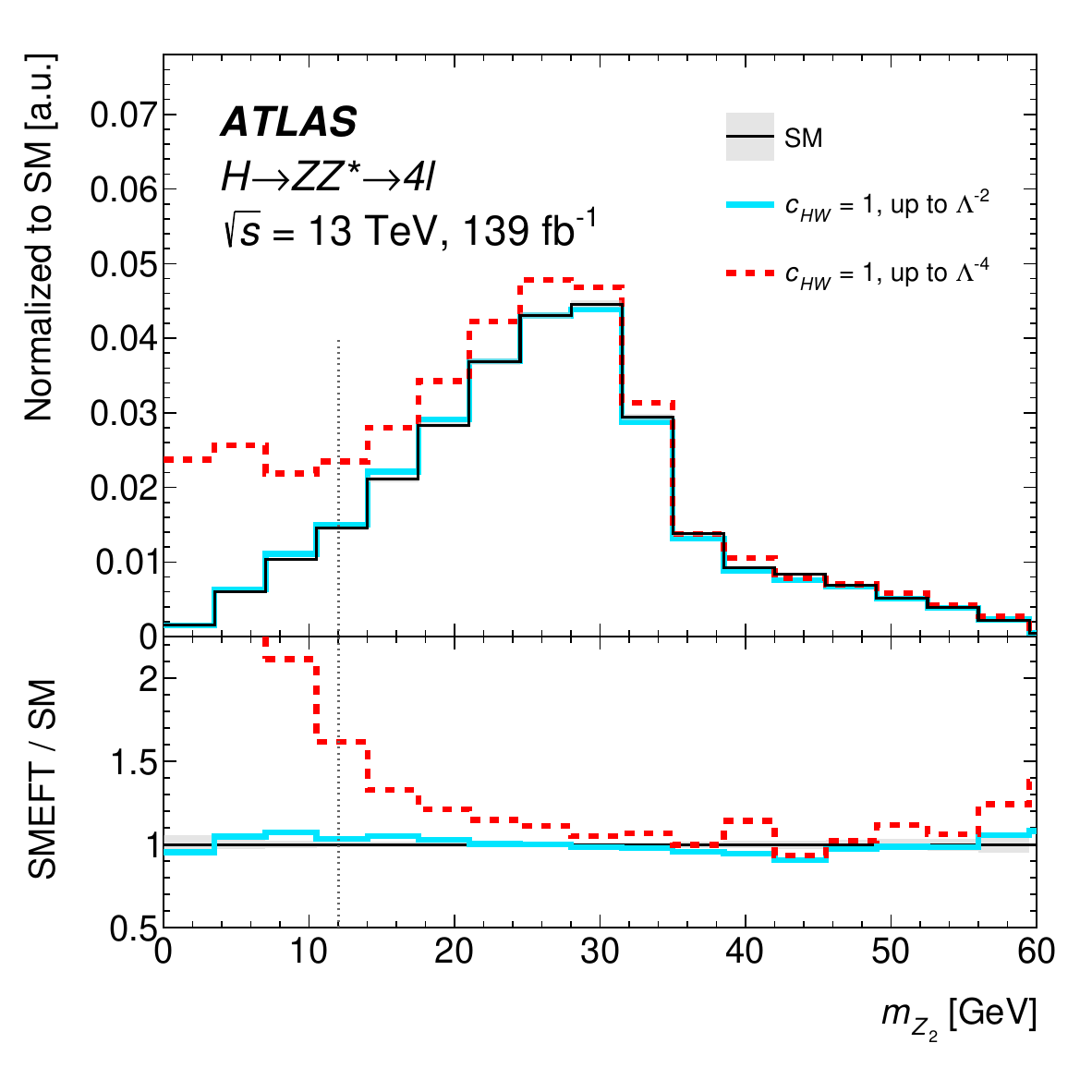}}
\caption{Illustration of the effect of a variation of the SMEFT
coefficient \cHW\ on kinematic observables used in the event
selection: (a) impact on the dilepton invariant mass in the
$H\to WW^*\to e\nu\mu\nu$ analysis, and
(b) impact on the invariant mass of the 3rd
and 4th \pT-ranked lepton in the $H\to ZZ^* \to 4\ell$ analysis.
The vertical dotted lines indicate the selection criteria applied
to the corresponding variables by the $H\to WW^*$ analysis
($10 < m_{\ell\ell} < 55$~\GeV) and by the $H\to ZZ^*$ analysis
($m_{Z_2}> 12$~\GeV).
The shaded bands show the statistical uncertainty of the SMEFT
simulation.
The value of the Wilson coefficient ($c_{HW}=1$) is significantly
larger than the sensitivity of this study, and is chosen
for illustration purposes.
}
\label{fig:acc_wwzz}
\end{figure}

Finally, if the signal yield distribution inside an analysis region $k$
deviates from the SM distribution encoded in $f_s^{r,k}$ in
Eq.~\eqref{eq:Lbasic}, the measured yield $s_k$ may be biased.
This effect is expected to be negligible for all regions where the
discriminant observable inside the region is the reconstructed invariant
mass of the Higgs boson, which is largely unaffected by SMEFT operators.
Regions that feature a multivariate discriminant as the observable can
potentially be more affected by SMEFT induced shape differences, but are
only used in regions with low signal yields
where it is assumed that any effect induced by signal shape deviations is
small compared to the statistical uncertainties.
For the high-yield \hww\ channel, where the transverse mass $m_\mathrm{T}$
is used as discriminant observable, which is sensitive to SMEFT operators,
it is verified that the effect of shape variations induced by the
SMEFT operators is negligible for the range of coefficient values that
are consistent with the observed measurement.


%

%
%

%
\subsection{Constraints from STXS measurements}
\label{sec:eft-results}

The aim of the EFT analysis is to obtain constraints on the Wilson
coefficients $c_j$ through a maximum likelihood analysis of the Higgs boson
STXS measurements shown in Figures~\ref{fig:stxs_muplot_WW_ZZ}--\ref{fig:stxs_muplot_bb_tautau_mumu}. All signal
strength modifiers $\mu^{i,k^\prime,X}$ of Eq.~\eqref{eq:Lsk} are directly expressed in terms of the coefficients $c_j$:
\begin{equation}
s_k(\vec{c},\vec{\theta})
=  \sum\limits_{i,k^\prime,X} \left( \mu^{i,k^\prime,X} \equiv \frac{(\sigmaTimesBR)_{\text{SMEFT}}^{i,k^\prime,H\to
X}(\vec{c})}{(\sigmaTimesBR)_\text{SM,(N(N))NLO}^{i,k^\prime,H\to
X}}  \right) \times \mathcal{L} \times
(\sigmaTimesBR)_\text{SM,(N(N))NLO}^{i,k^\prime, H \to X}(\vec{\theta}) \times \epsilon^{i,k^\prime,X}_k(\vec{\theta}),
\nonumber
\end{equation}
\noindent with $\mathcal{L},\epsilon^{i,k^\prime,X}_k(\vec{\theta})$ and $\vec{\theta}$ as defined in Eq.~\eqref{eq:Lsk}, and
where the signal cross-section
$(\sigmaTimesBR)^{i,k^\prime,H\to X}_\text{SMEFT}(\vec{c})$ is either taken
from the linear model of Eq.~\eqref{eq:eftlinmodel} or the
model including quadratic terms of Eq.~\eqref{eq:eftquadmodel}.

The available data samples contain insufficient information to constrain
all coefficients $\vec{c}$ listed in Table~\ref{tab:CoeffOperators}
simultaneously. A standard numerical joint maximum likelihood estimation of this set of parameters $\vec{c}$ will
not converge due to the presence of flat directions in the parameter space, corresponding to the
unconstrained degrees of freedom.

As degrees of freedom left unconstrained by the data do not necessarily
correspond to individual coefficients $c_j$, but may also be linear
combinations $\sum_j a_j c_j$, a modified basis is defined and used
in the following, to ensure the convergence of the numerical likelihood
maximisation procedure for the largest number of degrees of freedom
that can be probed by the current set of measurements.

\subsubsection{Sensitivity estimate and choice of parameters}
\label{sec:eft-details}

To determine a modified basis $\vec{c^{\prime}}$ that can be estimated from the
data, the SM expected covariance matrix $V_\text{STXS}$ of the measurement,
expressed in the STXS basis $\{\mu^{i,k^\prime, X}\}$,
is analysed.
From $V_\text{STXS}$, the Hessian matrix $V^{-1}_\text{STXS}$ is obtained,
and then rotated to the SMEFT basis $\{c_j\}$:
\begin{equation}
V^{-1}_\text{SMEFT} = P^T_{(i,k^\prime,X)\to(j)}\,V^{-1}_\text{STXS}\,P_{(i,k^\prime,X)\to(j)}.
\label{eq:stxs_rot}
\end{equation}
\noindent The expected Jacobian matrix $P_{(i,k^\prime,X)\to(j)}$
is based on the linearised SMEFT model of Eq.~\eqref{eq:eftlinmodel}
assuming small $c_i$:
\begin{equation}
P_{(i,k^\prime,X)\to(j)} = A^{\sigma_{i,k^\prime}}_j + A^{\Gamma^{H\to X}}_j-A^{\Gamma^H}_j,
\label{eq:pmatrix}
\end{equation}
\noindent where $A^{\sigma_{i,k^\prime}}_j,A^{\Gamma^{H\to X}}_j$ and
$A^{\Gamma^H}_j$ are the constant factors obtained from the
simulation.%

In the limit of Gaussian STXS measurements, the matrix
$V^{-1}_\text{SMEFT}$ represents the Fisher information matrix of its
linearised SMEFT model re-parameterisation, with the additional caveat that for large coefficient variations the representation of $V^{-1}_\text{SMEFT}$
may be suboptimal, as a consequence of the fully linearized form of
Eq.~\eqref{eq:pmatrix} that is only valid in the limit of small
coefficient values.
An eigenvalue decomposition of the Fisher
information matrix yields the eigenvectors $ev^{[i]}$ and their
corresponding eigenvalues $\lambda_i$
and variances $1/\lambda_i$.

Figure~\ref{fig:EVofEFTcovMatrix1} lists the eigenvectors
obtained from the expected measurements
accounting for the observed values of nuisance parameters, ranked by
eigenvalue and truncated to eigenvalues $\lambda_i \geq 0.01$,
corresponding to a
truncation at an estimated uncertainty of $\sigma(ev^{[i]})\leq 10$, well
beyond the  validity range of the Wilson coefficients of
$\mathcal{O}(1)$.

\begin{sidewaysfigure}[!htbp]
\includegraphics[width=1.02\textwidth]{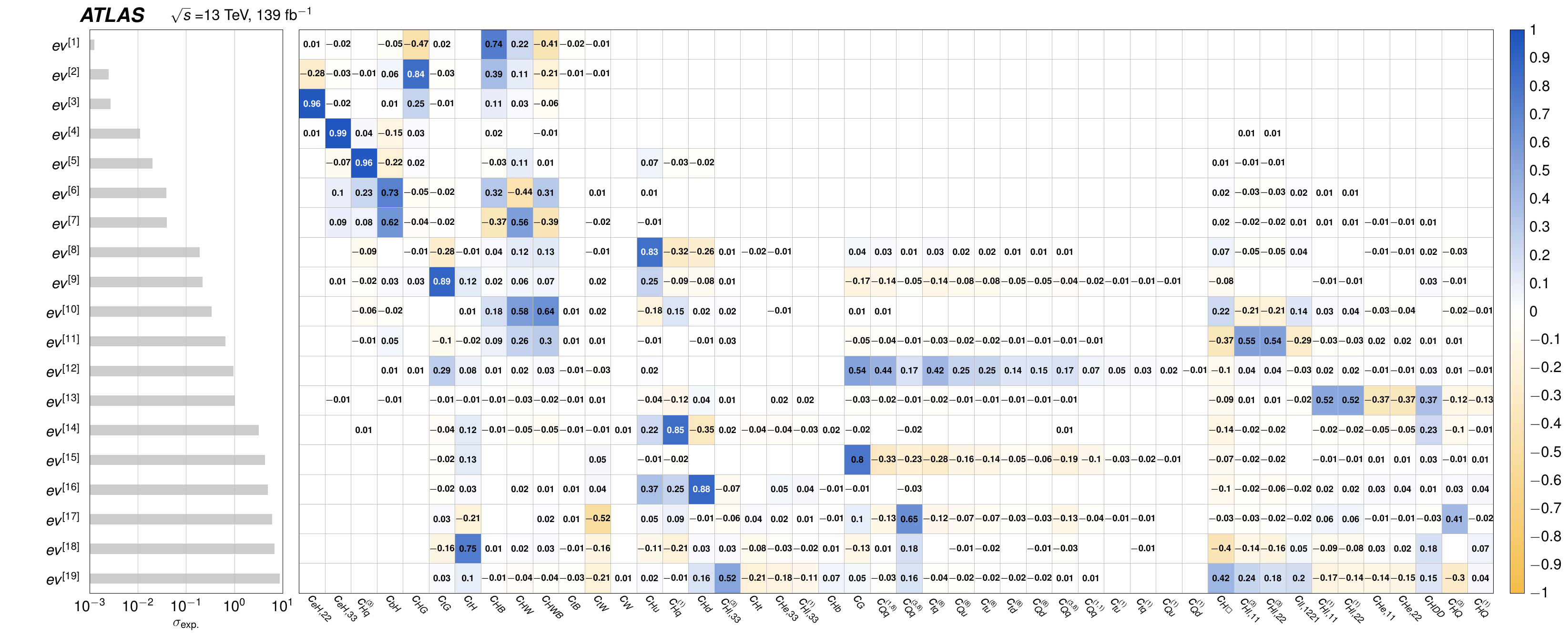}
\caption{
Eigenvectors $ev^{[i]}$ of the inverse EFT covariance matrix obtained by
propagating the SMEFT parameterisation to the covariance matrix
$V_\text{STXS}$ and requiring a significant
($\geq 0.01$) eigenvalue $\lambda$.
The eigenvectors are linear combinations $\sum A_{ij} c_j$ of the Wilson
coefficients $c_j$ shown at the bottom of the right panel. The elements
of the matrix shown in this panel correspond to the values of the $A_{ij}$
coefficients. Values of $|A_{ij}|<0.01$ are not shown.
The corresponding expected uncertainty
$\sigma_\mathrm{exp.} = 1/\sqrt{\lambda}$
for each eigenvector is shown in the left panel.
}
\label{fig:EVofEFTcovMatrix1}
\end{sidewaysfigure}

From the ranking shown in Figure~\ref{fig:EVofEFTcovMatrix1} and a survey of
the sensitivity of the STXS regions to the Wilson coefficients in the
linearised model, as shown in Figure~\ref{fig:bij_impact}, the
following observations are made:

\begin{itemize}
\item The coefficients $c_{eH,33}$, $c_{eH,22}$, and $c_{bH}$,
representing the Yukawa coupling modifiers of the $\Htt$, $\Hmm$,
and $\Hbb$
decays respectively, can be individually measured
from the corresponding Higgs boson channels that enter the
combination.
There are weak correlations of these operators with other SMEFT
operators,
primarily for $c_{bH}$, through their
contribution to the total Higgs boson width.
\item The coefficient $c_{Hq}^{(3)}$ affects both the $WH$ and $ZH$
production modes with an increasing impact as a function of the
transverse momentum $\pT^V$ of the $W$ and $Z$ bosons,
and is constrained almost exclusively by the $VH$,
$H\to b\bar{b}$ analysis.
\item The coefficients $c_{HG}$, $c_{tG}$ and $c_{tH}$ are constrained by
\ggF\ and $t\bar{t}H$ production. As the uncertainties of the constrained
directions span more than two orders of magnitude, it is beneficial to
represent them by decorrelated parameters, even though the rotation
matrix required to decorrelate them is close to an identity matrix.
\item The coefficients $c_{HB}$, $c_{HW}$, $c_{HWB}$, $c_{tB}$
and $c_{tW}$ are constrained in two directions through their
impacts on the branching ratios of the $H\to\gamma\gamma$ and
$H\to Z\gamma$ decays.
There is an additional small impact of these
operators on $\VBF$ and $\VH$ production, mildly constraining a third
direction in this parameter group.
\item The coefficients $c_{Hu}$, $c_{Hd}$, $c_{Ht}$, $c_{Hb}$ and $c_{Hq}^{(1)}$
mainly affect the neutral current interactions with quarks, while
$c_{He,33}$, $c_{Hl,33}^{(1)}$, $c_{Hl,33}^{(3)}$
mainly affect the $W$ and $Z$ vertices with third generation leptons.
Both sets of coefficients are primarily constrained by the
$VH$, $H\to b\bar{b}$ analysis. The small sensitivity to $c_{Ht}$ and
$c_{Hb}$ arises from the \pT-dependent corrections that they induce
in the subleading production mode $gg\to ZH$.
\item The coefficients of the four-fermion operators involving the
top-quark, i.e.,
$c_{Qq}^{(1,8)}$, $c_{Qq}^{(3,1)}$, $c_{tq}^{(8)}$, $c_{Qu}^{(8)}$, $c_{tu}^{(8)}$,
$c_{td}^{(8)}$, $c_{Qd}^{(8)}$, $c_{Qq}^{(3,8)}$, $c_{Qq}^{(1,1)}$, $c_{tu}^{(1)}$,
$c_{tq}^{(1)}$, $c_{Qu}^{(1)}$, and $c_{Qd}^{(1)}$,
as well as the trilinear gluon coupling $c_{G}$,
affect the \tttH\ and $tH$ production modes and are largely
degenerate.
The $\pT^H$ spectrum measured in the $H\to\gamma\gamma$ and $\Hbb$
channels constrains two linear combinations of these 14 coefficients. The
separate constraint on $tH$ production in the $\Hgg$ analysis
weakly constrains a third direction in this group.
\item The coefficients $c_{ll,1221}$, $c_{Hl,11}^{(3)}$,
and $c_{Hl,22}^{(3)}$ primarily cause a shift in the Fermi
constant, resulting in an overall normalisation factor across
different production modes. The coefficient $c_{H\Box}$ only affects
the measurements through a Higgs boson
propagator correction and thus acts similarly as an overall scale factor
in the observed cross-sections. Only a single linear combination of these
four operators can be constrained from the data.
\item The coefficients $c_{Hl,11}^{(1)}$, $c_{Hl,22}^{(1)}$,
$c_{He,11}$ and $c_{He,22}$ introduce anomalous $HZee$ and $HZ\mu\mu$ vertices
and are mainly constrained by the branching ratio of the Higgs boson
decay into four leptons (electrons or muons), together
with $c_{HDD}$ (responsible for an anomalous $HZZ$ vertex)
and $c_{HQ}^{(3)}$, $c_{HQ}^{(1)}$ (leading to $HZtt$ and $HZbb$
couplings that modify the $Z$ boson propagator, but to which
the Higgs boson measurements included here are not sensitive).
The $ZH$, $H\to b\bar{b}$ analysis also constrains the couplings
to first- and second-generation leptons such as
$c_{He,11}$, $c_{Hl,11}^{(1)}$, $c_{He,22}$, $c_{Hl,22}^{(1)}$,
but the constraints are significantly looser than those from
the $H\to ZZ^*\to 4\ell$ decay.
\end{itemize}

Based on these observations, a new fit basis $\vec{c^\prime}$ is defined
that achieves both fit stability and fit-parameter
interpretability.
This basis $\vec{c^\prime}$ is expressed in terms of single Warsaw basis
coefficients $c_j$ whenever possible, and in terms of linear combinations
(denoted by $e$) of coefficients when necessary:

\begin{align*}
\vec{c}  =  ~&  \{  \ceHtwo \} ~\cup                                                                &  \vec{c}^\prime =~   &    \{ \ceHtwo \} ~ \cup  \\
&  \{  \ceHthree \} ~\cup                                                              &                        & \{  \ceHthree \}~ \cup \\
&  \{  \cHqthree \} ~\cup                                                                &                        & \{ \cHqthree \}  ~\cup \\
&  \{  \cbH  \} ~\cup                                                                  &                        & \{  \cbH \} ~\cup \\
&  \{  c_{HG}, c_{tG}, c_{tH} \} ~\cup                                                   &  \rightarrow ~~~~~~    &  \{ \cGGFone, \cGGFtwo, \cGGFthree \}~ \cup \\
&  \{  c_{HB}, c_{HW}, c_{HWB}, c_{tB}, c_{tW} \} ~\cup   &  \rightarrow ~~~~~~           & \{ \cHZggone, \cHZggtwo, \cHZggthree \} ~\cup \\
&  \{  c_{Hu}, c_{Hq}^{(1)}, c_{Hd}, c_{Hl,33}^{(3)},                                        &                        &   \\
&  ~~~~~~~~~~~~~~~~c_{Ht}, c_{He,33}, c_{Hl,33}^{(1)}, c_{Hb} \} ~\cup &  \rightarrow ~~~~~~ & \{ \cZHone, \cZHtwo, \cZHthree, \cZHfour \} ~\cup \\
&  \{  c_{G}, c_{Qq}^{(1,8)}, c_{Qq}^{(3,1)}, c_{tq}^{(8)}, c_{Qu}^{(8)}, c_{tu}^{(8)}, c_{td}^{(8)},  & &  \\
&  ~~~~~~~~~~~~~~~c_{Qd}^{(8)}, c_{Qq}^{(3,8)}, c_{Qq}^{(1,1)}, c_{tu}^{(1)}, c_{tq}^{(1)}, c_{Qu}^{(1)}, c_{Qd}^{(1)}  \} ~\cup &  \rightarrow ~~~~~~  & \{ \ctopone, \ctoptwo, \ctopthree \} ~\cup \\
&  \{  c_{H\Box}, c_{Hl,11}^{(3)}, c_{Hl,22}^{(3)}, c_{ll,1221} \}  ~\cup                        & \rightarrow ~~~~~~&  \{  \cglobone\} ~\cup  \\
&  \{  c_{Hl,11}^{(1)}, c_{Hl,22}^{(1)}, c_{He,11}, c_{He,22}, c_{HDD}, c_{HQ}^{(3)}, c_{HQ}^{(1)} \} & \rightarrow ~~~~~~&  \{  \cHllllone\}. \\
\end{align*}

\noindent
Only subgroup eigenvectors with an expected uncertainty $\leq 10$ are retained as model parameters.
The remaining coefficients, to which the data are not really sensitive, are fixed to zero in the analysis.
The subgroup eigenvector directions relative to the
Warsaw basis are shown in Figure~\ref{fig:eft_parRotations}.
The names of the eigenvectors are chosen to reflect the production or decay process that dominates the experimental
sensitivity of the parameter group, as shown in Figure~\ref{fig:bij_impact}, although single parameters within some of the groups
sometimes have a dominant sensitivity to another process.

\begin{sidewaysfigure}[hbt]
\includegraphics[width=1.0\textwidth]{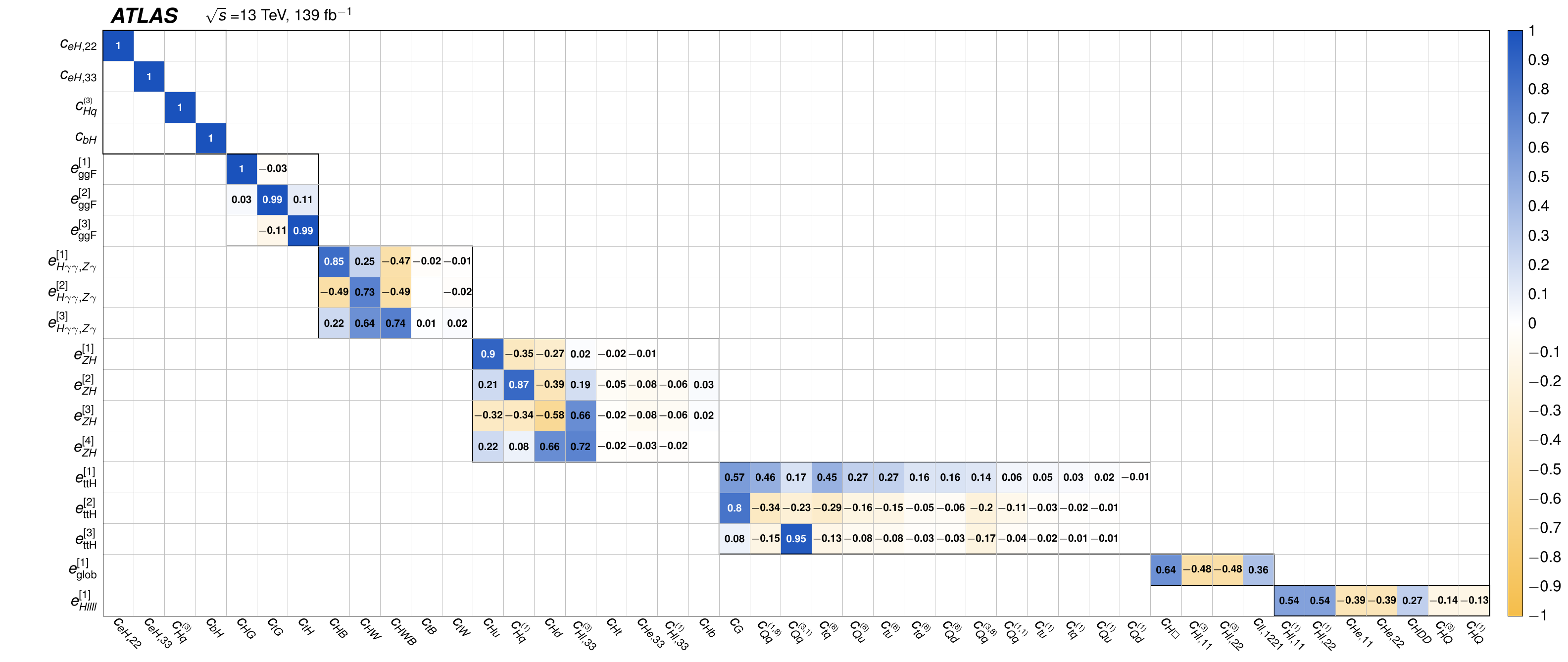}
\caption{Definition of the fit basis coefficients $\vec{c}^{\prime}$
in terms of the Warsaw basis coefficients $\vec{c}$.}
\label{fig:eft_parRotations}
\end{sidewaysfigure}

\noindent No separate optimisation of the parameter basis $\vec{c^\prime}$
is performed for the SMEFT model including quadratic terms of Eq.~\eqref{eq:eftquadmodel}
as the non-linear effects of this model are expected to vanish for small
$c_j$, and thus to asymptotically yield the same rotation matrix as
Eq.~\eqref{eq:pmatrix}.

The impacts of the variations of the coefficients of the rotated
basis $\vec{c}^\prime$ on the STXS regions and Higgs
boson branching ratios are shown in Figure~\ref{fig:eft_impact_ev},
with the same parameter grouping as shown in Figure~\ref{fig:bij_impact}.
The figure clearly demonstrates that the basis $\vec{c^{\prime}}$ represents
impact variations across regions that are much less correlated than those
represented by the Warsaw basis $\vec{c}$.
The magnitude of the parameter variations shown in
Figure~\ref{fig:eft_impact_ev} are set to the {\em expected uncertainty}
for each parameter in the linear SMEFT model, thus giving a realistic
indication of the magnitude of variations that can be constrained from
the data.
For completeness, the impact of the quadratic terms for the
same values of the parameter basis is overlaid, indicating the
relative magnitude of the quadratic terms at the expected sensitivity
level of the linear model. The impact of these expected sensitivities
for the measured SMEFT parameters is discussed in Section~\ref{sec:stxs-results}.

\begin{figure}[!hbtp]
\centering
\includegraphics[width=1\textwidth]{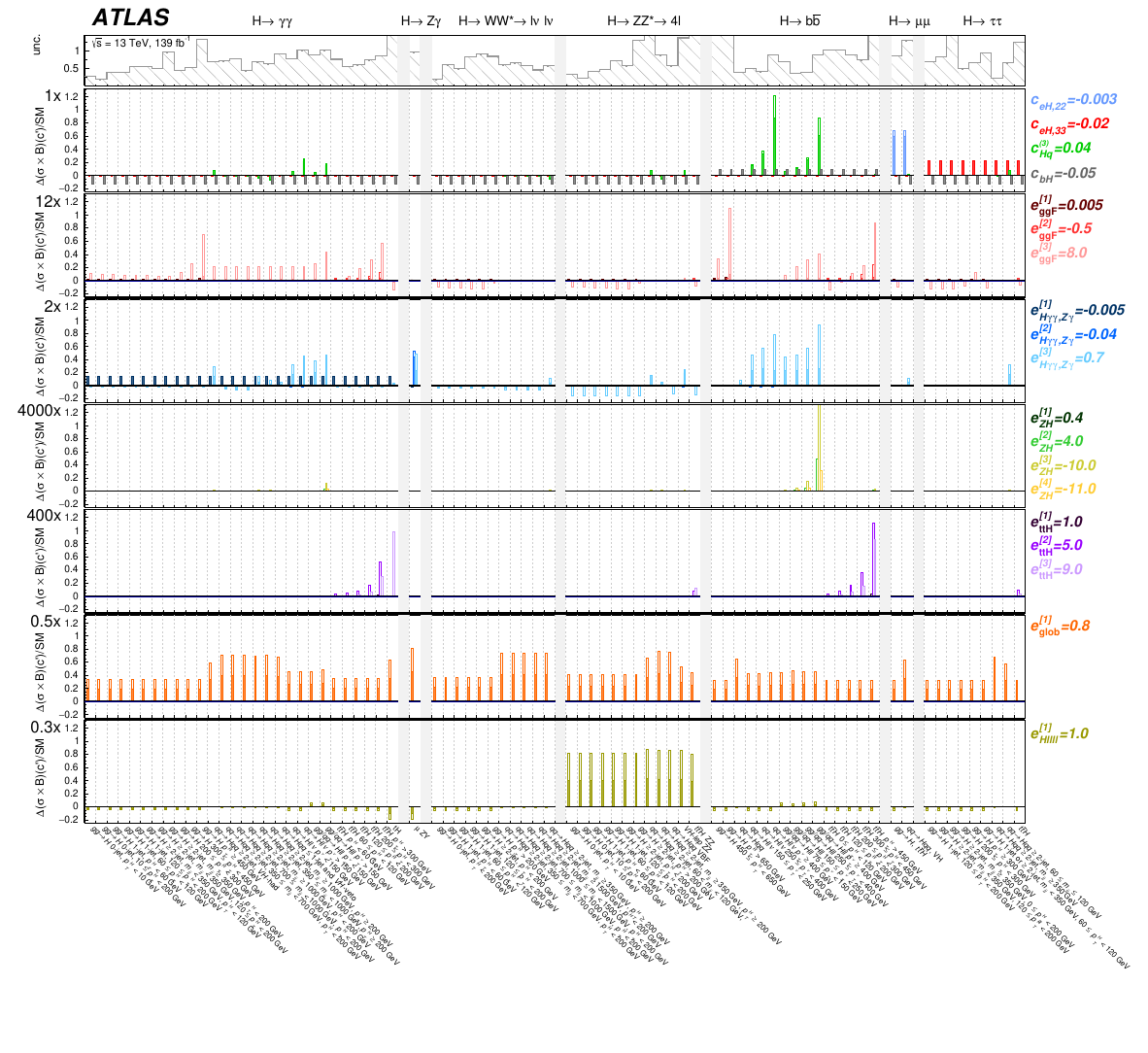}
\caption{Impact of coefficients of the rotated basis $\vec{c}^{\prime}$ on
the STXS regions, relative to the SM cross-section, under the
assumption of the linearised SMEFT model (shaded histogram) and the
SMEFT model including quadratic terms (open histogram).
For all coefficients, a variation equal to the
expected uncertainty when using the linearised model is shown.
Groupings in this figure are consistent with those
in Figure~\ref{fig:bij_impact}, but vertical axis scales differ
in order to completely show all quadratic impact terms,
which can occasionally be very large.
To judge the experimental sensitivity to constrain the operators
from the data in the listed STXS regions, the total
uncertainty in the corresponding regions is shown in the top panel.
For presentational clarity, the uncertainty of low
precision STXS regions
is clipped off in the plot.
}
\label{fig:eft_impact_ev}
\end{figure}

\FloatBarrier


%

\subsubsection{Results}

\paragraph{Linear model}
Figure~\ref{fig:EFT_lin_expobs} compares the expected and observed
results obtained using the linearised SMEFT model, showing good
agreement of the observed data with the SM expectation, corresponding
to a $p$-value of 94.5\%. The parameter value ranges shown in the
bottom panel correspond to 68\% and 95\% confidence level (CL)
intervals, where all other coefficients and nuisance parameters are profiled.
The observed uncertainties are generally about 10\% smaller than the
expected uncertainties, since the assumed Higgs boson width,
when computed with the observed parameter values, is smaller than its
SM expectation value.
This effect is mostly driven by a high observed value of $\cbH$, which
corresponds to a reduced value of $\Gamma^{H \to b\bar{b}}$.
This reduction of the observed linear model uncertainties is comparable
to that of the reduced observed uncertainties in the $\kappa$ model analysis
of the same data reported in Ref.~\cite{HIGG-2021-23}, due to an equivalently
low measured value ($<1$) of $\kappa_b$.

\begin{figure}[!htbp]
\centering
\includegraphics[width=\textwidth]{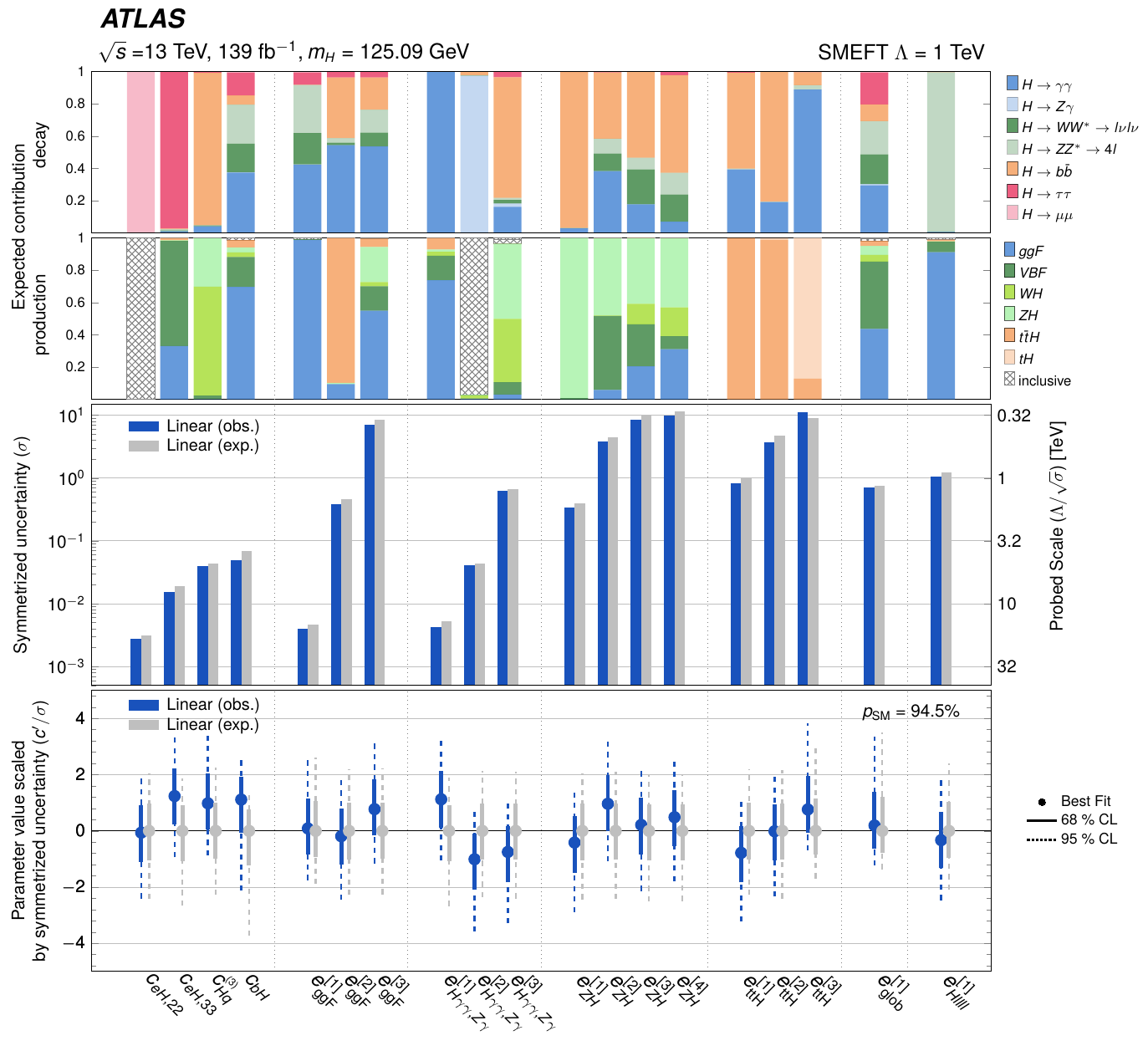}
\caption{Comparison of the expected (gray) and observed (blue)
parameters of the rotated basis $\vec{c}^{\prime}$ with the SMEFT
linearised model, where all other coefficients and nuisance
parameters are profiled.
The middle panel shows the symmetrised 68\% CL uncertainty $\sigma$
of each parameter measurement (left vertical axis) and the corresponding
energy scale $\Lambda/\sqrt{\sigma}$ that is probed
(right vertical axis).
The bottom panel shows the measured parameter
value (dot) and 68\% (solid line) and 95\% (dashed line) CL intervals, divided by
the symmetrised uncertainty shown in the middle panel. The $p$-value
for the compatibility of the data with the Standard Model expectation
(all coefficients vanishing) is 94.5\%.
The top panel shows the expected breakdown of contributions to
the sensitivity of each measurement from the various measured Higgs
boson decay and production modes. Green shades represent production and
decay modes with Higgs boson couplings to vector bosons, red shades
represent modes with Higgs boson couplings to fermions, and blue shades
represent modes with loop couplings. The production category
labelled `inc' collects final states for which the breakdown into
production modes is not available ($H\to \mu\mu$ and $H\to Z\gamma$).
The 95\% CL interval for \ctopthree\ is truncated at the boundary
at which the logarithm of the likelihood function becomes undefined,
resulting in a small undercoverage.
\label{fig:EFT_lin_expobs}}
\end{figure}

Figure~\ref{fig:EFT_lin_expobs} also shows the contribution of each
measured Higgs boson decay or production mode to the sensitivity of each
measurement of the coefficients of the rotated basis. The contribution
of a measurement $i$ to the sensitivity to a coefficient $c^\prime_j$
is determined from the ratio of the Fisher information ${\cal
I}_{i}(c^\prime_j)$ of that measurement to the sum of the Fisher
information  of all contributing components, i.e., $\sum_k {\cal
I}_{k}(c^\prime_j)$, where in all cases the Fisher information is
estimated as the inverse of the covariance matrix $V_\textrm{SMEFT}$,
as defined in Eq.~\eqref{eq:stxs_rot}.
This breakdown reveals that six parameters are (almost) exclusively
measured by a single decay mode: the Yukawa coefficients
\cHetwo\ and \cHethree  by the decay modes $H \to \mu\mu$ and
$H \to \tau\tau$ respectively, the coefficient \cHqthree by $H \to b\bar{b}$;
coefficients \cHZggone\ and \cHZggtwo\ by $H \to \gamma \gamma$ and $H \to Z
\gamma$ respectively and the coefficient \cHllllone\ by $H \to ZZ^*$. Due
to its large effect on the Higgs boson width, the measurement of the Yukawa
coefficient \cbH is not dominated by $H \to b\bar{b}$, but instead
constrained by a combination of measured decays. Similarly, five
parameters are measured (almost) exclusively by a single production
mode: coefficient \cGGFone\ by \ggF\ production, \cZHone\ by $ZH$
production, \ctopone\ and \ctoptwo\ by $ttH$ production at high-\pT
and medium-\pT respectively, and \ctopthree\ by $tH$ production. No
measured coefficient is predominantly sensitive to $H \to WW^*$
decays. The VBF or $WH$ production modes each only dominate a single parameter
measurement, \cHethree and \cHqthree, respectively.

Generally, the tightest constraints on SMEFT coefficients are observed
for processes where the SM amplitudes are suppressed by factors that
do not enter in SMEFT operators contributing to the same measured final
states.
The most prominent example of these is the \oHetwo operator, for which
the corresponding SM \hmm\ amplitude is so strongly suppressed by the
Yukawa factor that it is the most precisely measured coefficient, despite
being constrained only by the very low-yield \hmm\ channel.
Along similar lines, the precisely measured coefficients \cGGFone,
\cHZggone\ and \cHZggtwo\ benefit from suppressed SM amplitudes that
require loops at lowest order.

\paragraph{Analysis of parameter uncertainty sources}
Figure~\ref{fig:EFT_lin_statsys} illustrates the relative importance of
the various uncertainty components on the coefficients measured in the
linearised model, where the top panel shows the relative contributions
of the statistical and systematic components, and the bottom panel
shows the relative contributions of the experimental, signal theory
and background theory contributions to the total systematic uncertainty, where
signal theory systematic uncertainties include both total cross-section uncertainties
and acceptance uncertainties.
The breakdown of the total uncertainty into statistical and systematic
components is performed as follows.
The statistical uncertainty in each parameter is estimated from the 68\%
confidence interval obtained when fixing to their best-fit values
all nuisance parameters associated with systematic uncertainties, and
leaving unconstrained all other parameters.
The total systematic uncertainty is calculated as the square root of the
difference of the squares of the total uncertainty and the statistical uncertainty.
The contribution from a group of systematic uncertainties is estimated as
the difference of the squares of the total uncertainty and the uncertainty
from a fit in which the associated nuisance parameters are fixed to their
best-fit values.

\begin{figure}[!htbp]
\centering
\includegraphics[width=\textwidth]{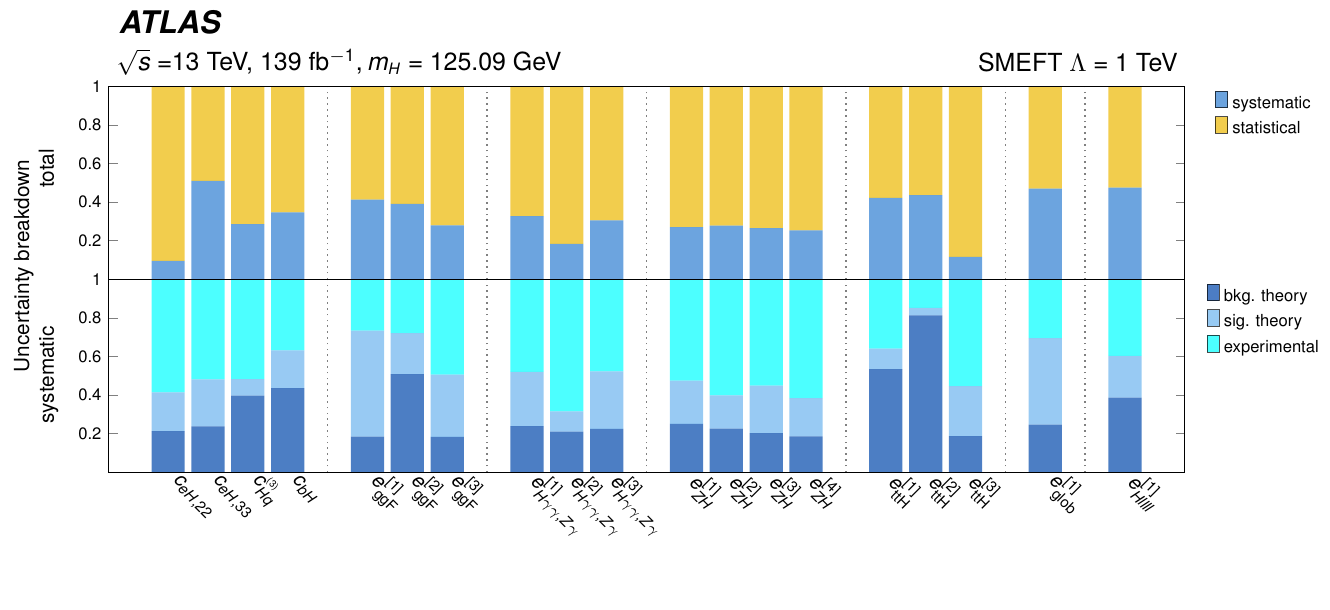}
\caption{Expected fractional contributions of the statistical
(orange) and systematic (blue) uncertainties to the total
uncertainty in the measurements of the parameters of the rotated
basis $\vec{c}^{\prime}$ with the SMEFT linearised model (top panel),
and the corresponding expected fractional contributions of experimental
(cyan), signal theory (light blue) and background theory (dark blue)
uncertainties to the total systematic uncertainty (bottom panel).
}
\label{fig:EFT_lin_statsys}
\end{figure}

For parameters \cHethree, \cglobone and \cHllllone, the fractional
contribution of systematic uncertainties is close to 50\%; for parameters
\cGGFone, \cGGFtwo, \ctopone\ and \ctoptwo, the fractional contribution
of systematic uncertainties is around 40\%; the remaining parameters
have smaller relative systematic uncertainties.
In particular, the uncertainties in the \cHetwo\ and \ctopthree\ parameters,
probed by the measurements of the rare processes $H \to \mu\mu$ decay
and $tH$ production, respectively, are almost completely dominated by the
statistical component.
The total systematic uncertainty in most parameters is dominated by
experimental systematic uncertainties, except
\cGGFone\ and \cglobone, where signal theory systematic uncertainties
dominate, and \cGGFtwo, \ctopone\ and \ctoptwo, where background theory
systematic uncertainties dominate.
The breakdown of the uncertainty components, including a breakdown of
the systematic uncertainty into experimental, signal theory and background
theory components, is also given in Tables~\ref{tab:EFT_resultsObs}
and~\ref{tab:EFT_resultsExp}.

\begin{table}[!htbp]
\caption{Summary of the observed measurements of the parameters
of the rotated basis $\vec{c}^{\prime}$
in the SMEFT linearised model, with a breakdown of the
total uncertainty into statistical and systematic components. A
further breakdown of the systematic uncertainty into experimental and
theoretical uncertainties in the signal and background is also
listed. All MC statistical uncertainties are classified as experimental
systematic uncertainties in the breakdown.
All uncertainties correspond to 68\% CL intervals.
The sum in quadrature of the systematic uncertainty components
may differ from the total systematic uncertainty due to correlations.
}
\label{tab:EFT_resultsObs}
\begin{center}
\renewcommand{\arraystretch}{1.4}
\begin{tabular}{  c S  l  ll  lll }
\toprule{Parameter  \hspace{1cm}}&  {Value}       &  {$\sigma_\textrm{tot}$} \hspace{1cm} & {$\sigma_\textrm{stat}$} & {$\sigma_\textrm{syst}$} \hspace{1cm} & {$\sigma_\textrm{syst}^\textrm{exp}$} & {$\sigma_\textrm{syst}^\textrm{th.sig}$} & {$\sigma_\textrm{syst}^\textrm{th.bkg}$} \\
\midrule
\ceHtwo & -0.0002  &  $0.0028$ &  $0.0026$ &  $0.0009$ &  $0.0006$ &  $0.00049$ &  $^{+0.0003 }_{-0.0005 }$  \\
\ceHthree & 0.019  &  $0.016$ &  $0.011$ &  $0.011$ &  $0.008$ &  $0.006$ &  $0.005$ \\
\cHjt & 0.039  &  $0.040$ &  $0.034$ &  $0.021$ &  $0.015$ &  $^{+0.011 }_{-0.005 }$  &  $0.012$ \\
\cbH & 0.055  &  $^{+0.039 }_{-0.058 }$  &  $^{+0.032 }_{-0.045 }$  &  $^{+0.023 }_{-0.037 }$  &  $^{+0.014 }_{-0.022 }$  &  $0.015$ &  $^{+0.013 }_{-0.025 }$  \\
\hline
\cGGFone & 0.0003  &  $0.0039$ &  $0.0029$ &  $0.0026$ &  $0.0013$ &  $0.0021$ &  $^{+0.0013 }_{-0.0008 }$  \\
\cGGFtwo & -0.07  &  $0.38$ &  $0.29$ &  $0.25$ &  $0.13$ &  $0.11$ &  $0.18$ \\
\cGGFthree & 5.5  &  $7.1$ &  $5.8$ &  $4.00$ &  $2.71$ &  $^{+2.8 }_{-2.1 }$  &  $1.6$ \\
\hline
\cHZggone & 0.0048  &  $0.0043$ &  $0.0034$ &  $0.0026$ &  $0.0017$ &  $0.0015$ &  $0.0012$ \\
\cHZggtwo & -0.042  &  $0.041$ &  $0.037$ &  $^{+0.015 }_{-0.022 }$  &  $0.014$ &  $^{+0.006 }_{-0.011 }$  &  $^{+0.006 }_{-0.010 }$  \\
\cHZggthree & -0.47  &  $0.62$ &  $0.51$ &  $0.35$ &  $0.24$ &  $0.20$ &  $0.16$ \\
\hline
\cZHone & -0.14   &  $0.34$ &  $0.29$ &  $0.19$ &  $0.16$ &  $0.09$ &  $^{+0.08 }_{-0.11 }$  \\
\cZHtwo & 3.7  &  $3.9$ &  $3.3$ &  $2.1$ &  $1.6$ &  $0.9$ &  $1.0$ \\
\cZHthree & 1.8  &  $8.5$ &  $7.2$ &  $4.5$ &  $3.4$ &  $2.2$ &  $2.0$ \\
\cZHfour & 4.8  &  $9.8$ &  $8.2$ &  $5.2$ &  $4.0$ &  $2.5$ &  $2.2$ \\
\hline
\ctopone & -0.65  &  $0.84$ &  $0.64$ &  $0.54$ &  $0.31$ &  $0.17$ &  $0.41$ \\
\ctoptwo & -0.05  &  $3.7$ &  $2.8$ &  $2.4$ &  $0.94$ &  $^{+0.3 }_{-0.6 }$ & $2.2$ \\
\ctopthree  & 8.6  &  $^{+13.4 }_{-9.2 }$  &  $^{+12.1 }_{-8.8 }$  &  $^{+5.73 }_{-2.91 }$  &  $^{+3.9 }_{-2.1 }$  &  $^{+3.8 }_{-1.5 }$ &  $^{+2.0 }_{-1.2 }$ \\
\hline
\cglobone & 0.13  &  $^{+0.83 }_{-0.57 }$  &  $^{+0.57 }_{-0.42 }$  &  $^{+0.60 }_{-0.38 }$  &  $^{+0.30 }_{-0.20 }$  &  $^{+0.42 }_{-0.29 }$  &  $^{+0.29 }_{-0.17 }$  \\
\hline
\cHllllone & -0.349  &  $1.1$ &  $0.76$ &  $0.74$ &  $0.44$ &  $0.38$ &  $0.46$ \\
\bottomrule
\end{tabular}
\end{center}
\end{table}

\begin{table}[!htbp]
\caption{Summary of the expected measurements of the parameters
of the rotated basis $\vec{c}^{\prime}$
with the SMEFT linearised model, with a breakdown of the
total uncertainty into a statistical and systematic component.
A further breakdown of the systematic uncertainty into experimental and
theoretical uncertainties in the signal and background is
also listed. All MC statistical uncertainties are classified as experimental
systematic uncertainties in the breakdown.
All uncertainties correspond to 68\% CL intervals.
The sum in quadrature of the systematic uncertainty components
may differ from the total systematic uncertainty due to correlations.
}
\label{tab:EFT_resultsExp}
\begin{center}
\renewcommand{\arraystretch}{1.4}
\begin{tabular}{  c c  l  ll  lll }
\toprule
Parameter  \hspace{1cm}&  Value   &  $\sigma_\textrm{tot}$ \hspace{1cm} & $\sigma_\textrm{stat}$ & $\sigma_\textrm{syst}$ \hspace{1cm} & $\sigma_\textrm{syst}^\textrm{exp}$ & $\sigma_\textrm{syst}^\textrm{th.sig}$ & $\sigma_\textrm{syst}^\textrm{th.bkg}$ \\
\midrule
\ceHtwo & 0.0  &  $0.0031$ &  $0.0030$ &  $0.0010$ &  $0.0007$ &  $^{+0.0004 }_{-0.0005 }$  &  $^{+0.0004 }_{-0.0006 }$  \\
\ceHthree & 0.0  &  $0.019$ &  $0.013$ &  $0.014$ &  $0.010$ &  $0.007$ &  $0.007$ \\
\cHjt & 0.0  &  $0.044$ &  $0.037$ &  $0.023$ &  $0.017$ &  $0.007$ &  $0.015$ \\
\cbH  & 0.0 &  $^{+0.05 }_{-0.09 }$  &  $^{+0.04 }_{-0.07 }$  &  $^{+0.03 }_{-0.06 }$  &  $^{+0.02 }_{-0.03 }$  &  $^{+0.01 }_{-0.02 }$  &  $^{+0.02 }_{-0.04 }$  \\
\hline
\cGGFone & 0.0  &  $0.0046$ &  $0.0035$ &  $0.0029$ &  $0.0015$ &  $0.0022$ &  $^{+0.0015 }_{-0.0011 }$  \\
\cGGFtwo & 0.0  &  $0.46$ &  $0.36$ &  $0.29$ &  $0.15$ &  $0.13$ &  $0.20$ \\
\cGGFthree & 0.0  &  $8.4$ &  $7.1$ &  $4.4$ &  $3.1$ &  $2.5$ &  $1.9$ \\
\hline
\cHZggone & 0.0  &  $0.0057$ &  $0.0043$ &  $0.0030$ &  $0.0021$ &  $0.0016$ &  $^{+0.0012 }_{-0.0018 }$  \\
\cHZggtwo & 0.0  &  $0.044$ &  $0.040$ &  $0.019$ &  $0.016$ &  $^{+0.005 }_{-0.007 }$  &  $0.009$ \\
\cHZggthree& 0.0 &  $0.68$ &  $0.56$ &  $0.37$ &  $0.25$ &  $^{+0.25 }_{-0.15 }$  &  $0.17$ \\
\hline
\cZHone & 0.0  &  $0.40$ &  $0.34$ &  $0.21$ &  $0.15$ &  $0.10$ &  $0.10$ \\
\cZHtwo & 0.0  &  $4.4$ &  $3.7$ &  $2.3$ &  $1.8$ &  $0.96$ &  $1.10$ \\
\cZHthree & 0.0  &  $9.8$ &  $8.4$ &  $5.0$ &  $3.7$ &  $2.5$ &  $2.3$ \\
\cZHfour & 0.0  &  $11$ &  $9.9$ &  $5.7$ &  $4.5$ &  $2.6$ &  $2.5$ \\
\hline
\ctopone & 0.0  &  $1.0$ &  $0.79$ &  $0.68$ &  $0.40$ &  $0.22$ &  $0.49$ \\
\ctoptwo & 0.0  &  $4.7$ &  $3.5$ &  $3.1$ &  $1.2$ &  $0.61$ &  $2.8$ \\
\ctopthree & 0.0  &  $^{+10.4 }_{-7.6 }$  &  $^{+10.0 }_{-7.0 }$  &  $3.02$ &  $2.25$ &  $^{+1.79 }_{-1.30 }$  &  $^{+0.97 }_{-1.51 }$  \\
\hline
\cglobone & 0.0  &  $^{+0.93 }_{-0.60 }$  &  $^{+0.64 }_{-0.46 }$  &  $^{+0.67 }_{-0.39 }$  &  $0.28$ &  $^{+0.49 }_{-0.23 }$  &  $^{+0.35 }_{-0.18 }$  \\
\hline
\cHllllone & 0.0  &  $1.23$ &  $0.89$ &  $0.85$ &  $0.54$ &  $0.39$ &  $0.53$ \\
\bottomrule
\end{tabular}
\end{center}
\end{table}

The expected and observed correlation matrices are shown in
Figure~\ref{fig:EFT_cormat_expobs}, and show a reasonably good decorrelation
achieved between the fit basis parameters: over half of the off-diagonal
elements are smaller than 0.1 and over 85\% are smaller than 0.3 in both
the observed and expected matrix.
A few exceptions stand out, notably the correlation between
\cGGFtwo\ and \ctopone\ and between \cGGFthree\ and \cHZggone,
which are caused by a common sensitivity to \ttH\ production
and \ggF\ \Hgg, respectively.

\begin{figure}[!htbp]
\centering
\subfloat[]{\includegraphics[width=0.65\textwidth,page=1]{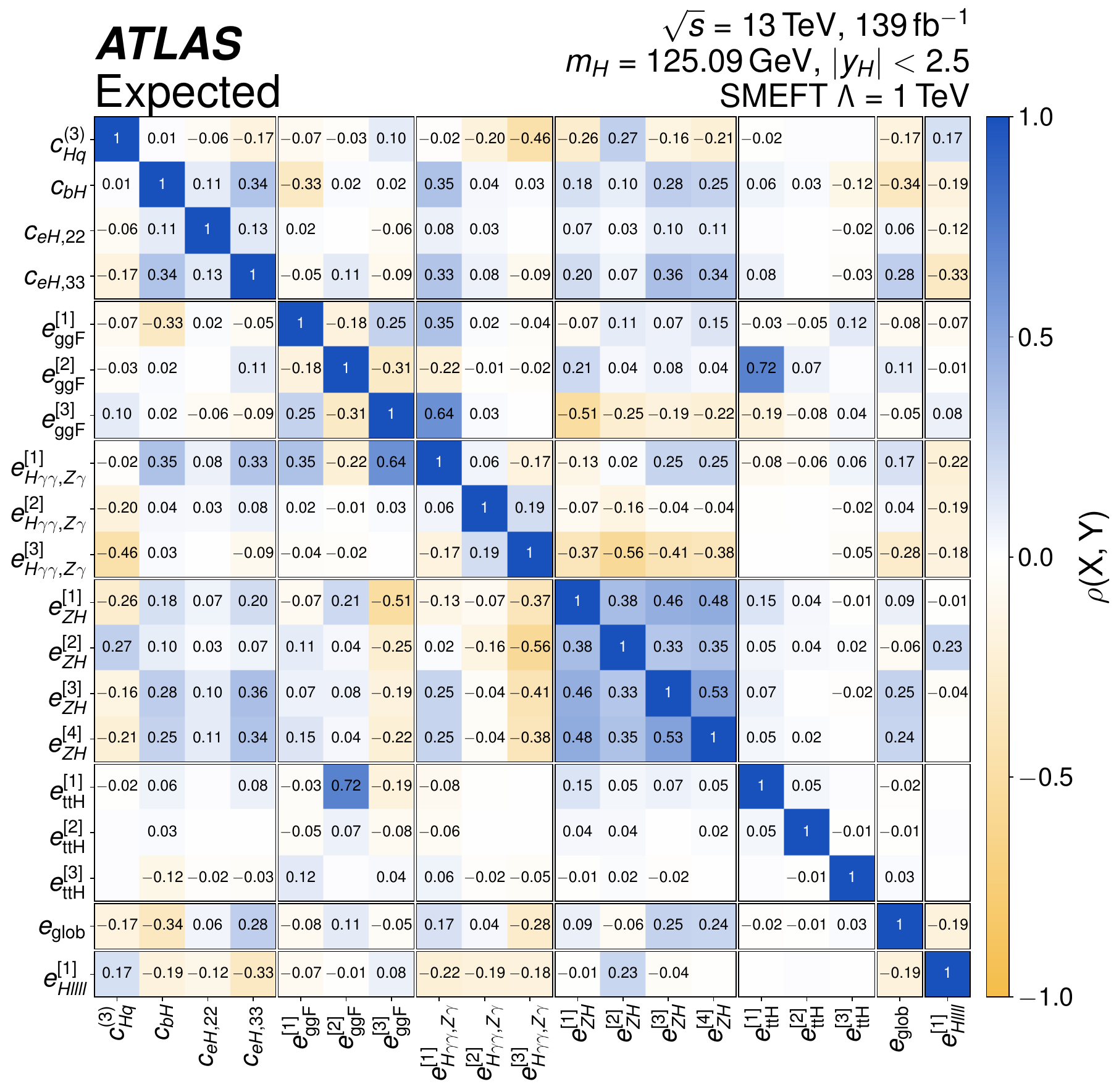}} \\
\subfloat[]{\includegraphics[width=0.65\textwidth,page=1]{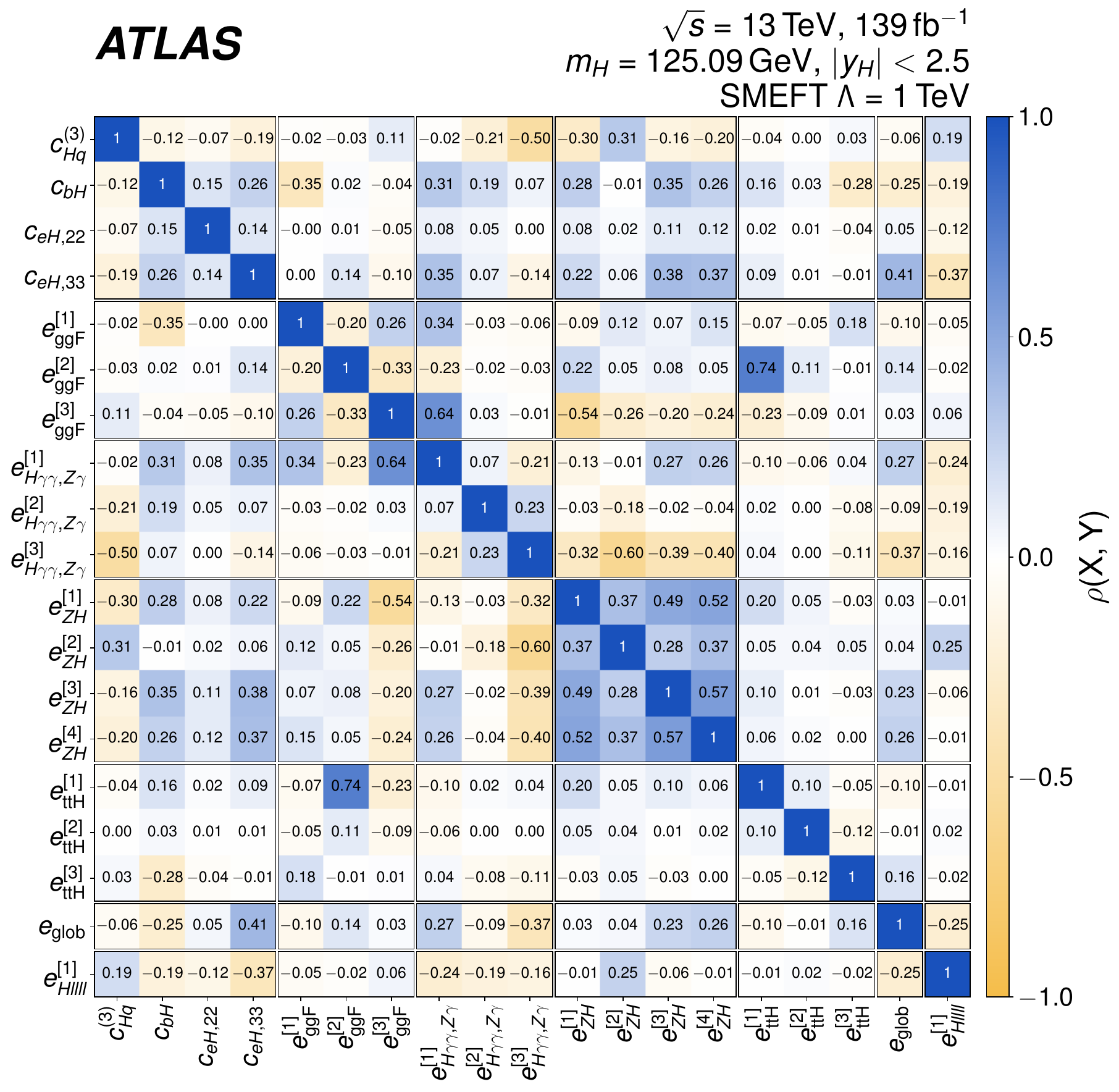}}
\caption{The (a) expected and (b) observed correlation matrix for the parameters of the rotated basis $\vec{c}^{\prime}$ obtained from a fit to the linearised SMEFT model. }
\label{fig:EFT_cormat_expobs}
\end{figure}

\paragraph{Model including quadratic terms}
Figure~\ref{fig:EFT_quad_expobs} compares the expected and observed
results obtained using the SMEFT model including quadratic terms,
again showing good agreement of the observed data with the SM
expectation, corresponding to a $p$-value of 98.2\%. For most
parameters, the observed uncertainty is noticeably smaller than
the expected uncertainty.
The cause of this discrepancy is related to appearance of quadratic
parameter terms in the cross-section predictions, which cause the
likelihood function to generally have multiple minima in each
parameter.
In the expected result, most nearby minima that are exactly
degenerate with the global minimum merge into a single wider
uncertainty interval.
Conversely, in the observed result, the generally non-zero observed
coefficient values lift some of these minima to become local minima,
thereby resulting in a narrower interval around the remaining global
minimum.
This effect can be clearly seen in more detail in the profile
likelihood scans of each parameter, shown in
Figures~\ref{fig:EFT_scans1D_obs_1} and~\ref{fig:EFT_scans1D_obs_2}.
Most secondary minima are raised by 1--2 units in $-2 \log L$, thus
narrowing the 68\% confidence intervals, which are based on a unit
threshold, but less so the 95\% confidence intervals based on a 3.84
unit threshold.
For all parameters with multiple solutions, the global minimum in the
observed result resolves to the minimum closest to a coefficient value of
zero, except parameter \cHZggone, where the minimum
furthest from zero becomes the global minimum.
For the parameter \ceHtwo, there are two exactly degenerate minima.
In that case, the minimum closest to that of the interpretation
with the linear model is quoted as the best-fit result.

\begin{figure}[!htbp]
\centering
\includegraphics[width=\textwidth]{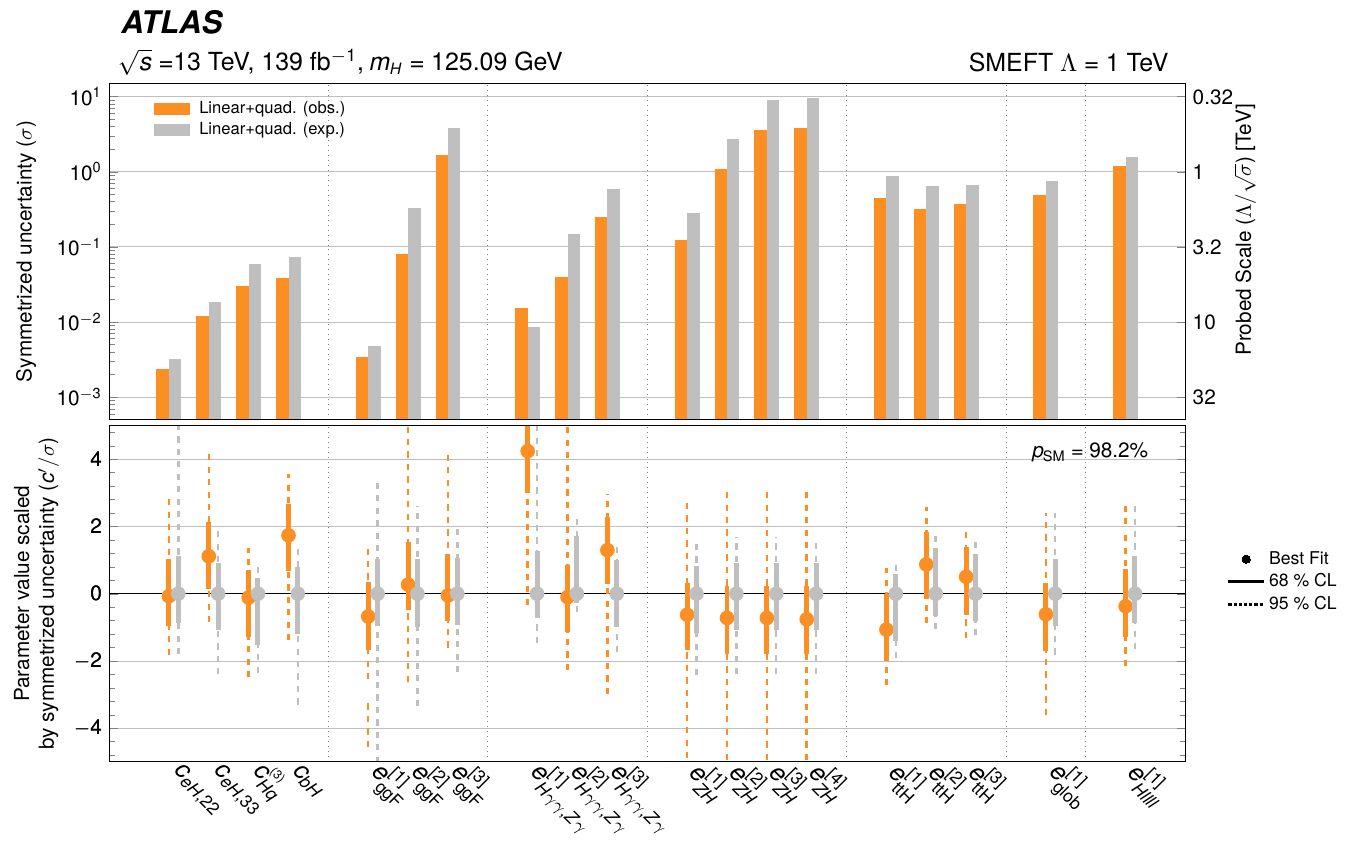}
\caption{Comparison of the expected (gray)
and observed (blue)
parameters of the rotated basis $\vec{c}^{\prime}$ with the SMEFT model
with quadratic terms,
where all other coefficients and nuisance parameters are profiled.
The top panel shows the symmetrised 68\% uncertainty $\sigma$ of each
parameter measurement (left vertical axis) and the corresponding
energy scale $\Lambda/\sqrt{\sigma}$ that is probed
(right vertical axis).
The bottom panel shows the measured parameter value and 68\% (solid) and
95\% (dashed) CL intervals, divided by the symmetrised uncertainty
shown in the top panel. Some of the observed intervals shown are clipped off.
Due to multiple minima, the likelihood curves from which these CL
intervals are derived are non parabolic.
For the parameter \ceHtwo, for which there are two degenerate minima
in the interpretation with the model including quadratic terms,
the point corresponds to the best-fit value that is closest to that
of the interpretation with the linear model.
\label{fig:EFT_quad_expobs}}
\end{figure}

Almost all profile likelihood functions show Gaussian behaviour
to a good approximation within the 95\% CL range of each parameter,
although this is masked by the appearance of degenerate solutions.
The only exception is the expected
profile likelihood for the \ctopthree\ parameter in the linearised
model, where a boundary occurs at $\ctopthree=-10$, just before the
likelihood threshold of the 95\% CL is reached. The yield
prediction in the high-$\pT$ region of the \ttH, \hgg\ channel becomes
negative for very negative values of \ctopthree, rendering the
likelihood undefined beyond that point. The 95\% CL interval for
\ctopthree\ shown in Figures~\ref{fig:EFT_lin_expobs} and
\ref{fig:EFT_obs_linquad} is truncated at this boundary,
resulting in a small undercoverage.

\subsubsection{Comparison between the linearised model and the model including quadratic terms}
\label{sec:stxs-results}
Figure~\ref{fig:EFT_obs_linquad} compares the results for the
linearised model and the model with quadratic terms for the observed
data and shows that for several of the coefficients the constraints
from the model with the quadratic terms are significantly stronger
than those observed for the linear model.
In most cases, the stronger constraints arise from the relatively
weak impact of the BSM--SM interference term on the cross-section
compared to the quadratic BSM terms in specific production or decay modes.

Figure~\ref{fig:eft_impact_ev} illustrates the impact of both the
linear model and the model with quadratic terms, where the impact strength
visualised for {\em both} models corresponds to the expected 68\%
uncertainty of each fit basis parameter $c^\prime$ under the linear
model.
The comparatively larger impact of the model including quadratic terms
at these chosen parameter values identifies STXS regions where quadratic terms
outweigh the linear terms at the measured scale and thus identifies
regions that predominantly provide the extra constraining power of the
models including quadratic terms.
The strongest observed sensitivity enhancements occur in \cGGFthree,
\ctoptwo\ and \ctopthree. For coefficient \cGGFthree,
the quadratic term significantly increases sensitivity in the \ggF, $ZH$
and \ttH\ production modes as well as in the $H \to \gamma \gamma$ decay
mode. For coefficients \ctoptwo\ and \ctopthree,
the quadratic term enhances sensitivity in
high-\pT \ttH\ production and \tH\ production,
respectively. Conversely, for the coefficients \cGGFtwo\ and \cHZggthree,
the effect of the increased impact of the quadratic terms
is limited, and the strong reduction in uncertainties is mostly driven
by the lifting of degenerated solutions in the observed case compared with
expected, as visible in the profile likelihood scans in
Figure~\ref{fig:EFT_scans1D_obs_1}.

\begin{figure}[!htbp]
\raggedright
\hspace{0.7cm}Operators in Warsaw basis: \ceHtwo, \ceHthree, \cHqthree and \cbH \\
\subfloat[]{\includegraphics[width=0.25\textwidth,page=1]{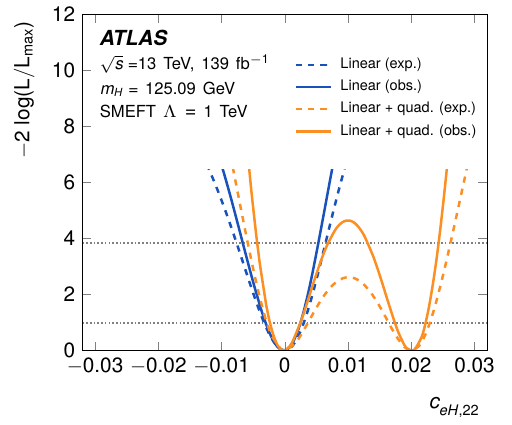}}
\subfloat[]{\includegraphics[width=0.25\textwidth,page=1]{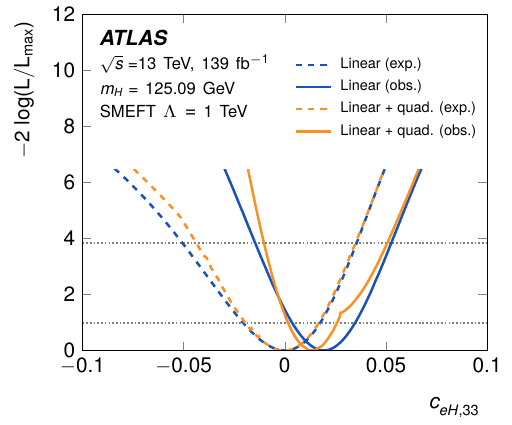}}
\subfloat[]{\includegraphics[width=0.25\textwidth,page=1]{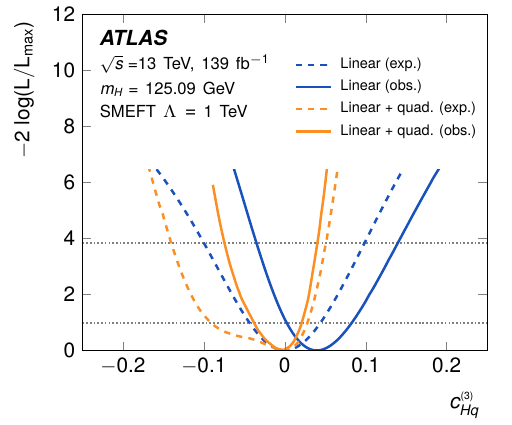}}
\subfloat[]{\includegraphics[width=0.25\textwidth,page=1]{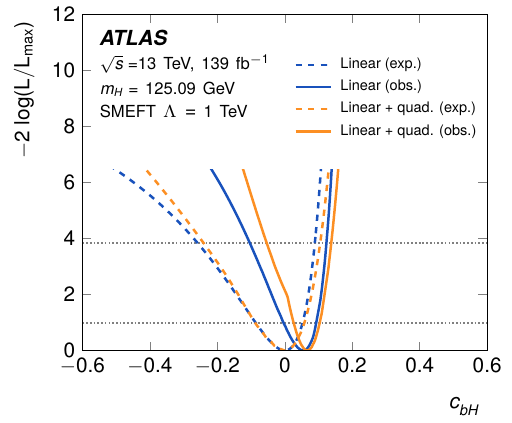}} \\
\hspace{0.7cm}Eigenvector group $ggF$  \\
\subfloat[]{\includegraphics[width=0.25\textwidth,page=1]{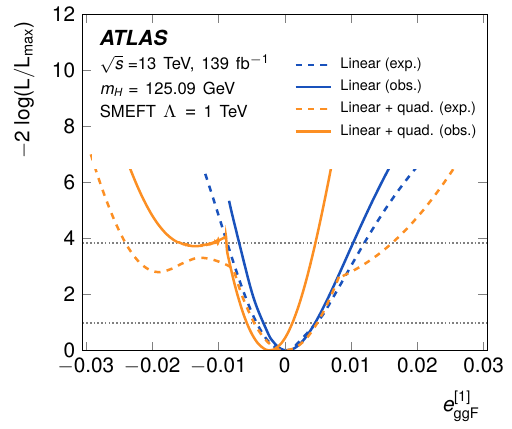}}
\subfloat[]{\includegraphics[width=0.25\textwidth,page=1]{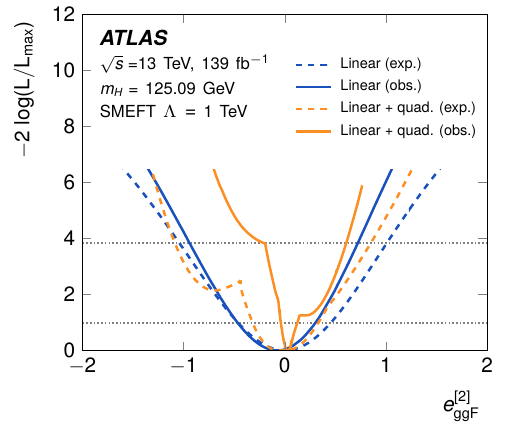}}
\subfloat[]{\includegraphics[width=0.25\textwidth,page=1]{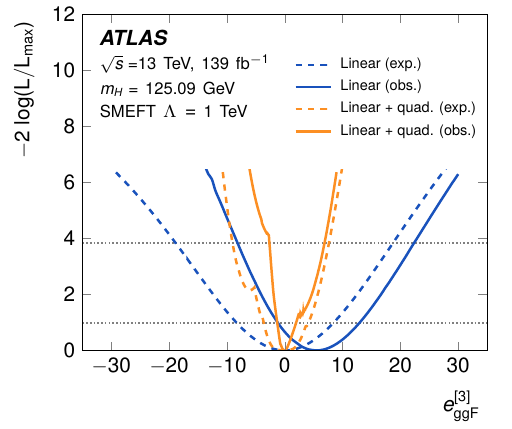}} \\
\hspace{0.7cm}Eigenvector group $H \to \gamma \gamma, Z \gamma$ \\
\subfloat[]{\includegraphics[width=0.25\textwidth,page=1]{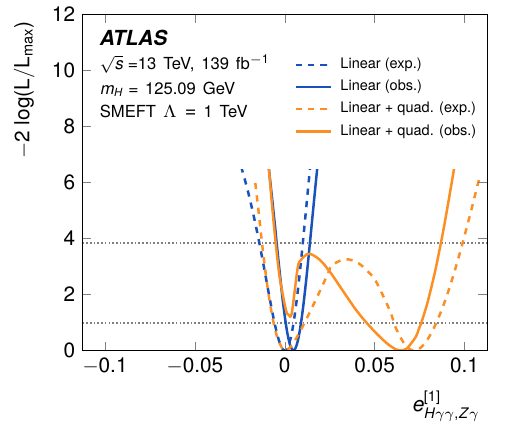}}
\subfloat[]{\includegraphics[width=0.25\textwidth,page=1]{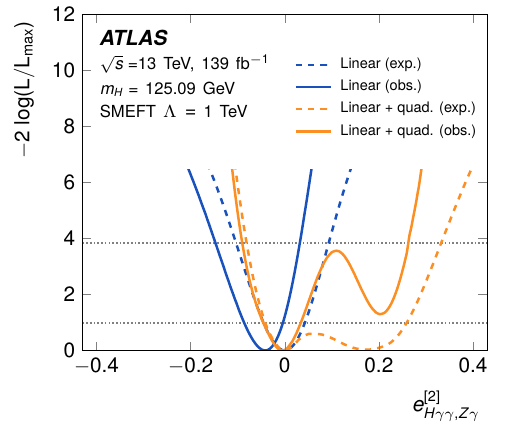}}
\subfloat[]{\includegraphics[width=0.25\textwidth,page=1]{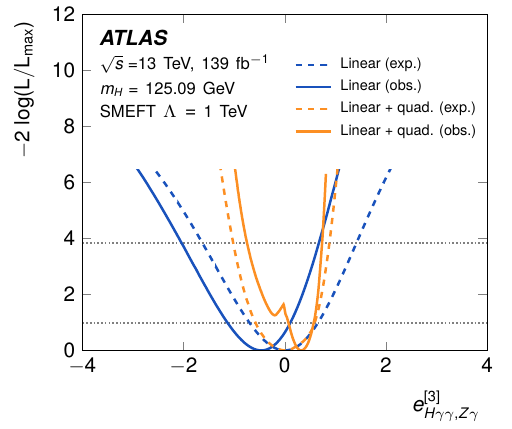}} \\
\hspace{0.7cm}Eigenvector group $ZH$  \\
\subfloat[]{\includegraphics[width=0.25\textwidth,page=1]{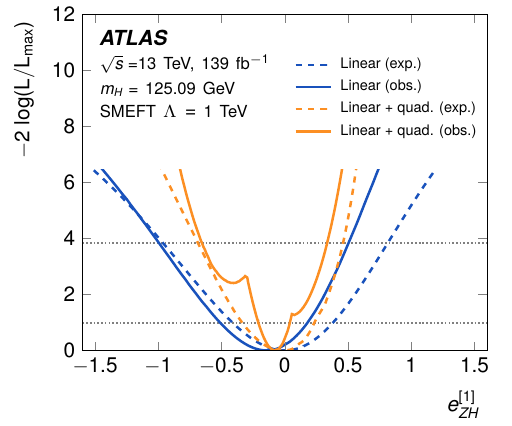}}
\subfloat[]{\includegraphics[width=0.25\textwidth,page=1]{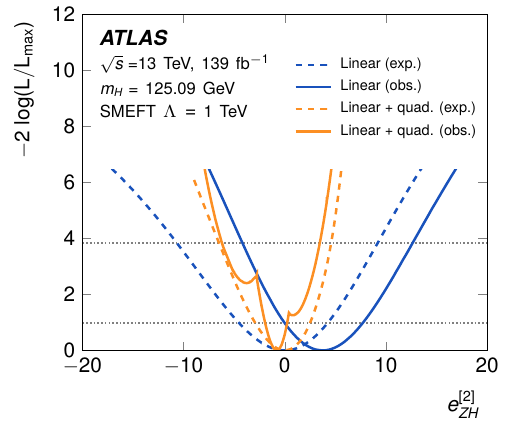}}
\subfloat[]{\includegraphics[width=0.25\textwidth,page=1]{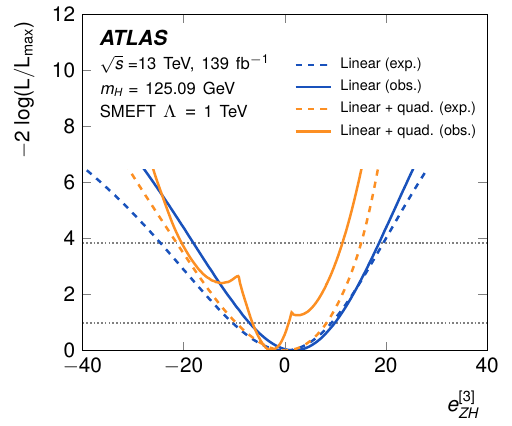}}
\subfloat[]{\includegraphics[width=0.25\textwidth,page=1]{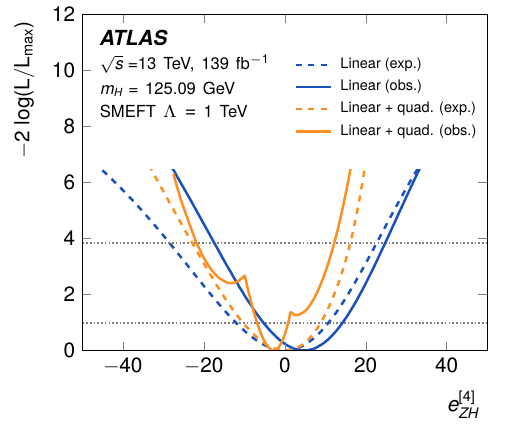}} \\
\caption{Profile likelihood scans of the fitted coefficients
of the rotated basis $\vec{c}^{\prime}$ obtained from fits to the
SM expectation (dashed lines) and to the observed data (solid lines),
based on the SMEFT linearised model (blue) and the SMEFT model
with quadratic terms (orange).
The horizontal dashed lines in each plot correspond to the
asymptotic threshold values for 68\% and 95\% CL intervals.
The profile likelihood scans of the SMEFT model with quadratic terms can
exhibit two minima in the scanned parameter, whereas the scans of the
linearised model have one minimum per parameter.
Multiple minima in the coefficients that are profiled
in each scan may furthermore result in discontinuous derivatives in
the profile likelihood (e.g., prominently visible in the observed
data in (e) and (l)).}
\label{fig:EFT_scans1D_obs_1}
\end{figure}

\begin{figure}[!htbp]
\raggedright
\hspace{0.7cm}Eigenvector group {\em top}  \\
\subfloat[]{\includegraphics[width=0.25\textwidth,page=1]{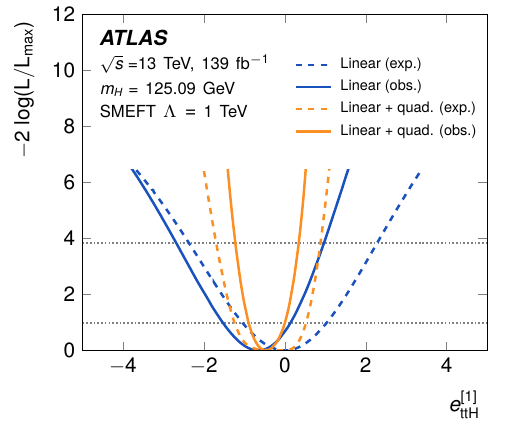}}
\subfloat[]{\includegraphics[width=0.25\textwidth,page=1]{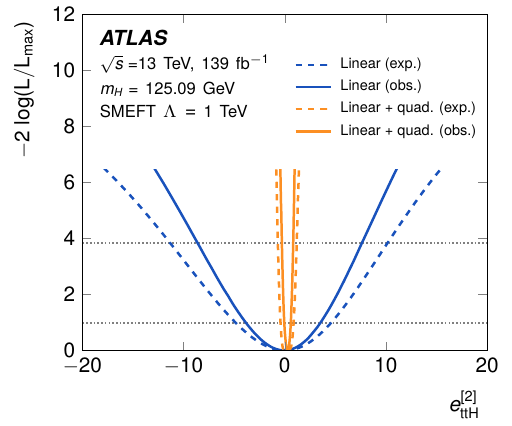}}
\subfloat[]{\includegraphics[width=0.25\textwidth,page=1]{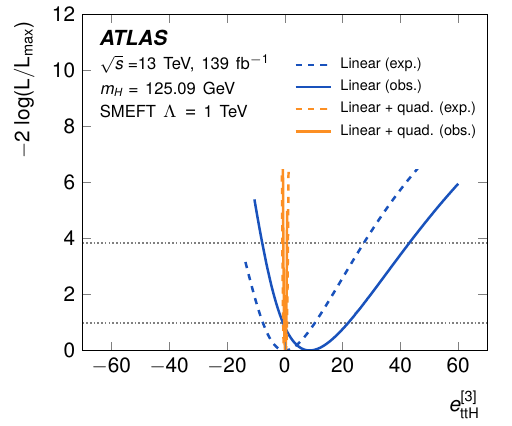}} \\
\hspace{0.7cm}Eigenvector group {\em overall normalisation} \\
\subfloat[]{\includegraphics[width=0.25\textwidth,page=1]{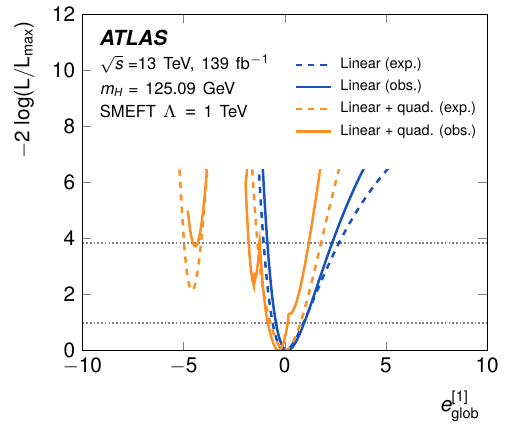}}  \\
\hspace{0.7cm}Eigenvector group $H \to ZZ^*$ \\
\subfloat[]{\includegraphics[width=0.25\textwidth,page=1]{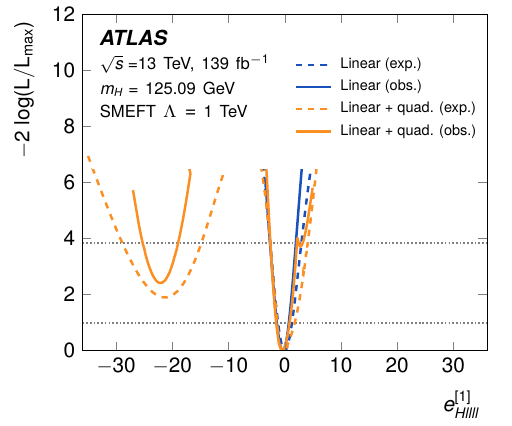}} \\
\caption{Profile likelihood scans of the fitted coefficients
of the rotated basis $\vec{c}^{\prime}$
obtained from fits to the SM expected (dashed lines)
and to the observed data (solid lines), based on the SMEFT
linearised model (blue) and the SMEFT model with quadratic terms
(orange).
The horizontal dashed lines in each plot correspond to the
asymptotic threshold values for 68\% and 95\% CL intervals.
The profile likelihood scans of the SMEFT model with quadratic
terms can exhibit two minima in the scanned parameter, whereas
the scans of the linearised model can only have one minimum per
parameter.
Multiple minima in the coefficients that are profiled
in each scan may furthermore result in discontinuous derivatives in
the profile likelihood (e.g., prominently visible in the observed
data in (d)).}
\label{fig:EFT_scans1D_obs_2}
\end{figure}

\begin{figure}[!htbp]
\centering
\includegraphics[width=\textwidth]{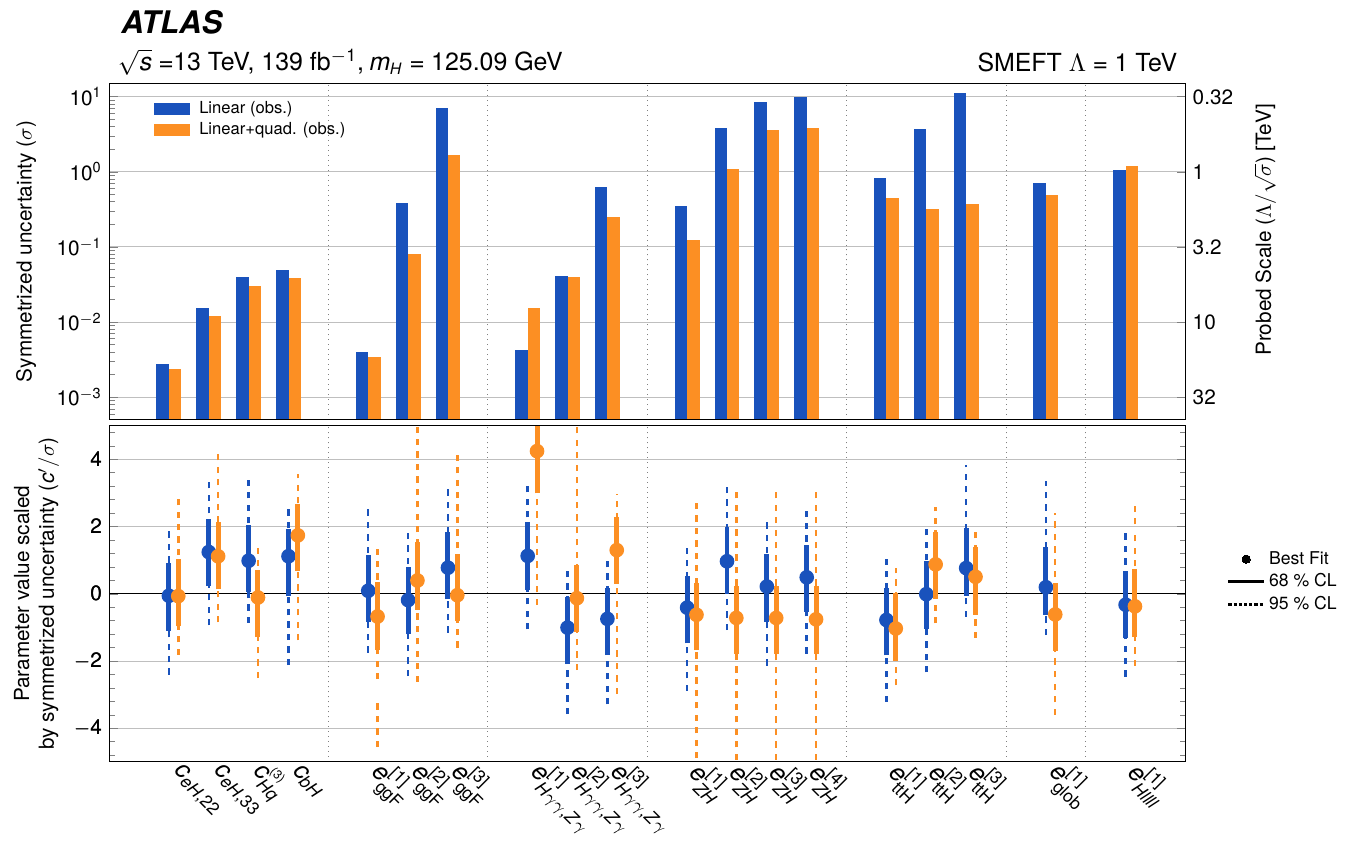}
\caption{Comparison of the observed parameters of the rotated
basis $\vec{c}^{\prime}$ with the SMEFT linearised model (blue)
and the model including quadratic terms (orange),
where all other coefficients and nuisance parameters are profiled.
The top panel shows the symmetrised 68\% CL uncertainty $\sigma$
of each parameter measurement (left vertical axis) and the corresponding
energy scale $\Lambda/\sqrt{\sigma}$ that is probed
(right vertical axis).
The bottom panel shows the measured parameter value and 68\% (solid) and
95\% (dashed) CL intervals, divided by the symmetrised uncertainty shown
in the top panel.
Some of the intervals shown for the model with quadratics terms are
clipped off.
Due to multiple minima the likelihood curves from which these CL intervals
are derived are non-parabolic.
The 95\% CL interval for \ctopthree\ is truncated at the boundary
at which the log likelihood becomes undefined, resulting in a small
undercoverage.
For the parameter \ceHtwo, for which there are two degenerate minima
in the interpretation with the model including quadratic terms,
the point corresponds to the best-fit value that is closest to that
of the interpretation with the linear model.
\label{fig:EFT_obs_linquad}}
\end{figure}


\FloatBarrier
%

\subsection{Constraints from fiducial differential measurements}
\label{sec:eft-results-diff}

Coefficients of SMEFT operators can alternatively be constrained from
fiducial cross-section measurements.
Fiducial differential cross-section measurements are, unlike STXS
measurements, explicitly unfolded to the particle level.
The finer granularity of kinematic regions defined in differential measurements,
compared to STXS measurements, offers an advantage in the analysing power of
differential distributions, but is counteracted by the inclusive treatment
of Higgs boson production processes, which are modelled separately in the STXS
approach.

For a comparison to the STXS approach, differential measurements in the \pt
of the Higgs boson are considered. Anomalous couplings of the Higgs boson
to gluons and top quarks, as well as between gluons and top quarks, can
affect the total Higgs boson production cross-section and its dependence on
the Higgs boson transverse momentum.

The expected deviations from the SM predictions due to these anomalous
couplings can be relatively large in high Higgs boson \pt regions, which
are also characterised by a better
signal-to-background ratio, making the $\pT^H$-differential cross-section
measurement more sensitive to these effects compared to a
measurement of the inclusive rate.

\subsubsection{Sensitivity estimate and choice of parameters}

In this study, constraints on anomalous Higgs boson couplings to gluons
and top quarks are set from the observed transverse momentum differential
spectra of the Higgs boson decay products in the \hfourl\ and \hgg\ decay
channels. The constraints are inferred using an
EFT approach in which the SM Lagrangian is augmented with the
three dimension-6 SMEFT operators
$\mathcal{O}_{HG}$ (Higgs-gluon point-like contact term),
$\mathcal{O}_{tH}$ (top quark Yukawa coupling modifier)
and $\mathcal{O}_{tG}$ (chromomagnetic dipole operator),
that are defined in Table~\ref{tab:CoeffOperators}.
The operator $\mathcal{O}_{HG}$ introduces an $Hgg$ contact
interaction that modifies the value and \pT-dependence of the
\ggF\ production cross-section, as well
as the $H\to gg$ partial decay width, thus affecting indirectly
the \hfourl\ and \hgg\ branching ratios.
The operator $\mathcal{O}_{tH}$ modifies the $t\bar{t}H$ vertex and
thus affects Higgs boson production through top-quark-loop mediated
gluon--gluon fusion or in association with a $t\bar{t}$ pair,
as well as the top-quark-loop amplitude
contributing to the $H\to\gamma\gamma$ partial width.
The operator ${O}_{tG}$ introduces a $t\bar{t}Hg$ vertex that
leads to additional contributions to the amplitude for \ggF\
or $t\bar{t}H$ Higgs boson production,
as well as for $H\to gg$ decays,
affecting in turn the branching ratios for both
$H\to\gamma\gamma$ and \hfourl.

The study presented in this section is performed in the scenario
in which only the SM--BSM interference effects, linear in the Wilson
coefficients, are considered.
For simplicity,
due to the statistical precision of the data sample and the
very large correlations
between the estimators for the three Wilson
coefficients, the constraints are initially set on one Wilson coefficient
at a time, while the values
of the remaining coefficients are assumed to be equal to zero.
Subsequently, similarly to the approach presented in
Section~\ref{sec:eft-details},
a rotation in the parameter space is performed to define a new
set of coefficients that are decorrelated and can be probed
simultaneously.
The new basis is formed by the three eigenvectors $ev^{[i]}$ ($i=1..3$)
of the Fisher information matrix, which are related to the
three Wilson coefficients by the following rotation:
\begin{align}
ev^{[1]} = \phantom{-} 0.999 c_{HG}  - 0.035 c_{tG} - 0.003 c_{tH}, \nonumber \\
ev^{[2]} = \phantom{-} 0.035 c_{HG}  + 0.978 c_{tG} + 0.205 c_{tH}, \nonumber \\
ev^{[3]} =          -  0.005 c_{HG}  - 0.205 c_{tG} + 0.979 c_{tH}. \nonumber
\end{align}
The impact of the operators defining the new basis
on the fiducial differential cross-sections is shown in
Figure~\ref{fig:aij_impact_diff_rotated}.

\begin{figure}[hbt]
\centering
\includegraphics[width=1\textwidth]{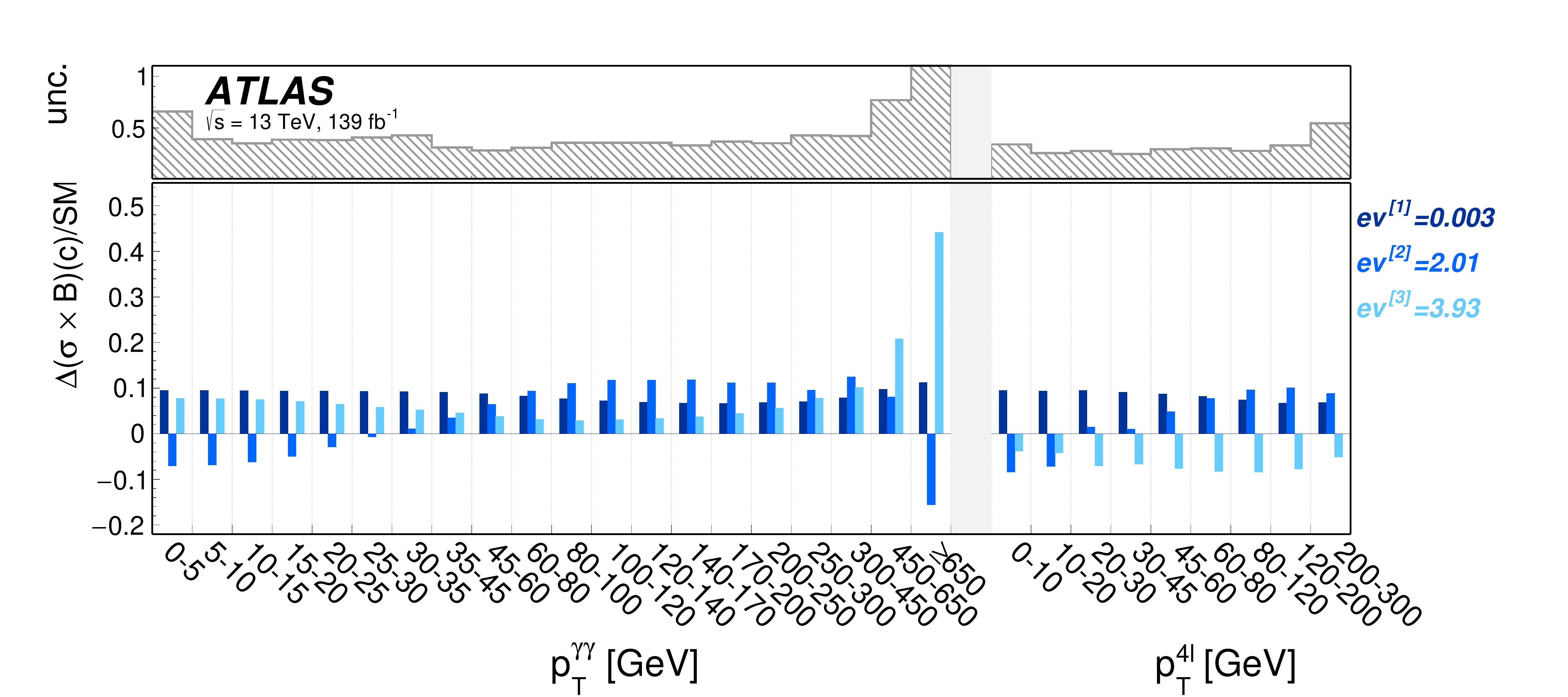}
\caption{Impact of the rotated SMEFT operators on the fiducial
cross-section differential in $\pT^H$ in \hgg~and
\hzz~decays, relative to the SM cross-section, under the
assumption of the linearised SMEFT model.
To judge the experimental sensitivity to constrain the operators
from the data in the listed fiducial regions, the total
uncertainty in the corresponding regions is shown in the top panel.
For presentational clarity, the uncertainty
of low precision fiducial regions is clipped off in the plot.
}
\label{fig:aij_impact_diff_rotated}
\end{figure}

\subsubsection{Results}

The statistical interpretation is performed using a joint likelihood
model of the data in the two decay channels, built,
as described in Section~\ref{sec:sigyieldparam},
from the observed and expected values of the fiducial differential
cross-sections
in each bin of the transverse momentum of the Higgs boson decay products
for both $H\to 4\ell$ and $H\to \gamma\gamma$ decays.
In the likelihood function, the theoretical prediction for each
decay channel and transverse momentum interval is parameterised
as a function of the Wilson coefficients,
as summarised in Section~\ref{sec:eft-intro}.
The procedure assumes common higher-order corrections for
the SMEFT and SM processes.
The theoretical uncertainties are taken from the SM calculations
in Section~\ref{sec:mc}, with the uncertainty associated with
the SMEFT scaling assumed to be negligible.

Figure~\ref{fig:compare_results} shows the 68\% CL one-dimensional
intervals in the linearised model
for the rotated parameters $ev^{[i]}$ with the other two parameters profiled.
The constraint on the first rotated parameter, which is almost
aligned with $c_{HG}$, is of the order of 1\%, while the constraints
on the other two rotated parameters are $\mathcal{O}(1)$.
Figure~\ref{fig:compare_results} compares these results to those obtained
from the STXS measurements performed with the same decay channels using
a linearised model in which all Wilson coefficients except for $c_{tH}$,
$c_{tG}$ and $c_{HG}$ are set to zero.
The comparison illustrates that the differential cross-section measurements,
which have a finer granularity compared to the STXS measurements but probe
the distribution of a single observable inclusively in production mode,
have less constraining power than the STXS measurements, which use information
on the particles produced with the Higgs boson decay products to separate the
different production modes whose cross-sections are affected in different
ways by the different operators, as shown in Figure~\ref{fig:bij_impact}.
In particular, the large sensitivity difference observed for $ev^{[2]}$
originates from the separate measurements of \ggF\ and \ttH\ production
in the STXS framework.

\begin{figure}[!htbp]
\centering
\subfloat[]{\includegraphics[width=0.7\textwidth,page=1]{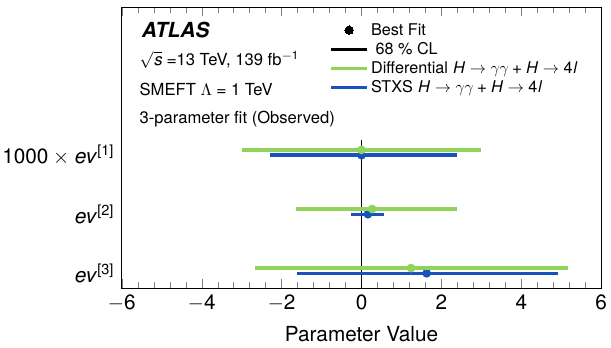}} \\
\subfloat[]{\includegraphics[width=0.7\textwidth,page=1]{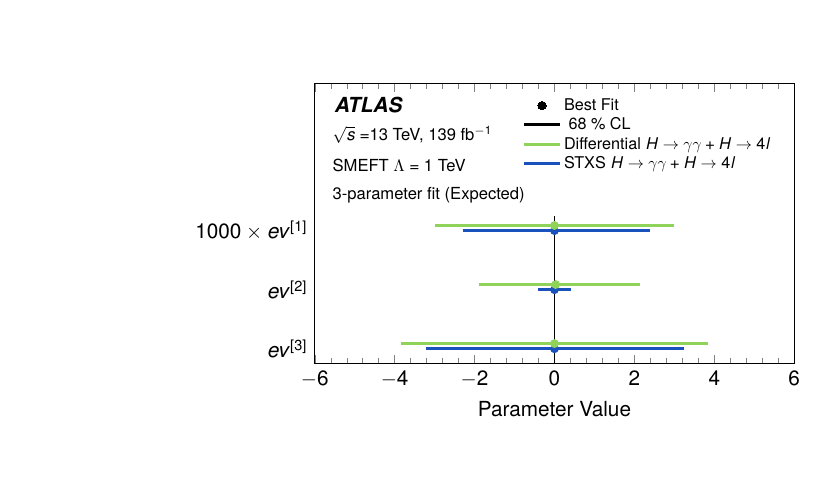}}
\caption{The (a) observed and (b) expected 68\% CL intervals
on the three rotated parameters $ev^{[i]}$ obtained with the SMEFT
linearised model using either STXS (blue) or fiducial
$\pT$-differential cross-section measurements (green)
in the $H\to\gamma\gamma$ and $H\to 4\ell$ decay channels.
The parameter $ev^{[1]}$ is almost aligned with $c_{HG}$, which
mainly affects \ggF\ production, while the parameters $ev^{[2]}$ and
$ev^{[3]}$ are close to $c_{tG}$ and $c_{tH}$ respectively, which
impact both \ggF\ and $t \bar{t} H$ production.
}
\label{fig:compare_results}
\end{figure}


\clearpage

\section{Interpretations based on UV-complete BSM models}
\label{sec:bsm}

In this section, the measurements of Higgs boson cross-sections
and decay rates are interpreted in two UV-complete extensions of the
SM: a two-Higgs-doublet model near the alignment limit
(Section~\ref{sec:2hdm}), and eight benchmark scenarios of the minimal
supersymmetric extension of the SM (Section~\ref{sec:mssm}).


%

\subsection{Constraints on two-Higgs-doublet models}
\label{sec:2hdm}
In two-Higgs-doublet models, the SM Higgs sector with one
doublet of scalar complex fields $\Phi_1$ is extended by introducing
a second doublet $\Phi_2$~\cite{Lee:1973iz,Gunion:2002zf,Branco:2011iw}.
The models considered here assume CP conservation and a (softly broken)
$Z_2$ discrete symmetry that forbids quartic terms of the scalar field
potential $V(\Phi_1,\Phi_2)$ that contain odd powers of either
$\Phi_1$ or $\Phi_2$.
The vacuum expectation values $v_{1,2}$ of the scalar doublets
$\Phi_{1,2}$ that minimise $V$ are related by the sum rule
$v_1^2+v_2^2 = v^2$.
These can be assumed to be both real and non-negative without loss of
generality.
Electroweak symmetry breaking leads to five physical scalar Higgs fields:
two neutral CP-even Higgs bosons $h$ and $H$,
one neutral CP-odd Higgs boson $A$, and two charged Higgs bosons $H^{\pm}$.
For the results presented in this section, the observed Higgs boson is
identified with the lightest CP-even neutral
scalar particle $h$.

The $Z_2$ symmetry of the potential forbids tree-level flavour-changing
neutral currents~\cite{Glashow:1976nt,Paschos:1976ay}, which are strongly
constrained by existing data, and implies that all fermions with the
same quantum numbers couple to only one Higgs doublet.
Depending on which Higgs doublet couples to the three groups of elementary
fermions (up-type and down-type quarks, and leptons), four types of 2HDM
are defined:
\begin{itemize}
\item Type I:
All fermions couple to the same Higgs doublet.
\item Type II:
One Higgs doublet couples to up-type quarks while the other
one couples to down-type quarks and charged leptons.
\item Lepton-specific:
One Higgs doublet couples to leptons while the other one couples
to up- and down-type quarks.
\item Flipped:
One Higgs doublet couples to down-type quarks while the other one
couples to up-type quarks and leptons.
\end{itemize}

The Higgs sector of these models contains seven free physical
parameters: four Higgs boson masses ($m_h$, $m_H$, $m_A$ and $m_{H^{\pm}}$),
two mixing angles $\alpha$ and $\beta$,
and the coefficient $m^2_{12}$ of the
$(\Phi_1^\dagger\Phi_2 + \Phi_2^\dagger\Phi_1)$ term in the Higgs field
potential that softly breaks the $Z_2$ symmetry.
The angle $\beta$ is
defined as
$\tan\beta = \frac{v_2}{v_1}$ and can be assumed to be in the first
quadrant.
The mixing angle $\alpha$ of the neutral CP-even Higgs sector
is defined, modulo a phase, equal to $\pi$.
It can thus be assumed
that $0\leq \beta-\alpha \leq\pi$.

The values of $\alpha$ and $\beta$ determine the couplings of the Higgs
fields to the vector bosons and to the fermions.
In this section, the \textit{decoupling limit} is assumed, in which
$m_H \gg v$.
This in turn implies the {\em alignment
limit}, $|\cos(\beta-\alpha)| \ll 1$,
in which the light scalar Higgs boson $h$
has SM-like couplings.
The mass of the light Higgs boson $m_h$ is assumed to be close to 125~\GeV,
while all other Higgs bosons are assumed to be significantly heavier
(and their masses are assumed to be degenerate, $m_A \sim m_{H^\pm} \sim m_H ~\sim M$).
Near the alignment limit, i.e., for $|\cos(\beta-\alpha)|$ small,
the modifications of the tree-level couplings of the $h$ boson with respect
to the SM predictions follow the expressions summarised in
Table~\ref{tab:2hdm_kappa}~\cite{Branco:2011iw}.

\begin{table}[h!]
\centering
\caption{Multiplicative factors predicted in the four 2HDM scenarios
near the alignment limit, as a function of \tb and $\cos(\beta-\alpha)$,
for the Higgs boson couplings to up-type
quarks (1st row), down-type quarks (2nd row), charged leptons
(3rd row), vector bosons (4th row), and to itself (5th row).
The symbol \cba stands for \cbma, while \sba stands for \sbma.
The definition of the parameter $\bar{m}$ is given in the text.}
\label{tab:2hdm_kappa}
\begin{tabular}{ccccc}
\toprule
Coupling & Type I & Type II & Lepton-specific & Flipped \\
\midrule
$u$, $c$, $t$ & \multicolumn{4}{c}{$\sba+\cba/\tb$} \\
$d$, $s$, $b$ & $\sba+\cba/\tb$ & $\sba-\cba\times\tb$ & $\sba+\cba /    \tb$ & $\sba-\cba\times\tb$ \\
$e$, $\mu$, $\tau$  & $\sba+\cba/\tb$ & $\sba-\cba\times\tb$ & $\sba-\cba\times\tb$ & $\sba+\cba/\tb$ \\
$W$, $Z$ & \multicolumn{4}{c}{\sba} \\
$h$ & \multicolumn{4}{c}{$\sba^3 + \left(3- 2\frac{\bar{m}^2}{m_{h}^2}\right)\cba^2 \sba + 2 \cot\left(2\beta\right)\left(1-\frac{\bar{m}^2}{m_{h}^2}\right)\cba^3$} \\
\bottomrule
\end{tabular}
\end{table}

In addition to the impact of the tree-level coupling modifications,
the production and decay rates of the $h$
boson are modified through NLO EW
corrections involving the trilinear $hhh$ coupling
$\lambda$~\cite{Monti:2803606}.
In the 2HDM framework, this self-coupling is modified with respect
to the SM expectation $\lambda_\mathrm{SM}$ by
the scale factor (denoted by $\kappa_\lambda$) shown in the last row of
Table~\ref{tab:2hdm_kappa}~\cite{Arco:2020ucn}.
The parameter $\bar{m}$ is
$\bar{m}^2 = \frac{m_{12}^2}{\sin\beta\cos\beta} = m_A^2 + \lambda_5 v^2$,
where $\lambda_5$ is the coefficient of the $(\Phi_1^\dagger\Phi_2)^2$ term
of the Higgs potential.
Near the alignment limit considered, the value of $\bar{m}$ is close to
that of $m_A$ ($\lambda_5 v^2 \ll m_A^2$),
and a value $\bar{m}=m_A=1$~\TeV\ ($\lambda_5=0$) is assumed in this section
in all scenarios in which the effect of the self-coupling is considered
in the calculation of limits on \cbma and \tb.

In the following sections, limits on the 2HDM parameters \cbma and \tb
are inferred from studies of the ensemble of Higgs boson production and
decay rate measurements in two distinct scenarios: first by comparing the
measured rates for each production mode with inclusive 2HDM predictions
expressed in the `$\kappa$-framework', and then by comparing the more
fine-grained measurements described in Section~\ref{sec:input}
with the 2HDM predictions expressed in the linearised statistical model of the
EFT-based parameterisation described in Section~\ref{sec:eft}.

\subsubsection{Constraints based on the $\kappa$-framework}
\label{sssec:2hdm_kappa}
In the $\kappa$-framework~\cite{Heinemeyer:2013tqa}, the Higgs boson
production-mode cross-sections and decay branching ratios are parameterised in terms of
multiplicative coupling strength modifiers $\kappa$~\cite{HIGG-2021-23}.
In the model considered in this study, the loop-induced processes
(gluon--gluon fusion $gg\to H$ and $gg\to ZH$ production,
as well as $H\to \gamma\gamma$,
$H\to Z\gamma$ and $H\to gg$ decays) are expressed in terms of the
strength factors for the couplings to the SM particles inside the loop.
The model also assumes that there are no invisible or undetected Higgs boson
decays beyond the SM\@.
The impact of the trilinear $hhh$ coupling modifier $\kappa_\lambda$ on the
Higgs boson production and decay rates via NLO EW corrections
to the LO amplitudes is also included in the parameterisations.
Constraints on $\kappa_\lambda$ from searches for Higgs boson pair
production~\cite{HDBS-2022-03} are not taken into account.
In the likelihood function in Eq.~(\ref{eq:Lbasic}), the
signal strengths $\mu^{i,X}_{k}$ are re-parametrised as
$\mu^{i,X}_{k} = \mu^{i,X}\left(\{\vec{\kappa}(\tb,\cbma)\}\right)$
using the relations listed in Table~\ref{tab:2hdm_kappa}.
Confidence regions for $\tan\beta$ and $\cos(\beta - \alpha)$
can thus be inferred from the
combined measurements of Higgs boson production and decay rates.

The resulting 95\% CL contours in the (\tb,~\cbma) plane
for $0.1<\tan\beta<10$ are shown in Figure~\ref{fig:2hdm_kappa}.
All models exhibit similar exclusion regions in the (\tb, \cbma) plane
at low values $(\lesssim 1)$ of \tb, where only a narrow region of
\cbma around zero is consistent with the measured values of the Higgs boson
production and decay rates.
The interval of allowed values of \cbma increases in size with increasing \tb,
up to a total width of about 0.1--0.2 for $\tan\beta = 1$, depending on the model.
For higher values of \tb, in the models in which at least one of the
coupling strength modifiers is enhanced by a factor \tb, i.e.,
all models except Type-I, the allowed range of \cbma around zero shrinks as
\tb gets larger than one. This trend is observed also for the region
$\tan\beta > 10$, not shown in Figure~\ref{fig:2hdm_kappa}.
For the Type-I model, the allowed range of \cbma
further increases with \tb, reaching a width of around 0.5 for $\tb=10$.
For $\tan\beta > 10$, the allowed range of \cbma\ remains approximately
constant.
This includes regions of the parameter space that deviate
significantly from the alignment-limit hypothesis.
When the constraint from the trilinear $hhh$ coupling is included,
the width of the 95\% CL interval for $\cbma$ at large \tb
for the Type-I model is reduced by about 50\%.
An additional feature of the Type-I model result is that in the large
\tb region, for positive $\cbma$, the observed exclusion region is
significantly larger than the expected one. This derives from the fact
that the ATLAS measurement of Higgs boson production and decay rates
favours values of the coupling strength modifiers to $b$-quarks,
$t$-quarks and $\tau$-leptons smaller than one and of the couplings to
$W$ and $Z$ bosons larger than one ($\kappa_F = 0.95\pm 0.05$
and $\kappa_V = 1.035\pm 0.031$ when
assuming that all fermions share the same modifier $\kappa_F$ and
that $\kappa_W=\kappa_Z=\kappa_V$~\cite{HIGG-2021-23}).
In this scenario, the model predicts
$\sin(\beta-\alpha) = \kappa_V \approx 1$ and
$\cos(\beta-\alpha) = \tb (\kappa_F - \sin(\beta-\alpha)) \approx \tb(\kappa_F - 1)$,
disfavouring positive values of \cbma for $\kappa_F<1$.

\begin{figure}[!htbp]
\centering
\subfloat[]{\includegraphics[width=0.48\textwidth,page=1]{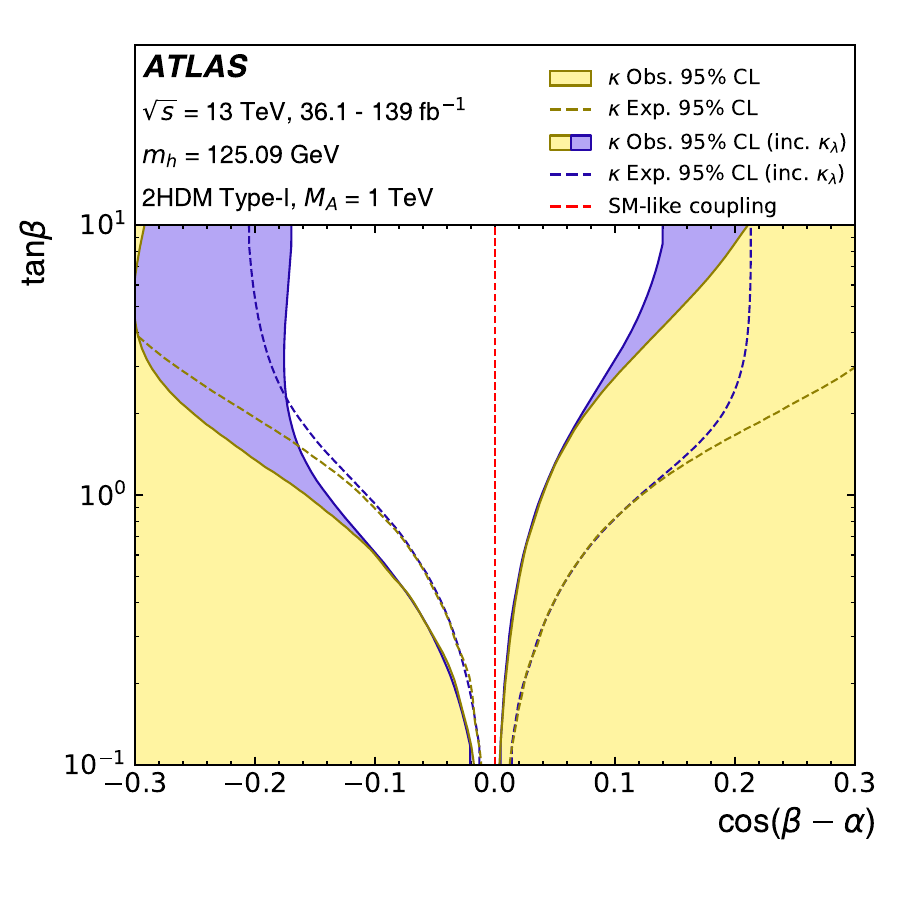}}
\subfloat[]{\includegraphics[width=0.48\textwidth,page=1]{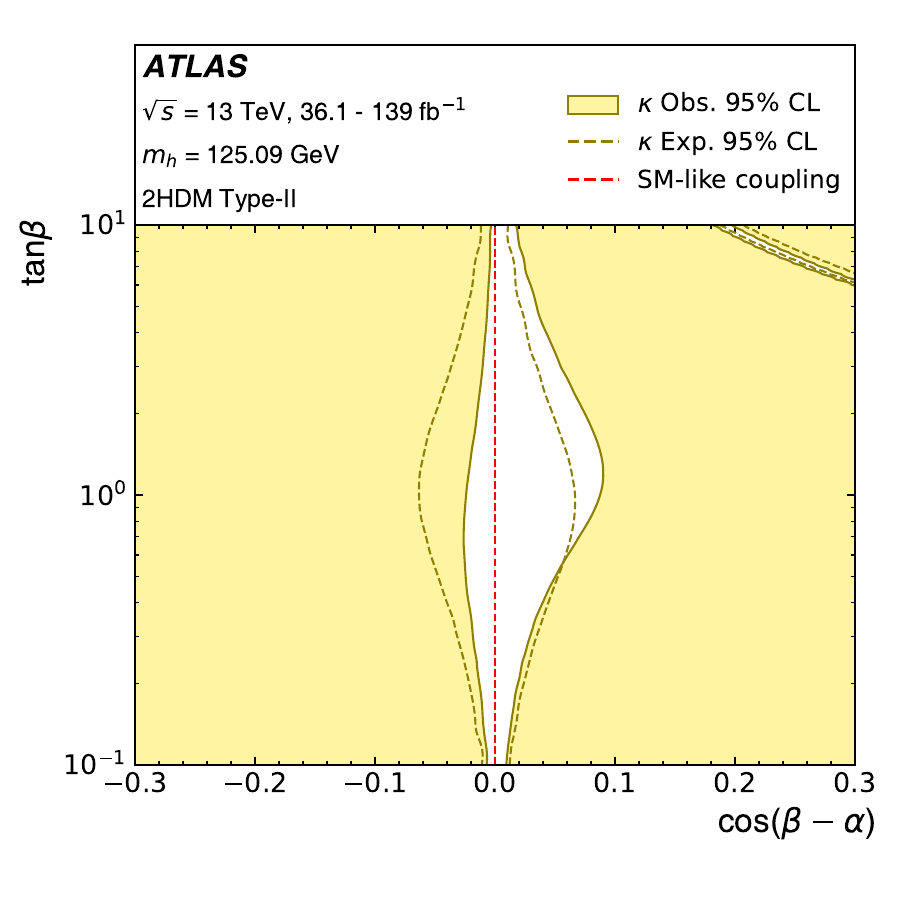}}\\
\subfloat[]{\includegraphics[width=0.48\textwidth,page=1]{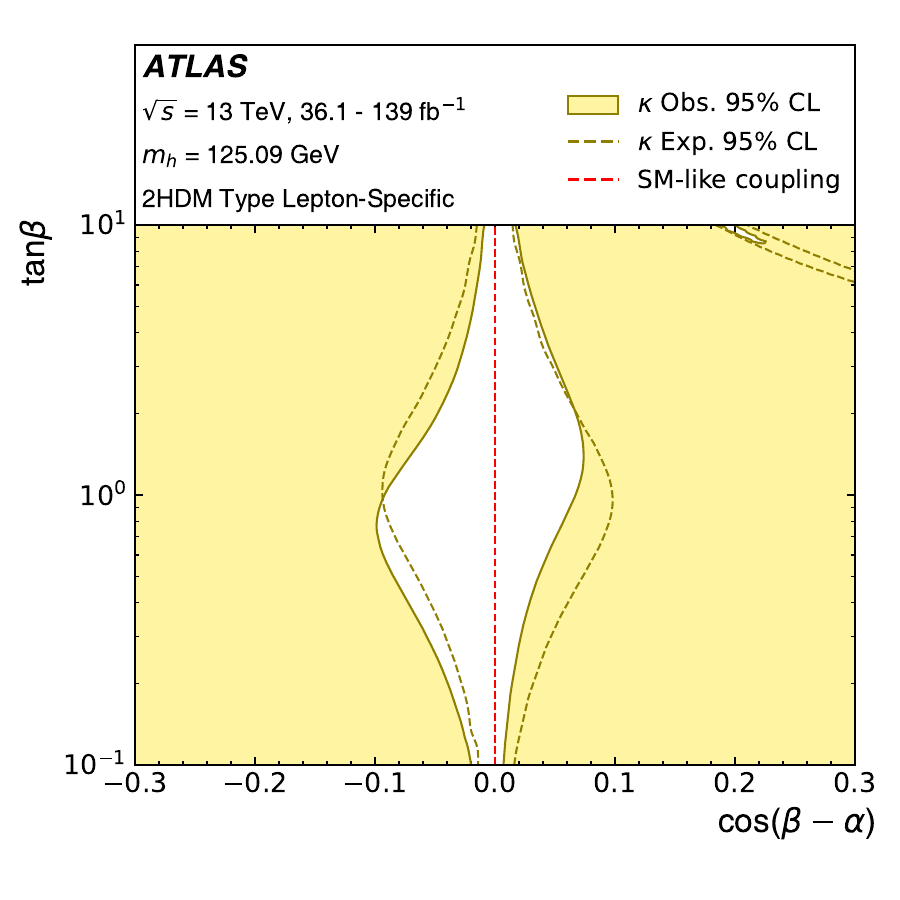}}
\subfloat[]{\includegraphics[width=0.48\textwidth,page=1]{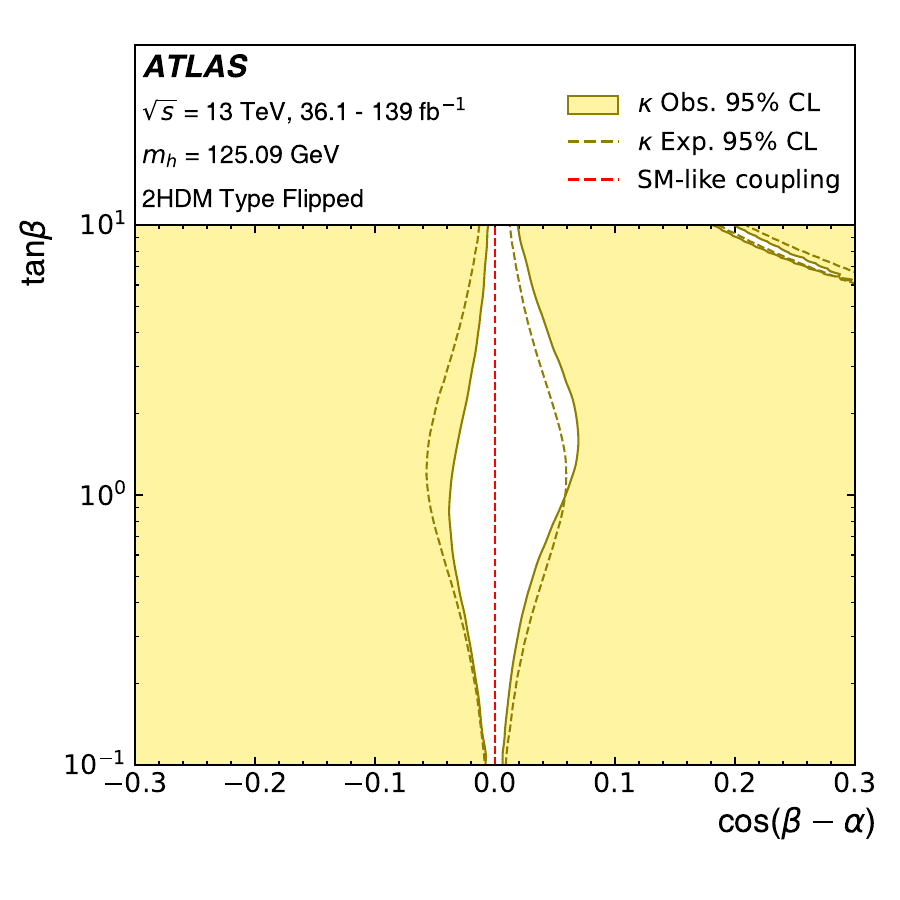}}
\caption{Regions of the 2HDM (\tb, \cbma) parameter plane excluded
at 95\% CL (light yellow filled regions)
in the $\kappa$-framework-based approach
by the measured rates of
Higgs boson production and decays in the
(a) Type-I, (b) Type-II, (c) Lepton-specific and (d) Flipped models.
The dark yellow dashed lines show the borders of the corresponding expected
exclusion regions for the SM hypothesis.
For the Type-I model, the observed and expected regions
excluded at 95\% CL when the $\kappa_{\lambda}$ constraint is
considered are also shown (solid and dashed blue lines).
The red dashed lines
correspond to the exact SM alignment limit ($\cbma=0$).
Results derived in the $\kappa$-framework assume the decoupling limit, i.e., that the masses of the non-SM-like Higgs bosons are large compared to the SM vev,
which implies the assumption of the alignment limit, i.e., $|\cos(\beta-\alpha)|\ll 1$.
}
\label{fig:2hdm_kappa}
\end{figure}

A second small region of allowed \tb and \cbma values for large
\tb and $\cbma\approx 0.2$ is present in the Type-II, Lepton-specific,
and Flipped models.
This corresponds to regions with $\cos(\beta + \alpha)\approx 0$, for
which some of the fermion couplings have the same magnitude as in the
SM but the opposite sign. In particular, this corresponds to a
negative sign of the lepton couplings in the
Lepton-specific model, of the down-type quark couplings in the Flipped
model, or both in the Type-II model.
Since the sign of these couplings is not sufficiently well constrained
by current experimental measurements, this region is not
excluded.
The same region is however not allowed in the Type-I model, in which
the coupling strength modifiers of down-type quarks and leptons have the same
sign (and value) as for up-type quarks.
In that case, a negative sign of the top-quark coupling strength
modifier $\kappa_t$ is experimentally excluded by the measurement of
$H\to\gamma\gamma$ decays, whose branching ratio depends linearly on
$\kappa_t\kappa_W$ because of the interference between
$W$~boson- and top-quark-mediated loop amplitudes.

\subsubsection{Constraints using an EFT-based approach}
The modifications to the tree-level Higgs boson couplings to fermions
near the alignment limit described in Section~\ref{sec:2hdm}
can be generated by the dimension-6 operators
$\mathcal{O}_{uH}$, $\mathcal{O}_{bH}$, $\mathcal{O}_{tH}$ and $\mathcal{O}_{eH}$ of the SMEFT
Lagrangian in the `top' flavour scheme, with the following matching between the Wilson
coefficients and the 2HDM parameters~\cite{Dawson:2020oco}:

\begin{eqnarray}
\frac{v^2c_{i H}}{\Lambda^2} &=& -Y_{i}\eta_{i}\frac{\cbma}{\tan\beta}, \label{eq:CXH_2HDM}
\end{eqnarray}
with $i=u,b,t,\tau,\mu$, and $c_{\mu H}$ and $c_{\tau H}$ referring to the \ceHtwo and \ceHthree coefficients, respectively.
In these formulae, computed to first order in $\cos(\beta - \alpha)$,
$Y_i=\sqrt{2}m_i/v$ are the SM Higgs boson couplings
and the values of $\eta_i$ depend on the type of model
($\eta_b=-\tan^2\beta$ in the Type-II and Flipped models, as
are $\eta_\mu$ and $\eta_\tau$ in the Type-II and the Lepton-specific models; $\eta_{b,t,\mu,\tau}=1$
otherwise).
The corrections to the $HVV$ vertices are quadratic in
$\cos(\beta-\alpha)$ and are neglected.
For the EFT to be valid, $\Lambda \gg v$
and therefore $|\cbma|\propto \frac{v^2}{\Lambda^2}$ should
be small, close to the alignment limit.

The operator $\mathcal{O}_H$ is also generated in the
models under study, with a
coefficient that scales with $\cos^2(\beta-\alpha)$ but which
can be significantly enhanced if the other scalar states are
much heavier than the SM-like Higgs boson.
In that case, $c_H$ is proportional to $(M/v)^2$~\cite{Dawson:2020oco}:
\begin{equation}
\frac{v^2c_{H}}{\Lambda^2}  = \cbma^2 \left(\frac{M}{v}\right)^2. \nonumber
\end{equation}
Non-zero values of $c_{H}$ modify the trilinear $hhh$ coupling $\lambda$ by a
scale factor~\cite{Dawson:2022cmu}:
\begin{equation}
\kappa_\lambda = 1 - \frac{v^2c_H}{\lambda_\mathrm{SM}\Lambda^2} = 1 - \frac{\cbma^2}{\lambda_\mathrm{SM}} \left(\frac{M}{v}\right)^2.
\nonumber
\end{equation}
This in turn affects Higgs boson production and decay rates
via NLO EW corrections to the LO amplitudes,
whose effect is included in the parameterisations used to
obtain the results.
As in Section~\ref{sssec:2hdm_kappa},
constraints on $\kappa_\lambda$ from searches for Higgs boson pair
production are not taken into account.
In the following, a value of $M=1$~\TeV\ is assumed for the masses
of the heavy scalar particles,
and thus $\kappa_\lambda \approx 1 - \left(11 \cbma \right)^2$.
Given the large predicted deviations of $\kappa_\lambda$ from one
for $\cbma$ different from zero, including  $\kappa_\lambda$
in the interpretation is expected to affect significantly,
for $\tan\beta \gtrsim 1$, the results for the Type-I model,
for which the impact of $\cbma \neq 0$ on the other Wilson coefficients,
$c_{bH}$, $c_{tH}$ and $c_{\tau H}$, is suppressed by a $1/\tan\beta$ factor.
Exclusion limits for all other types of 2HDM models are expected
to be largely unaffected.

Constraints on the values of $\tb$ and $\cos(\beta-\alpha)$ can thus
be inferred from those set on
$c_{bH}$, $c_{tH}$, $c_{\tau H}$ and $c_{H}$
using the SMEFT interpretation of the combined measurement of production
and decay rates in STXS bins.
The likelihood model defined in Section~\ref{sec:eft-results} is used,
expressing the four Wilson coefficients
$c_{bH}$, $c_{tH}$, $c_{\tau H}$ and $c_{H}$
in terms of the 2HDM parameters as in
Eq.~\eqref{eq:CXH_2HDM}
and fixing all the other Wilson coefficients to zero.
The results
are shown in Figure~\ref{fig:2hdm_eft}.

\begin{figure}[!htbp]
\centering
\subfloat[]{\includegraphics[width=.48\textwidth]{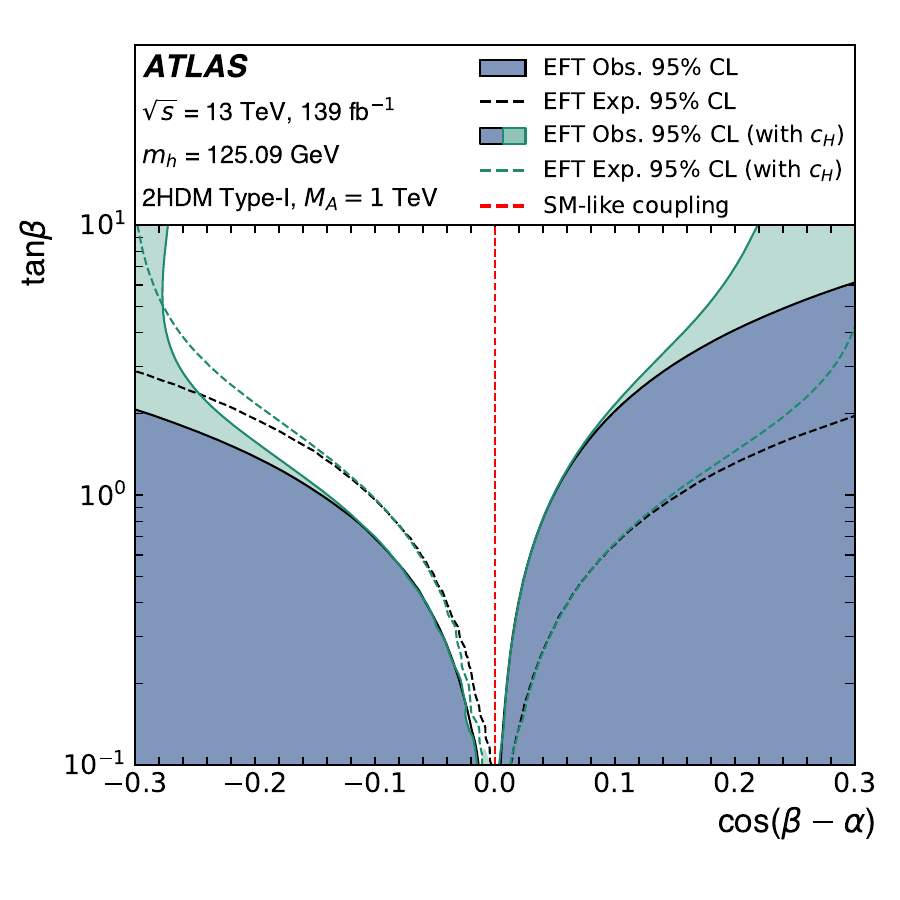}}
\subfloat[]{\includegraphics[width=.48\textwidth]{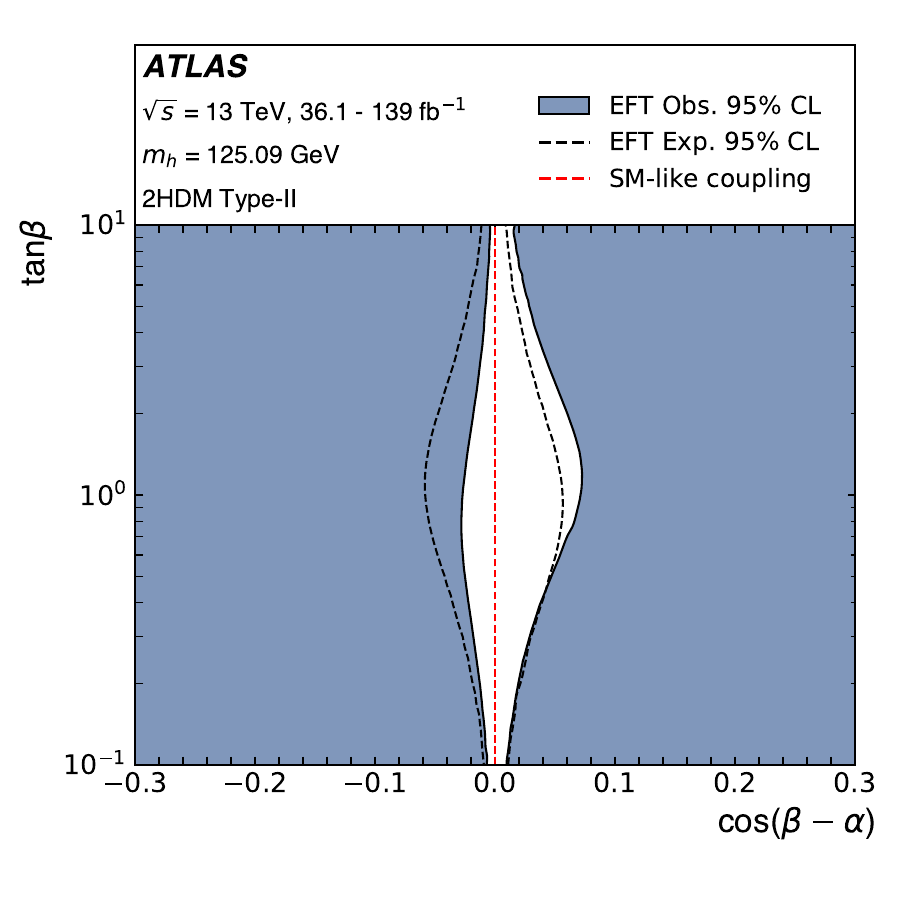}}\\
\subfloat[]{\includegraphics[width=.48\textwidth]{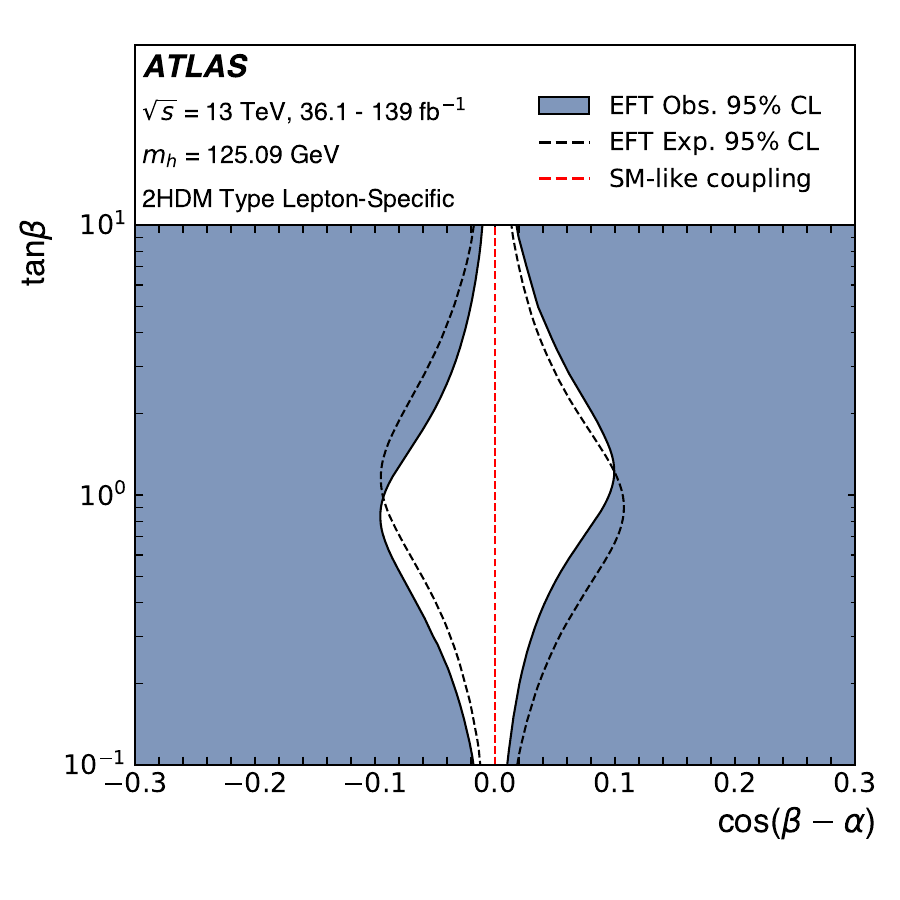}}
\subfloat[]{\includegraphics[width=.48\textwidth]{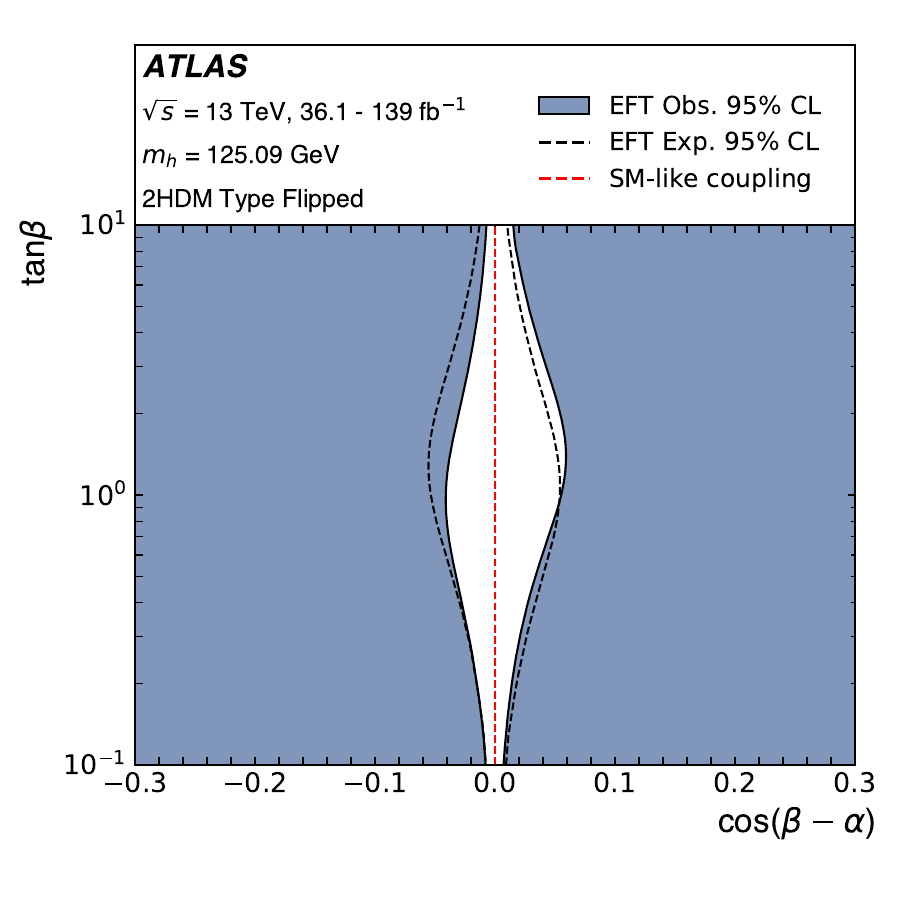}}
\caption{Regions of the 2HDM (\tb, \cbma) parameter plane excluded
at 95\% CL (blue filled regions) in the EFT-based approach
by the measured rates of
Higgs boson production and decays in the
(a) Type-I, (b) Type-II, (c) Lepton-specific and (d) Flipped models.
The dashed black lines show the borders of the corresponding expected
exclusion regions for the SM hypothesis.
For the Type-I model, the observed and expected regions
excluded at 95\% CL when the $c_H$ constraint is
considered are also shown (solid and dashed green lines).
Results are derived assuming that $|\cos(\beta-\alpha)|\ll 1$,
near the alignment limit represented by the red dashed lines,
and that the masses of the non-SM-like Higgs bosons are
large compared to the SM vev. In particular, the limits on the
Type-I model at high $\tan\beta$ are likely outside the validity range of the assumption made in the model.
}
\label{fig:2hdm_eft}
\end{figure}

A comparison between the excluded regions from the two approaches is shown
in Figure~\ref{fig:compare_kappa_smeft}.
In the regions where the assumptions used in this study are valid,
the excluded regions are very similar in the two approaches.
In the Type-I model for large values of \tb, the EFT-based
approach leads to looser constraints on \cbma than the
$\kappa$-framework-based approach.
This difference stems from the fact that the EFT-based approach
(i) does not exploit the constraints from the $HVV$
couplings, that only enter at dimension-8 in the SMEFT expansion
and are not considered here, and (ii) retains only terms
of $\mathcal{O}(\cbma)$ in the expansion of $\kappa_\lambda$,
while in the $\kappa$-framework-based approach the constraint
$\kappa_V = \sin(\beta-\alpha)$
and the full dependence of $\kappa_\lambda$ on \cbma are considered.
However, part of the region of Type-I model parameter space allowed
in the EFT-based approach but not in that based on the $\kappa$-framework
is inconsistent with the
alignment limit hypothesis of $|\cos(\beta-\alpha)| \ll 1$.

\begin{figure}[!htbp]
\centering
\subfloat[]{\includegraphics[width=.48\textwidth]{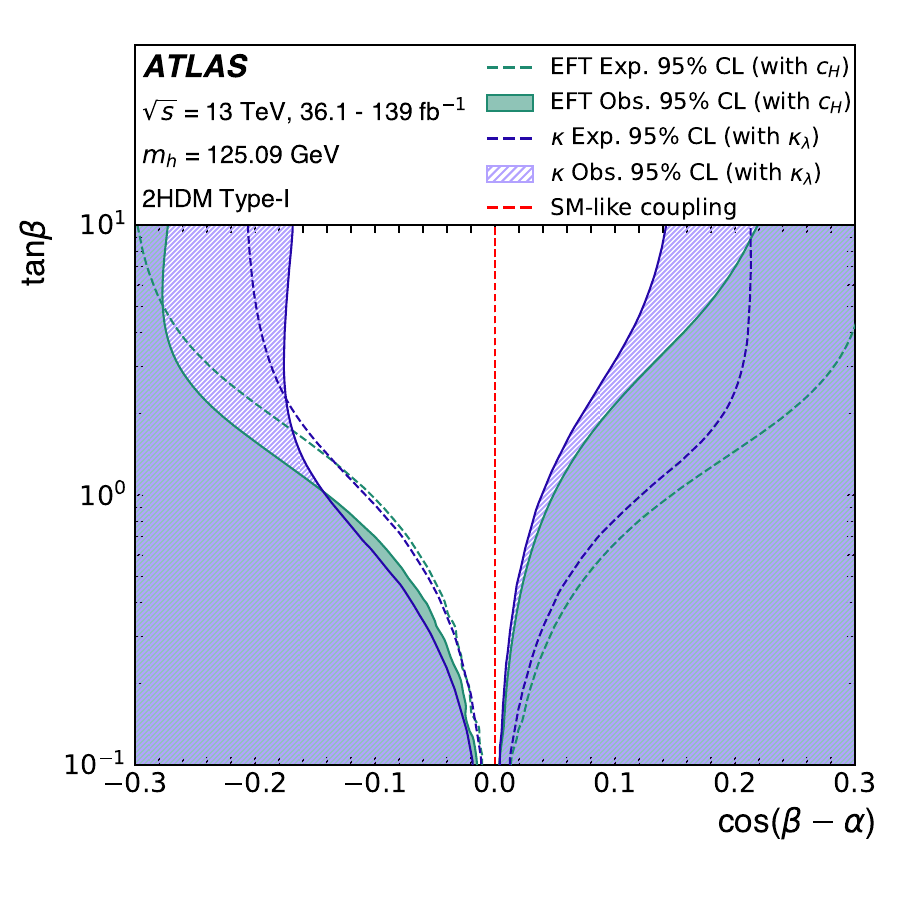}}
\subfloat[]{\includegraphics[width=.48\textwidth]{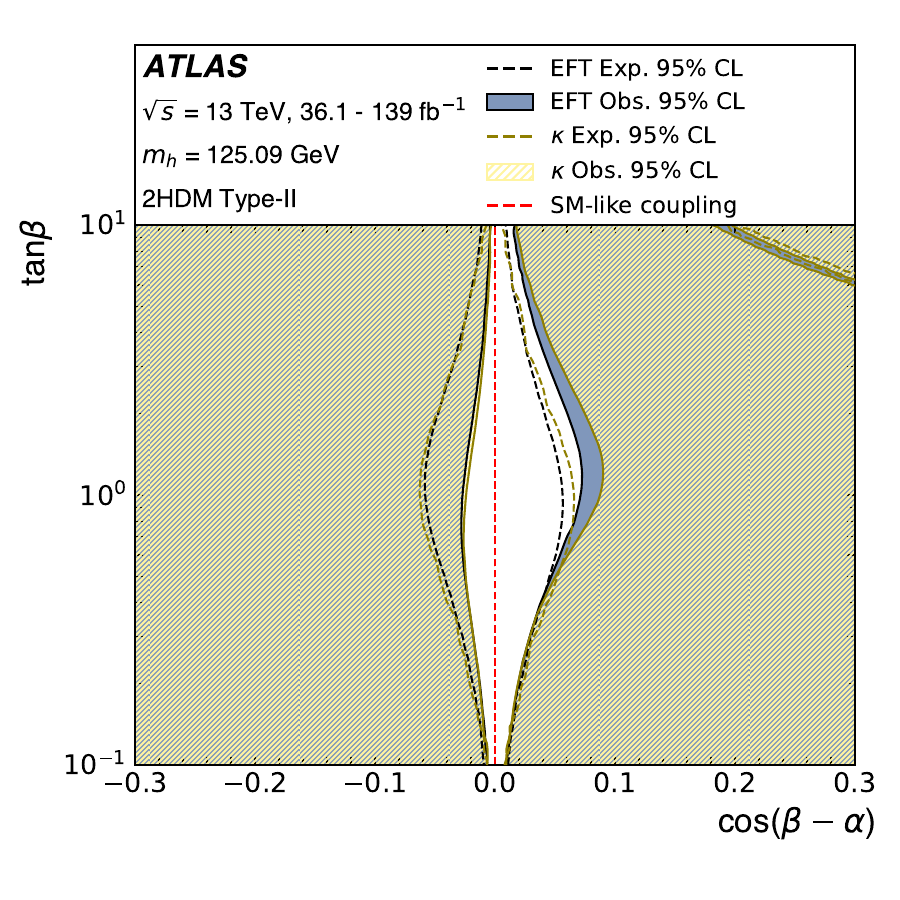}}\\
\subfloat[]{\includegraphics[width=.48\textwidth]{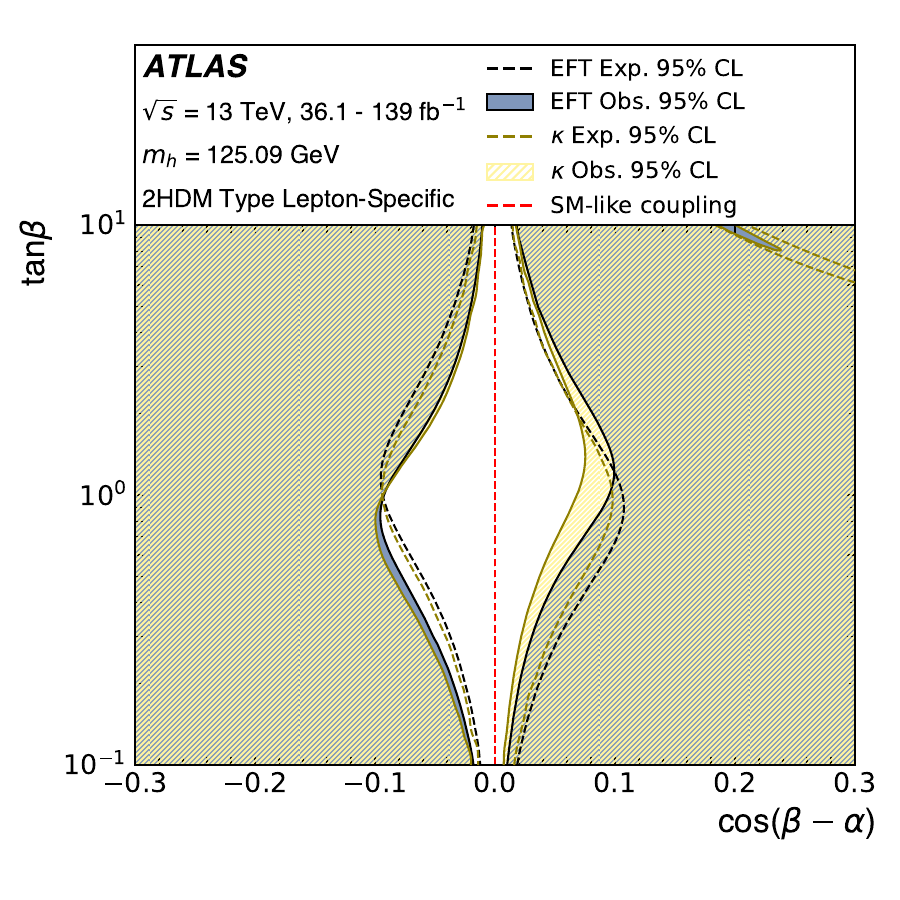}}
\subfloat[]{\includegraphics[width=.48\textwidth]{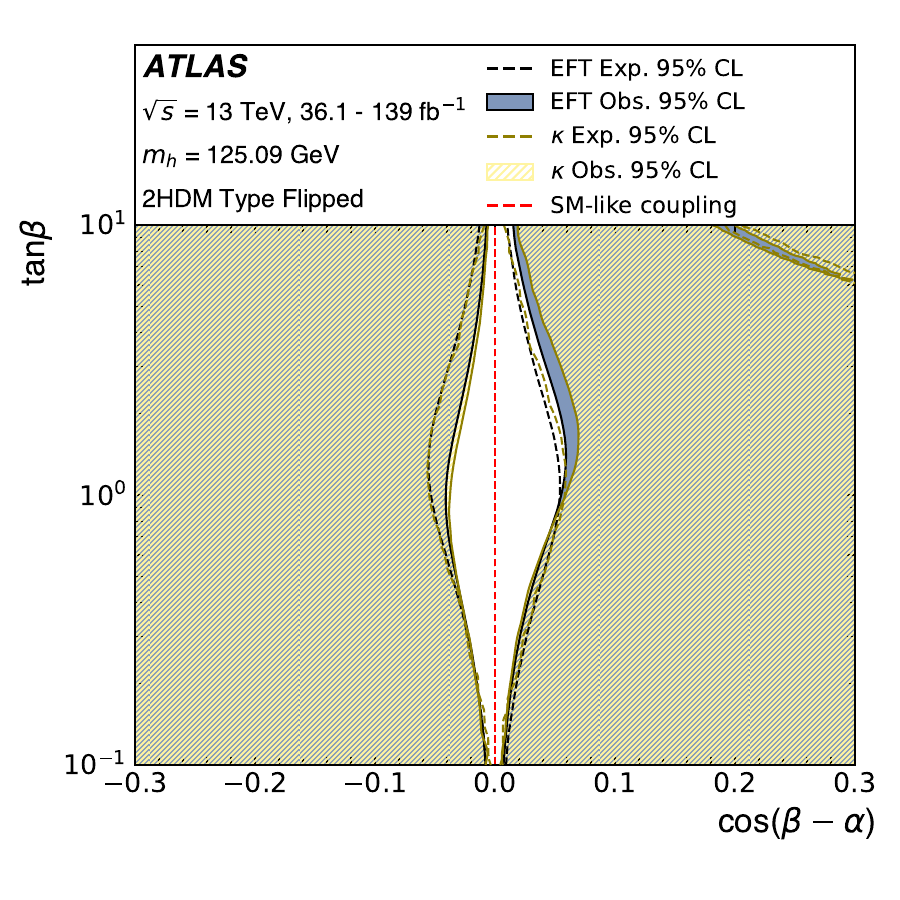}}
\caption{Comparison of the constraints from the approaches based on the
$\kappa$- and EFT-frameworks
in the \tb, \cbma\ plane in the (a) Type-I,
(b) Type-II, (c) Lepton-specific and (d) Flipped models.
The $\kappa_{\lambda}$ constraint is included in the Type-I model
interpretation.
Solid (dashed) lines indicate the observed (expected) constraints.
Results are derived assuming that $|\cos(\beta-\alpha)|\ll 1$,
near the alignment limit represented by the red dashed lines,
and that the masses of the non-SM-like Higgs bosons are
large compared to the SM vev. In particular, the EFT limits on the Type-I model at high $\tan\beta$
are likely outside the validity range of the assumption made in the model.
}
\label{fig:compare_kappa_smeft}
\end{figure}

A significant difference between the $\tb$, $\cbma$ constraints
from the approach presented in this section compared to those obtained using
the $\kappa$-framework-based approach is the absence of the
second small allowed region around $\tb \approx 10$, $\cbma\approx 0.2$.
This difference follows from the fact that in the EFT-based approach
only dimension-6 terms are considered in the SMEFT Lagrangian
and a linear expansion of $\sigma \times B$,
which does not include terms of second order in the Wilson coefficients,
is performed.
The likelihood function in the EFT-based approach is thus Gaussian
and has a single maximum~\cite{Belusca-Maito:2016dqe}.
As a consequence, the region with a flipped coupling sign does not appear.

\FloatBarrier


%

\subsection{Constraints on the MSSM}
\label{sec:mssm}

Supersymmetry (SUSY)~\cite{Miyazawa:1966, Ramond:1971gb, Golfand:1971iw,
Neveu:1971rx, Neveu:1971iv, Gervais:1971ji, Volkov:1973ix, Wess:1973kz,
Wess:1974tw}, a theoretically motivated framework for extending the
SM, was conceived to address some of the SM's
unanswered questions. In particular, SUSY offers a solution to the hierarchy
problem~\cite{Weinberg:1975gm,Gildener:1976ai,Weinberg:1979bn,Susskind:1978ms},
which is related to the fine tuning needed to obtain the relatively small
value of the Higgs boson mass compared to the Planck scale.
Supersymmetry can also provide credible dark
matter candidates~\cite{Goldberg:1983nd, Ellis:1983ew} and can improve
the unification of the electroweak and strong
interactions~\cite{Dimopoulos:1981yj, Sakai:1981gr, Ibanez:1981yh,
Einhorn:1981sx,Marciano:1981un,Giunti:1991ta,Ellis:1990wk,Amaldi:1991cn,
Langacker:1991an}.

Measurements of Higgs production and decay are interpreted in the context of
the minimal supersymmetric extension of the SM
(MSSM)~\cite{Fayet:1976et,
Fayet:1977yc,Farrar:1978xj,Fayet:1979sa,Dimopoulos:1981zb}.
The Higgs sector of the MSSM is a 2HDM with Type-II Yukawa couplings.
At tree level, all Higgs boson masses and couplings can be expressed in
terms of two parameters, usually chosen to be \tb and the mass of the
pseudoscalar Higgs boson, $m_A$.
Radiative corrections, including contributions from amplitudes mediated
by loops of SM particles and their superpartners, lead to modifications
of the tree-level relations and can raise the lightest CP-even Higgs boson
mass $m_h$, bounded at tree level by $m_h \leq m_Z|\cos 2\beta|$,
to the level of the mass of the Higgs boson discovered by the
ATLAS and CMS collaborations.
The MSSM predicts a supersymmetric partner for each SM state.
In addition to the supersymmetric spin-one partners of the SM fermions
(sfermions) and spin-half partners of the gluons (gluinos), there are a
total of eight spin-half partners of the electroweak gauge and Higgs
bosons: the neutral bino (superpartner of the $U(1)$ gauge field), the winos,
which are a charged pair and a neutral particle (superpartners
of the $SU(2)_L$ gauge fields) and the higgsinos, which are two neutral
particles and a charged pair (superpartners of the Higgs fields).
Mixing between the bino, the neutral wino and the neutral
higgsinos leads to four neutralinos, while charginos arise from mixing between
the charged winos and the charged higgsinos.

In the MSSM, SUSY is explicitly broken by some soft SUSY breaking
operators resulting from a separate sector with unspecified dynamics.
As a consequence, the MSSM has over a hundred parameters that describe
the pattern of sparticle masses and their decays, and the values of the
quantum corrections that they induce vary significantly as a function
of these parameters.
The parameter space is thus too large to be probed exhaustively
through a relatively small set of Higgs boson production and decay rate
measurements. Therefore, several {\em benchmark
scenarios} have been proposed in which the values of the MSSM parameters
other than \tb and $m_A$ are fixed to provide a wide variety of
phenomenologies of the MSSM Higgs sector.
With only two free parameters left, \tb and $m_A$, an MSSM interpretation
of the Higgs boson production and decay rate measurements is simplified.

\subsubsection{MSSM benchmark scenarios}

In this study, the measured production and decay rates of the observed
Higgs boson with mass close to 125~\GeV\ are compared with the predictions of
eight benchmark scenarios~\cite{Bahl:2018zmf,Bahl:2019ago,Djouadi:2013uqa},
assuming that the observed boson is the lightest CP-even
Higgs boson $h$ of the MSSM\@.
In each benchmark scenario, constraints are set on $m_A$ (or $m_{H^\pm}$) and
$\tan\beta$ after fixing the values of all the other MSSM parameters.

One of the models considered is the hMSSM~\cite{Djouadi:2013uqa}, which was also
investigated in our previous publication~\cite{HIGG-2018-57}.
The use of this model as a benchmark MSSM scenario
suffers from theoretical limitations in regions with small
$m_A$ or large $\tan\beta$ because its assumptions
may not hold in these regions of parameter space~\cite{Bahl:2018zmf}.
For this reason, seven more recent benchmark MSSM
scenarios~\cite{Bahl:2018zmf,Bahl:2019ago} were also explored.
These include two new scenarios~\cite{Bahl:2019ago}
that were recently designed to be compatible with a low value of
$\tan \beta$ ($< 10$). This parameter space region is otherwise excluded
in most MSSM scenarios, which predict for $\tan \beta \lesssim 10$
a mass of the lightest neutral
Higgs boson significantly lower than the observed Higgs boson mass
of $125.09\pm 0.24$~\GeV~\cite{HIGG-2014-14}.

The scenarios that are considered can be briefly summarised as follows,
with details provided in the references listed in
Table~\ref{tab:mssm_benchmarks}:
\begin{table}[!htbp]
\caption{Benchmark scenarios considered for the MSSM interpretation.}
\label{tab:mssm_benchmarks}
\centering
\begin{tabular}{lr}
\toprule
Scenario & Reference \\
\midrule
\mssmMh        & \cite{Bahl:2018zmf} \\
\mssmMhtau     & \cite{Bahl:2018zmf} \\
\mssmMhchi     & \cite{Bahl:2018zmf} \\
\mssmMhalign   & \cite{Bahl:2018zmf} \\
$M_{h_1}^{125}(\mathrm{CPV})$ & \cite{Bahl:2018zmf} \\
\mssmMhEFT     & \cite{Bahl:2019ago} \\
\mssmMhEFTchi  & \cite{Bahl:2019ago} \\
hMSSM          & \cite{Djouadi:2013uqa} \\
\bottomrule
\end{tabular}
\end{table}
\begin{itemize}
\item \textbf{\mssmMh scenario:}
All superparticles are assumed to be so heavy that they affect the
production and decay rates of the MSSM Higgs bosons very slightly.
The loop-induced contributions to the couplings of the light CP-even
scalar are small.
The largest ones, enhanced with $\tb$, affect the $b$-quark
Yukawa coupling but cancel out in the decoupling limit $m_A \gg m_Z$.
\item \textbf{\mssmMhtau scenario:}
The parameters related to the stop, sbottom and gluino masses are the same
as in the previous case, but the soft-SUSY-breaking masses and trilinear
interaction term for the staus are significantly smaller.
The resulting light staus and light wino-like charginos can alter
the amplitude for the decay of the SM-like Higgs boson into photons,
with contributions that scale as $1/\tb$ and $\tan^2\beta$, respectively.
As in the previous scenario, at low $m_A$ the $hbb$ coupling is enhanced,
leading to a suppression of the branching ratios for all the other
decay channels of the SM-like Higgs boson.

\item \textbf{\mssmMhchi scenario:}
The SUSY parameters in this scenario lead to all charginos and neutralinos
being relatively light, with significant
higgsino--gaugino mixing, and to a compressed electroweakino mass spectrum.
The \tb-enhanced corrections to the $hbb$ coupling are suppressed.
At low values of \tb, the partial width for the SM-like Higgs
boson decay into two photons is significantly increased.
The corresponding branching ratio for low \tb is larger (smaller)
than the SM prediction for large (small) values of $m_A$.

\item \textbf{\mssmMhalign scenario:}
In the `alignment without decoupling' scenario, for a given value of
$\tan\beta$, one of the two neutral CP-even scalars has
SM-like couplings independent of the mass spectrum of the remaining
Higgs bosons. In particular, for $\tan\beta\approx 7$ the lightest
scalar Higgs boson $h$ is SM-like, for relatively low values of $m_A$.

\item \textbf{$M_{h_1}^{125}(\mathrm{CPV})$ scenario:}
This scenario is characterised by CP violation in the Higgs sector,
where mixing between the neutral CP-even scalar states $h$ and $H$
and the CP-odd scalar state $A$ is induced by a non-zero phase
$\phi_{A_t}$ in the soft-SUSY-breaking Higgs--stop interaction term.
This causes significant interference effects in the production and
decay of the two heavier neutral states, which leads to weaker
constraints from searches for resonances decaying into $\tau^+\tau^-$.
The relevant parameters are chosen such that the strongest interference
region is located near the exclusion contour of the corresponding scenario
with real parameters, the mass of the SM-like scalar is near the observed value,
and all electric dipole moments are within the experimentally allowed ranges.

\item \textbf{\mssmMhEFT scenario:}
In the previous scenarios,
the sfermions have masses close to the \TeV\ scale.
As a consequence, values of $\tan\beta<5$ are automatically excluded
as they would lead to values of the Higgs boson mass $m_h$ that are
significantly lower than the experimental value of 125~\GeV.
To circumvent this problem,
in this scenario all SUSY particles,
including sfermions, are relatively heavy, with masses close to the
SUSY scale $m_{\textrm{SUSY}}$, which is adjusted dynamically
from $6~\TeV$ to $10^{13}~\TeV$ to yield a Higgs boson mass of 125~\GeV
for values of $\tan\beta$ between one and ten.
As all SUSY particles are so heavy that
production and decays of the MSSM Higgs bosons are only mildly
affected by their presence, their contribution to the Higgs boson
properties is calculated with an effective field theory.
As in most of the previous scenarios,
for low values of $m_A$ the $hbb$ coupling is enhanced with
respect to the SM prediction and reduces to the SM value for large
$m_A$. At very small $\tan\beta$, the model predicts a small
suppression of the $htt$ coupling, leading in turn to a slight
suppression of the \ggF\ production cross-section.

\item \textbf{\mssmMhEFTchi scenario:}
Similarly to the previous scenario, the SUSY scale is adjusted dynamically
to yield a light Higgs boson mass of $m_h \approx 125~\text{\GeV}$ at each
point of the SUSY parameter space.
In contrast to the \mssmMhEFT scenario, however, this scenario
features relatively light neutralinos and charginos.
As a consequence, loop-induced amplitudes mediated by these light
states can induce modifications of the light Higgs boson properties
such as an enhancement of its partial decay width to two photons,
in particular for low $\tan\beta$.
The model therefore predicts a branching ratio for $h\to\gamma\gamma$
significantly larger than the SM prediction for large $m_A$ and
low $\tan\beta$.
At low values of $m_A$, the enhancement of the $hbb$ coupling
leads instead to an increase of the $h\to b\bar{b}$ branching
ratio and a decrease of the $h\to\gamma\gamma$ branching ratio
relative to their SM expectations.
\item \textbf{hMSSM scenario:}
In this scenario, the lightest Higgs boson $h$ is assumed to
be the particle discovered by the ATLAS and CMS collaborations in 2012.
Its mass $m_h$ is assumed to be equal to the experimental value of 125.09~\GeV,
and it is used to fix the values of the dominant radiative corrections to the
CP-even Higgs mass matrix from the stop--top sector.
This hypothesis determines all other parameters of the Higgs sector,
including $\alpha$, $m_H$, $m_{H^\pm}$, and the couplings between the Higgs
bosons and the other particles, in terms of only the two parameters
$m_A$ and $\tan\beta$.
Due to its assumptions, the hMSSM is a good approximation of the MSSM
only for moderate values of $\tan \beta$,
$1 \lesssim \tb \lesssim 10$~\cite{HIGG-2018-57}.
\end{itemize}

\subsubsection{Interpretation procedure}

Since Higgs boson production cross-sections in the MSSM are not
calculated for the full STXS Stage-1.2 phase space partitioning of
production cross-sections, only inclusive production-mode
cross-sections are used in this analysis.
The signal strength modifiers $\mu^{i,X}$ in the signal yield expression of
Eq.~\eqref{eq:Lsk} are then written as:
\begin{equation}
\label{eq:nexp_mssm}
\mu^{i,X}(m_A,\tan\beta) = \frac{\sigma^{i}_\mathrm{(h)MSSM}(m_A,\tan\beta)}{\sigma^i_\text{SM}} \cdot \frac{\BR^{X}_\mathrm{(h)MSSM}(m_A,\tan\beta)}{\BR^X_\text{SM}} \equiv r^i(m_A,\tan\beta)\cdot r^X(m_A,\tan\beta),
\end{equation}
\noindent where the index $i$ enumerates the
production processes (\ggF, \bbH, \ttH, \VBF, \qqZH, \ggZH, \qqWH).

The values of the production cross-section scale factors $r^i$ and
branching ratio scale factors $r^X$ are obtained using the procedures
defined by the LHC Higgs Cross-section Working Group, described in~\cite{deFlorian:2016spz},
as well as the calculations detailed in~\cite{Bagnaschi:2791954}.
For all scenarios except the hMSSM,
Higgs boson masses and mixing (and effective Yukawa couplings) are
calculated with \textsc{FeynHiggs}~\cite{Heinemeyer:1998yj, Heinemeyer:1998np,
Degrassi:2002fi, Frank:2006yh, Hahn:2013ria, Bahl:2016brp,
Bahl:2017aev}, and branching ratios are obtained combining the most precise
estimates of \textsc{FeynHiggs},
\textsc{HDECAY}~\cite{Djouadi:1997yw, Djouadi:2018xqq}
and \textsc{PROPHECY4f}~\cite{Bredenstein:2006rh, Bredenstein:2006ha}.
In the hMSSM scenario, the Higgs boson mass is fixed at 125~\GeV, the mixing
is calculated analytically, and the branching ratios are calculated using
the $\kappa$-framework.
Inclusive cross-sections for \ggF{} production are obtained with
\textsc{SusHi}~\cite{Harlander:2012pb, Harlander:2016hcx}, which includes NLO
QCD corrections~\cite{Spira:1995rr}, NNLO QCD corrections for the
top-quark contribution in the effective theory of a heavy top
quark~\cite{Harlander:2002wh, Anastasiou:2002yz,
Ravindran:2003um,Harlander:2002vv, Anastasiou:2002wq} and EW
corrections involving light quarks~\cite{Aglietti:2004nj, Bonciani:2010ms}.
For the SM-like Higgs boson, \textsc{SusHi}
adds N$^3$LO corrections in the effective
theory of a heavy top quark in a threshold
expansion~\cite{Anastasiou:2014lda, Anastasiou:2015yha,
Anastasiou:2016cez}.
Cross-sections for $b$-quark-initiated Higgs boson
production rely on matched predictions~\cite{Bonvini:2015pxa,
Bonvini:2016fgf, Forte:2015hba, Forte:2016sja}, which are based on the
five flavour NNLO QCD calculation~\cite{Harlander:2003ai} and the
four flavour NLO QCD calculation~\cite{Dittmaier:2003ej,
Dawson:2003kb}.
The values of the cross-sections of rare production modes, such
as $gg\to Zh$, $pp\to thW$ and $pp \to thq$,
which are not calculated in Refs.~\cite{TwikiMSSM,Bagnaschi:2791954},
are determined through their relations, within the $\kappa$-framework,
to $\kappa_t$, $\kappa_b$, $\kappa_W$, $\kappa_Z$,
which are determined as either
$\kappa_t^2 = \sigma_\mathrm{MSSM}^{t\bar{t}h}/\sigma_\mathrm{SM}^{t\bar{t}H}$
or
$\kappa_x^2 = \Gamma^{h\to xx}_\text{MSSM} / \Gamma^{h\to xx}_\text{SM}$,
where $x=b,W,Z$:
\begin{eqnarray}
r^{gg\to Zh} & =&  2.456 \kappa_Z^{2}  + 0.456 \kappa_t^{2}
- 1.903 \kappa_Z  \kappa_t - 0.011 \kappa_Z \kappa_b
+ 0.003 \kappa_t \kappa_b , \nonumber \\
r^{thW} &=& 2.909 \kappa_t^{2} + 2.310 \kappa_W^{2} - 4.220 \kappa_t \kappa_W , \nonumber \\
r^{thq} &=& 2.633 \kappa_t^{2} + 3.578 \kappa_W^{2} - 5.211 \kappa_t \kappa_W .\nonumber
\end{eqnarray}

As the cross-section and branching ratio scale factors $r^i$ and $r^X$ are only
calculated for a discrete set of points in the two-dimensional benchmark
parameter space, a likelihood based on Eq.~\eqref{eq:nexp_mssm}
is not differentially expressed in terms of its theory parameters
($m_A,\tan \beta$). Instead, the values of the profile likelihood ratio are
calculated at fixed points in the $(m_A,\tan \beta)$ plane, profiling only
the nuisance parameters, and then numerically interpolated.
Regions of the benchmark scenario parameter space for which the mass
$m_h$ of the lightest CP-even Higgs boson $h$ predicted by the MSSM differs
from the experimental value of 125.09~\GeV\ by more than 3~\GeV, which is
the estimated bound of allowed values of the calculated Higgs boson mass in
\textsc{FeynHiggs}~\cite{Degrassi:2002fi,Allanach:2004rh},
are considered as excluded.


\FloatBarrier
%

\subsubsection{Results}

Figure\,\ref{fig:mh125set1} shows the observed and expected 95\% CL exclusion
limits of the MSSM in the two-dimensional plane ($m_A$, $\tan\beta$) for
the \mssmMh, \mssmMhchi, \mssmMhtau, and \mssmMhalign{} benchmark scenarios.
For all four scenarios, the regions excluded by the Higgs boson mass requirement
($|m_h-125.09~\GeV|<3~\GeV$) are separately indicated with gray shaded
areas. For completeness, the constraints from previous searches for
$H/A\to\tau\tau$~\cite{HDBS-2018-46} and $H^+ \to t\bar{b}$~\cite{HDBS-2018-51}
are also overlaid.

\begin{figure}[tbp]
\centering
\subfloat[]{\includegraphics[width=0.49\textwidth]{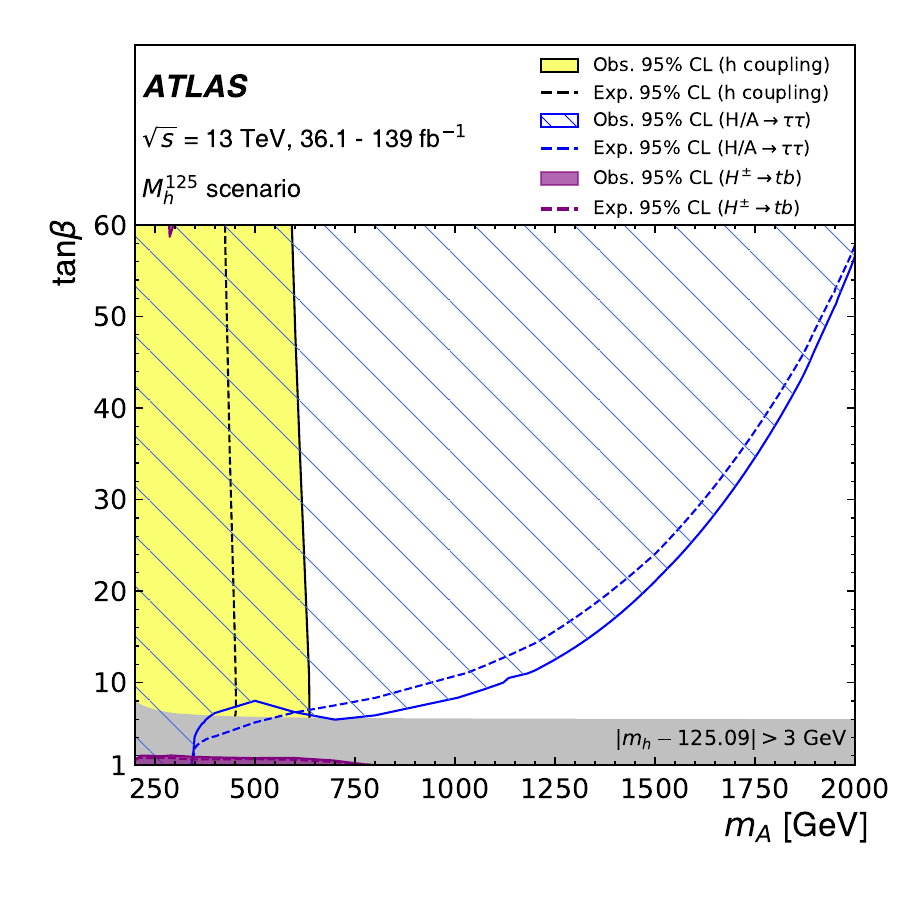}}
\subfloat[]{\includegraphics[width=0.49\textwidth]{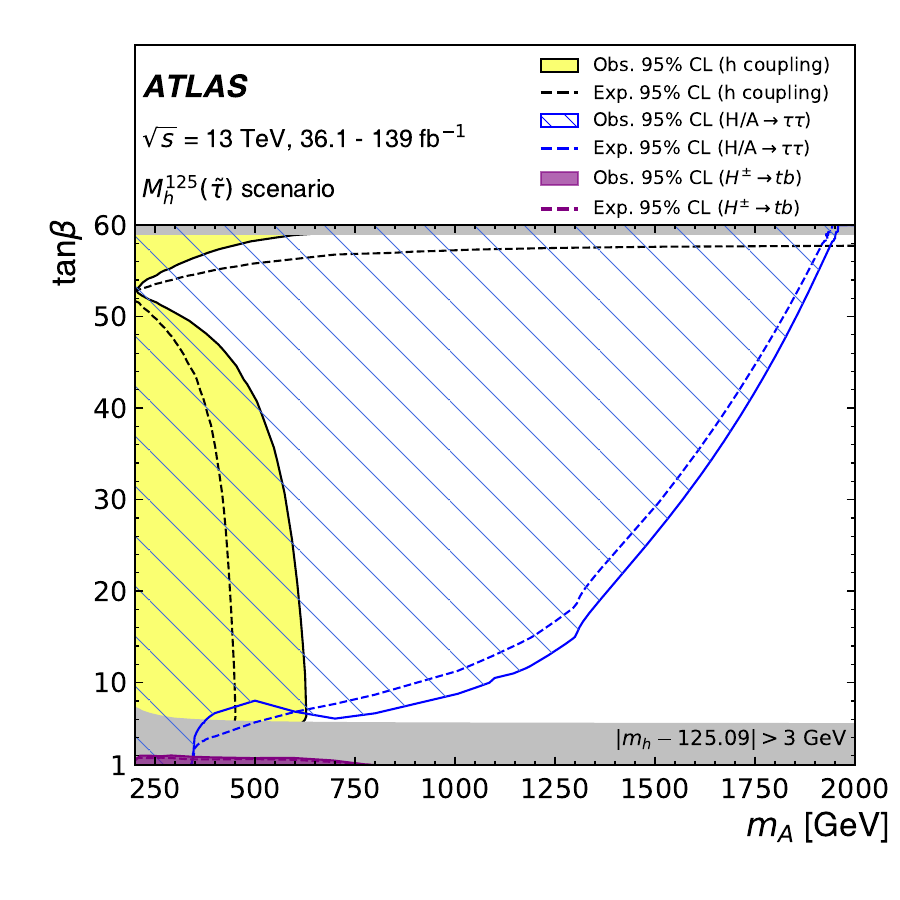}} \\
\subfloat[]{\includegraphics[width=0.49\textwidth]{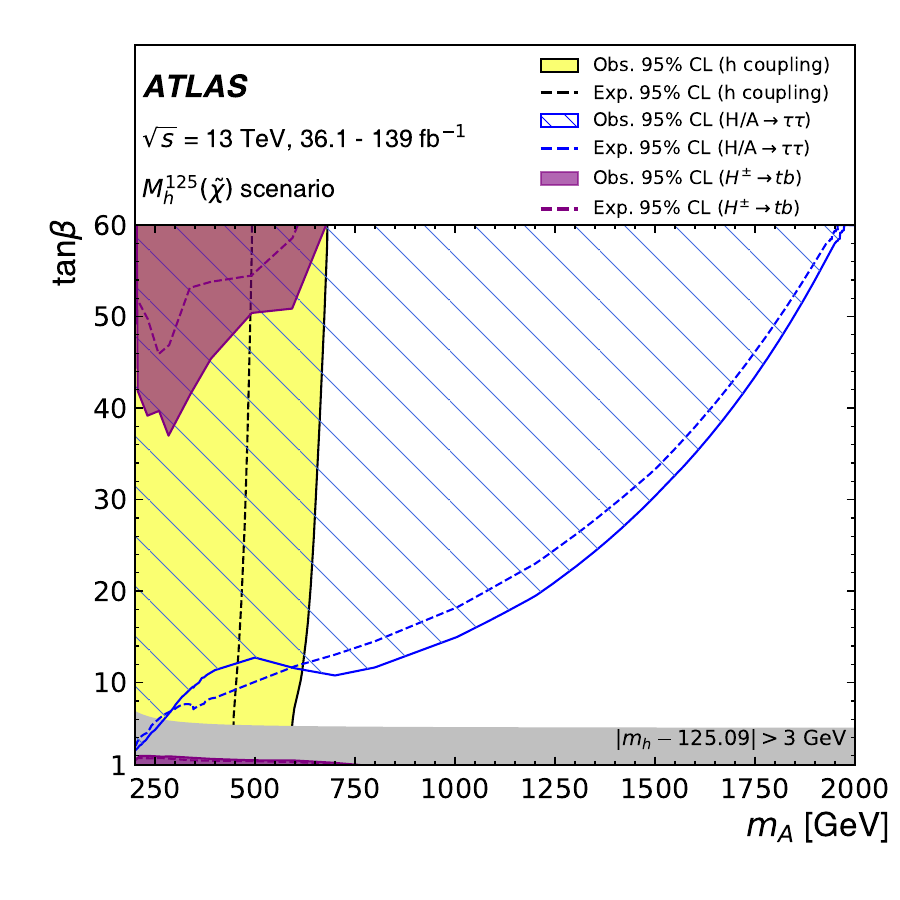}}
\subfloat[]{\includegraphics[width=0.49\textwidth]{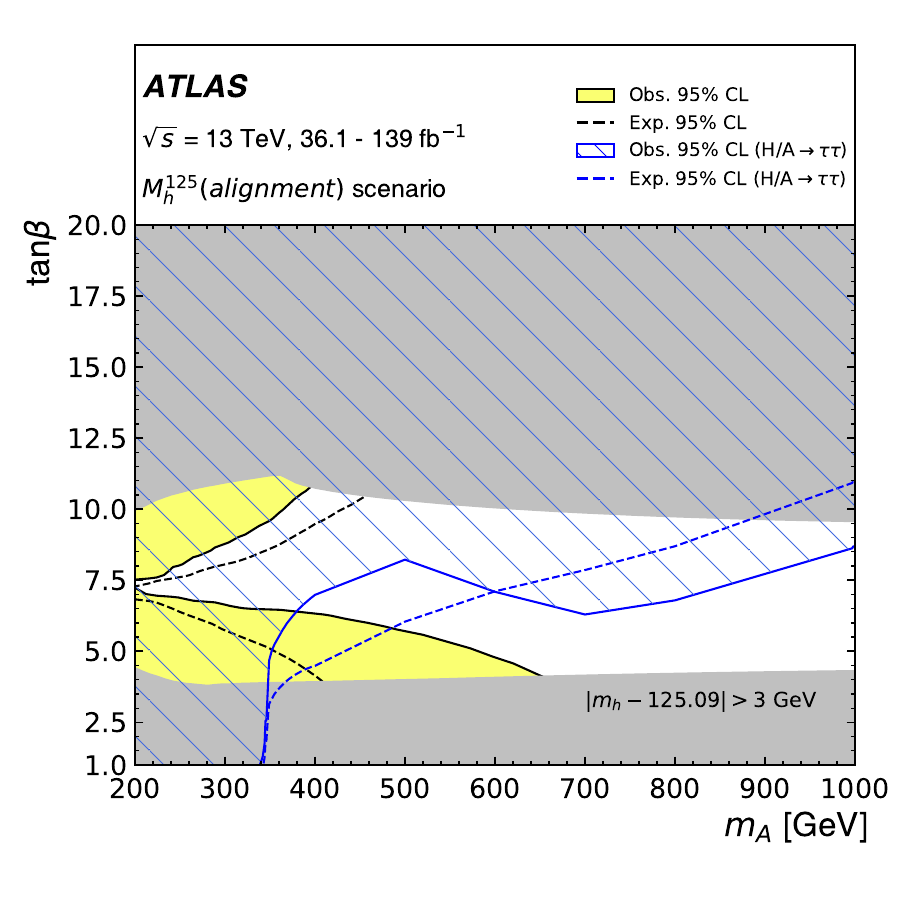}}
\caption{Observed (solid yellow area) and expected (dashed line)
exclusion contours at 95\% CL in the $(m_A,\tan\beta)$ plane for the
(a) \mssmMh{}, (b) \mssmMhtau{},
(c) \mssmMhchi{} and  (d) \mssmMhalign{}  scenarios.
The parameter space excluded by the condition $|m_h-125.09~\GeV|<3$~\GeV\
is marked in gray.
For comparison, the parameter space excluded by the searches for
$H/A\to \tau\tau$~\cite{HDBS-2018-46}
and for $H^+ \to t\bar{b}$~\cite{HDBS-2018-51} are overlaid in blue
and purple, respectively.
For the \mssmMhalign{} scenario, no constraints from the search for
$H^+ \to t\bar{b}$ are available.
}
\label{fig:mh125set1}
\end{figure}

In the \mssmMh{} and \mssmMhchi{} scenarios, the low $m_A$ region,
in which an enhancement of the $h \to b\bar{b}$ branching ratio
and a suppression of the branching ratios of the other decay
modes such as $h\to\gamma\gamma$ are predicted,
is disfavoured.
Since the data marginally prefer
$B(h \to b\bar{b})/B_\mathrm{SM}(h\to b\bar{b})<1$
and $B(h \to \gamma\gamma)/B_\mathrm{SM}(h\to \gamma\gamma)>1$,
the observed constraints are tighter than those expected under
the SM hypothesis.
In the \mssmMhtau{} scenario, the region at low $m_A$ and $\tan\beta<53$ is
excluded for the same reason.
The region $\tan\beta>53$, in which the $\tilde{\tau}$ loop reduces the
$hbb$ coupling, leading to an enhanced prediction of
$B(h\to \gamma\gamma)$, is also largely excluded.
In the \mssmMhalign{} scenario, the limit of alignment without decoupling is
only realised for $\tan\beta \approx 7$ and $m_A \approx 200~\GeV$.
For larger values of $m_A$, MSSM couplings are more similar to SM couplings,
causing the allowed $\tan\beta$ region to broaden.

In all four benchmark scenarios, the MSSM analysis generally excludes
the low $m_A$ regime for most of the scanned $\tan \beta$
range, while the
requirement $|m_h-125.09~\GeV|<3$~\GeV\ excludes the low
$\tan \beta$ range for all scanned values of $m_A$,
a range not fully covered by the direct searches for
$H/A\to \tau\tau$ and $H^+\to t\bar{b}$ decays.
The exclusion range due to the $m_h$ requirement is qualitatively different in the \mssmMhalign{}
scenario, as this scenario can only realized for a value of $\tan(\beta) \approx 7$.

Figure\,\ref{fig:mh125set2} shows the observed and expected 95\% CL exclusion
limits on the MSSM in the two-dimensional plane of
$m_{H^\pm}$ versus $\tan\beta$
for the \mssmMhCPV scenario and of
$m_A$ versus $\tan\beta$ for the \mssmMhEFT{}, \mssmMhEFTchi{}
and hMSSM scenarios.
The condition $|m_h-125.09~\GeV|<3~\GeV$ excludes values of
$\tb \lesssim 6$ for all values of $m_{H^\pm}$ in the \mssmMhCPV scenario,
while no part of the parameter space is excluded in the
\mssmMhEFT{} and \mssmMhEFTchi{} benchmark scenarios,
in which $m_h=125.09$~\GeV can be achieved through the flexible
$m_\text{SUSY}$ scale.
In these four models, the low $m_A$ region,
where an enhancement of the $hbb$ coupling is predicted, is excluded,
and the excluded region is larger than expected in the SM hypothesis
for the same reason as for the previous four models.
The similarity of the limits in the  \mssmMhEFT{} and hMSSM scenarios stems
from the fact that the \mssmMhEFT{} is a concrete realisation of an
MSSM scenario that fulfils the assumptions of the hMSSM approach~\cite{Bahl:2019ago}.

\begin{figure}[tbp]
\centering
\subfloat[]{\includegraphics[width=0.49\textwidth]{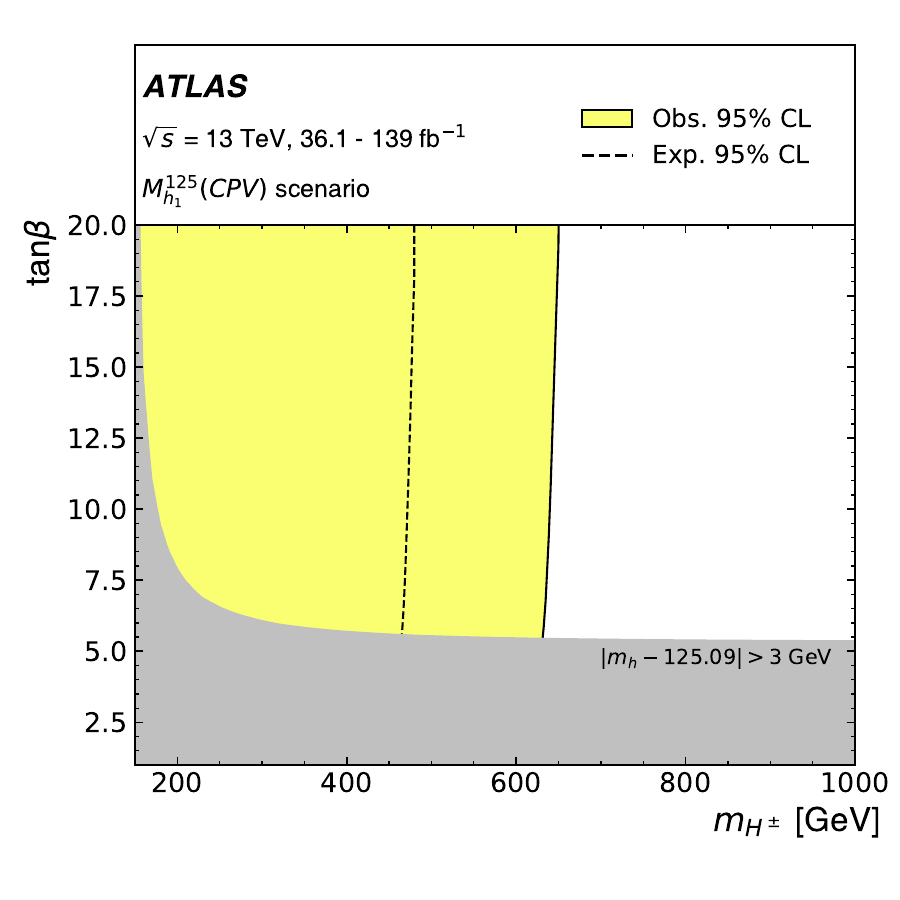}}
\subfloat[]{\includegraphics[width=0.49\textwidth]{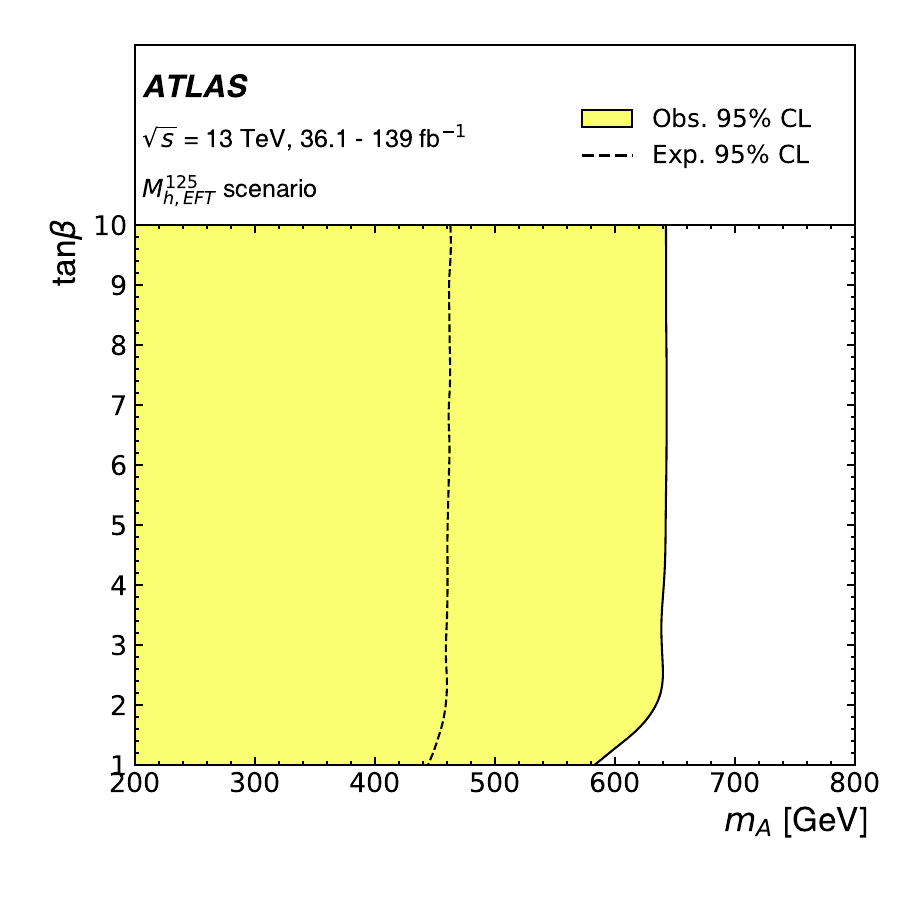}} \\
\subfloat[]{\includegraphics[width=0.49\textwidth]{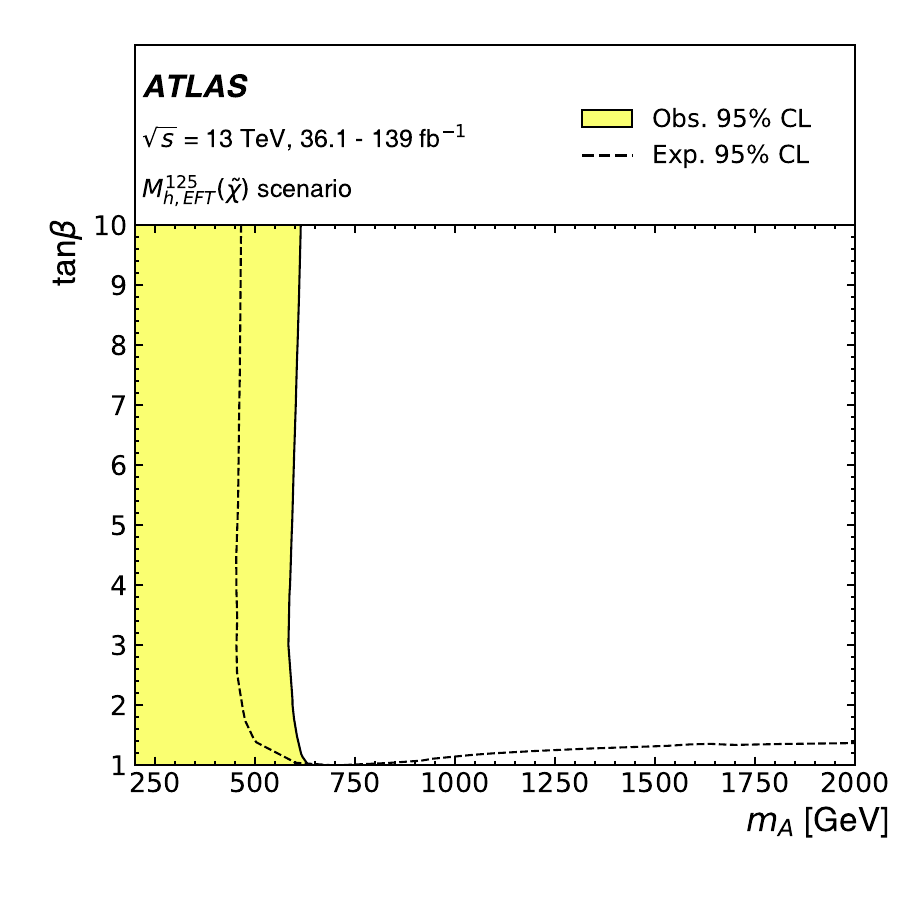}}
\subfloat[]{\includegraphics[width=0.49\textwidth]{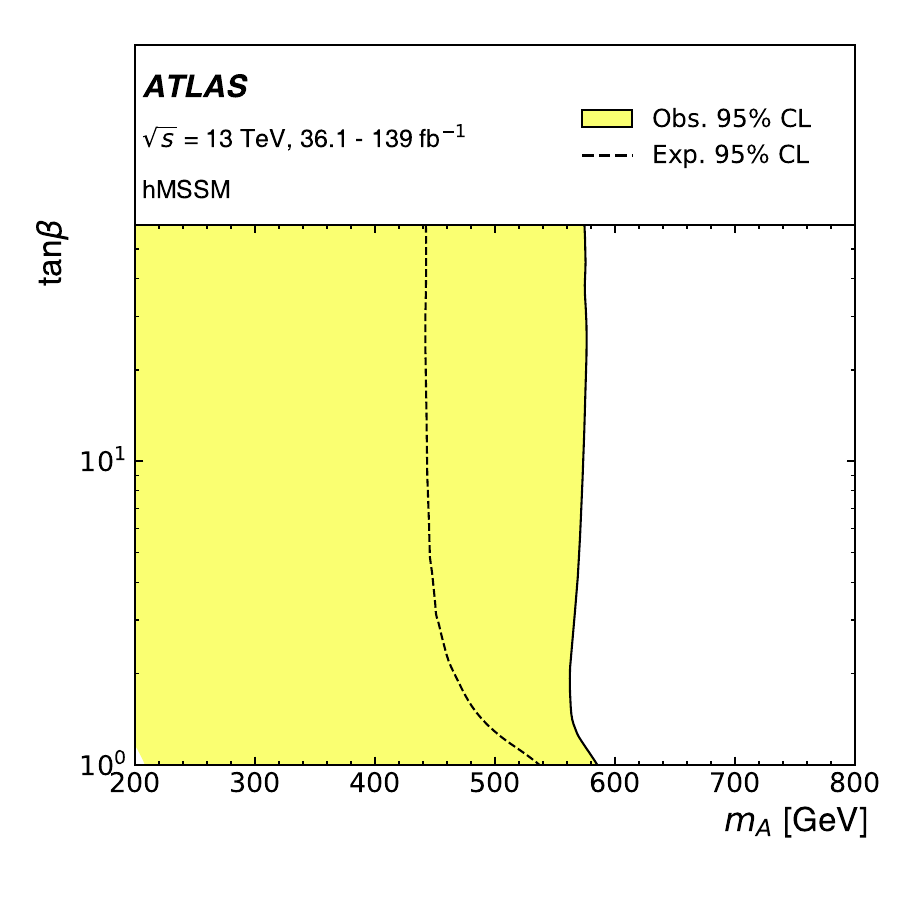}}
\caption{Observed (solid yellow area) and expected (dashed line)
exclusion contours at 95\% CL in (a) the
$(m_{H^\pm},\tan\beta)$ plane for the $M_{h_1}^{125}(\mathrm{CPV})$
scenario and in the $(m_A,\tan\beta)$ plane for the
(b) \mssmMhEFT{}, (c) \mssmMhEFTchi{} and (d) hMSSM scenarios.
In (a), the parameter space excluded by the condition
$|m_h-125.09~\GeV|<3$~\GeV is marked in gray.
}
\label{fig:mh125set2}
\end{figure}

%
%
%
%
%
%
%
%
%
%
%
%
%
%


\FloatBarrier

\section{Conclusions}
\label{sec:conclusions}
%

Novel interpretations of the recent combined ATLAS measurements of Higgs boson
production and decay rates and simplified template cross-sections
in several final states are presented.
Constraints on linear combinations of Wilson coefficients
corresponding to SM effective field theory operators in the Warsaw
basis are reported.
In this model-independent parameterisation of BSM physics effects,
no significant deviations from the SM are observed.
A comparison of results interpreted with a linearised SMEFT model
that only considers terms suppressed by up to a factor $\Lambda^{-2}$,
and a variant including quadratic terms that considers all available terms
including those with suppression factor $\Lambda^{-4}$,
shows that the effect of operators suppressed by $\Lambda^{-4}$
can significantly affect
constraints on Wilson coefficients for a mass scale of $\Lambda=1$~\TeV.

Constraints on a subset of these operators have also been set through
a joint interpretation of the fiducial cross-section measurements of
Higgs boson production as a function of transverse momentum in the
$H\to\gamma\gamma$ and $H\to ZZ^*\to 4\ell$ final states.
Three linear combinations of the Wilson coefficients of operators
affecting these spectra can be simultaneously constrained.
For two of these combinations, the constraints are similar to those
that can be derived from the interpretation of Higgs boson simplified
template cross-section measurements in the same channels.
The constraint on the top--gluon coupling is significantly stronger
in the interpretation of simplified template cross-sections.

Finally, the Higgs boson production and decay rate measurements
are interpreted in the context of two-Higgs-doublet models
and of eight benchmark scenarios of the minimal supersymmetric
model.
Constraints are set on the $(\cos(\beta-\alpha), \tan\beta)$
plane in 2HDM Type-I, Type-II, Lepton-specific and Flipped models,
and on the parameters $(m_A,\tan\beta)$ of the MSSM in eight
benchmark scenarios.
These results are complementary to limits from direct searches for
additional Higgs bosons.


\FloatBarrier

\section*{Acknowledgements}
%

%
%

%
%

We thank CERN for the very successful operation of the LHC and its injectors, as well as the support staff at
CERN and at our institutions worldwide without whom ATLAS could not be operated efficiently.

The crucial computing support from all WLCG partners is acknowledged gratefully, in particular from CERN, the ATLAS Tier-1 facilities at TRIUMF/SFU (Canada), NDGF (Denmark, Norway, Sweden), CC-IN2P3 (France), KIT/GridKA (Germany), INFN-CNAF (Italy), NL-T1 (Netherlands), PIC (Spain), RAL (UK) and BNL (USA), the Tier-2 facilities worldwide and large non-WLCG resource providers. Major contributors of computing resources are listed in Ref.~\cite{ATL-SOFT-PUB-2023-001}.

We gratefully acknowledge the support of ANPCyT, Argentina; YerPhI, Armenia; ARC, Australia; BMWFW and FWF, Austria; ANAS, Azerbaijan; CNPq and FAPESP, Brazil; NSERC, NRC and CFI, Canada; CERN; ANID, Chile; CAS, MOST and NSFC, China; Minciencias, Colombia; MEYS CR, Czech Republic; DNRF and DNSRC, Denmark; IN2P3-CNRS and CEA-DRF/IRFU, France; SRNSFG, Georgia; BMBF, HGF and MPG, Germany; GSRI, Greece; RGC and Hong Kong SAR, China; ISF and Benoziyo Center, Israel; INFN, Italy; MEXT and JSPS, Japan; CNRST, Morocco; NWO, Netherlands; RCN, Norway; MEiN, Poland; FCT, Portugal; MNE/IFA, Romania; MESTD, Serbia; MSSR, Slovakia; ARIS and MVZI, Slovenia; DSI/NRF, South Africa; MICINN, Spain; SRC and Wallenberg Foundation, Sweden; SERI, SNSF and Cantons of Bern and Geneva, Switzerland; NSTC, Taipei; TENMAK, T\"urkiye; STFC/UKRI, United Kingdom; DOE and NSF, United States of America.

Individual groups and members have received support from BCKDF, CANARIE, CRC and DRAC, Canada; CERN-CZ, PRIMUS 21/SCI/017 and UNCE SCI/013, Czech Republic; COST, ERC, ERDF, Horizon 2020, ICSC-NextGenerationEU and Marie Sk{\l}odowska-Curie Actions, European Union; Investissements d'Avenir Labex, Investissements d'Avenir Idex and ANR, France; DFG and AvH Foundation, Germany; Herakleitos, Thales and Aristeia programmes co-financed by EU-ESF and the Greek NSRF, Greece; BSF-NSF and MINERVA, Israel; Norwegian Financial Mechanism 2014-2021, Norway; NCN and NAWA, Poland; La Caixa Banking Foundation, CERCA Programme Generalitat de Catalunya and PROMETEO and GenT Programmes Generalitat Valenciana, Spain; G\"{o}ran Gustafssons Stiftelse, Sweden; The Royal Society and Leverhulme Trust, United Kingdom.

In addition, individual members wish to acknowledge support from Chile: Agencia Nacional de Investigaci\'on y Desarrollo (FONDECYT 1190886, FONDECYT 1210400, FONDECYT 1230812, FONDECYT 1230987); China: National Natural Science Foundation of China (NSFC - 12175119, NSFC 12275265, NSFC-12075060); Czech Republic: PRIMUS Research Programme (PRIMUS/21/SCI/017); EU: H2020 European Research Council (ERC - 101002463); European Union: European Research Council (ERC - 948254), Horizon 2020 Framework Programme (MUCCA - CHIST-ERA-19-XAI-00), European Union, Future Artificial Intelligence Research (FAIR-NextGenerationEU PE00000013), Italian Center for High Performance Computing, Big Data and Quantum Computing (ICSC, NextGenerationEU), Marie Sklodowska-Curie Actions (EU H2020 MSC IF GRANT NO 101033496); France: Agence Nationale de la Recherche (ANR-20-CE31-0013, ANR-21-CE31-0013, ANR-21-CE31-0022, ANR-22-EDIR-0002), Investissements d'Avenir Idex (ANR-11-LABX-0012), Investissements d'Avenir Labex (ANR-11-LABX-0012); Germany: Baden-Württemberg Stiftung (BW Stiftung-Postdoc Eliteprogramme), Deutsche Forschungsgemeinschaft (DFG - CR 312/5-1); Italy: Istituto Nazionale di Fisica Nucleare (FELLINI G.A. n. 754496, ICSC, NextGenerationEU); Japan: Japan Society for the Promotion of Science (JSPS KAKENHI JP21H05085, JSPS KAKENHI JP22H01227, JSPS KAKENHI JP22H04944); Netherlands: Netherlands Organisation for Scientific Research (NWO Veni 2020 - VI.Veni.202.179); Norway: Research Council of Norway (RCN-314472); Poland: Polish National Agency for Academic Exchange (PPN/PPO/2020/1/00002/U/00001), Polish National Science Centre (NCN 2021/42/E/ST2/00350, NCN OPUS nr 2022/47/B/ST2/03059, NCN UMO-2019/34/E/ST2/00393, UMO-2020/37/B/ST2/01043, UMO-2021/40/C/ST2/00187); Slovenia: Slovenian Research Agency (ARIS grant J1-3010); Spain: BBVA Foundation (LEO22-1-603), Generalitat Valenciana (Artemisa, FEDER, IDIFEDER/2018/048), La Caixa Banking Foundation (LCF/BQ/PI20/11760025), Ministry of Science and Innovation (MCIN \& NextGenEU PCI2022-135018-2, MICIN \& FEDER PID2021-125273NB, RYC2019-028510-I, RYC2020-030254-I, RYC2021-031273-I, RYC2022-038164-I), PROMETEO and GenT Programmes Generalitat Valenciana (CIDEGENT/2019/023, CIDEGENT/2019/027); Sweden: Swedish Research Council (VR 2018-00482, VR 2022-03845, VR 2022-04683, VR grant 2021-03651), Knut and Alice Wallenberg Foundation (KAW 2017.0100, KAW 2018.0157, KAW 2018.0458, KAW 2019.0447); Switzerland: Swiss National Science Foundation (SNSF - PCEFP2\_194658); United Kingdom: Leverhulme Trust (Leverhulme Trust RPG-2020-004); United States of America: U.S. Department of Energy (ECA DE-AC02-76SF00515), Neubauer Family Foundation.

%
%


%
\clearpage
\appendix
\part*{Appendix: Validity of Gaussian approximation of STXS measurements in the SMEFT interpretation }
\addcontentsline{toc}{part}{Appendix: Validity of Gaussian approximation of STXS measurements in SMEFT interpretation }
\renewcommand{\baselinestretch}{1.0}

The STXS SMEFT interpretation model constructs the likelihood function in terms of the fit basis parameters $c^\prime$ from three ingredients:

\begin{enumerate}
\item The likelihood function of the STXS measurements, measuring 78 cross-section parameters, and featuring over a thousand nuisance parameters expressing systematic uncertainties.
\item A specification of the impact of all relevant CP-even Warsaw basis SMEFT operators on the SM predictions of these cross-sections, including acceptance corrections where necessary,
i.e., Eqs.~\eqref{eq:eftlinmodel} and \eqref{eq:eftquadmodel}, using the coefficients $A_j^{\sigma_{i,k^\prime}}$, $A_j^{\Gamma^{H\to X}}$,  $B_{jl}^{\sigma_{i,k^\prime}}$ and $B_{jl}^{\Gamma^{H\to X}}$ as defined in Eqs.~\eqref{eq:EFTparam_production1}-\eqref{eq:EFTtotalwidth}, and visualised in Figure~\ref{fig:bij_impact}.
\item A rotation matrix from the 50 Warsaw basis parameters $\vec{c}$, which cannot be unambiguously constrained from the STXS data, to a fit basis $\vec{c^\prime}$ for which all 19 parameters can be simultaneously constrained
from the data, as shown in Figure~\ref{fig:eft_parRotations}.
\end{enumerate}

An important upside of a likelihood model built this way is that all
non-Gaussian effects in the STXS measurements are accounted for, and
that the correlated effects of all experimental and theoretical
systematic uncertainties are accounted at the level of detail
implemented in the original underlying measurements. A downside of
this likelihood model is that the required complexity of its
implementation prevents it from easily being made available.

An alternative likelihood function, based on a multivariate Gaussian
approximation of the STXS measurements instead of the full
measurement, with an identical SMEFT re-parameterisation, can be
constructed in a relatively straightforward way from the information
provided and represents a reasonably good approximation of
the full likelihood. A comparison of the results obtained with this
simplified Gaussian variant of the likelihood model and those obtained
with the full likelihood function is shown in
Figure~\ref{fig:EFT_mvg_lin} for the linear SMEFT model and in
Figure \ref{fig:EFT_mvg_quad} for the SMEFT model including the
quadratic terms. Both comparisons indicate generally good agreement
between expected and observed parameter estimates and their 68\% and
95\% uncertainty intervals.

\clearpage
\clearpage

\begin{figure}[p]
\centering
\includegraphics[width=0.9\textwidth]{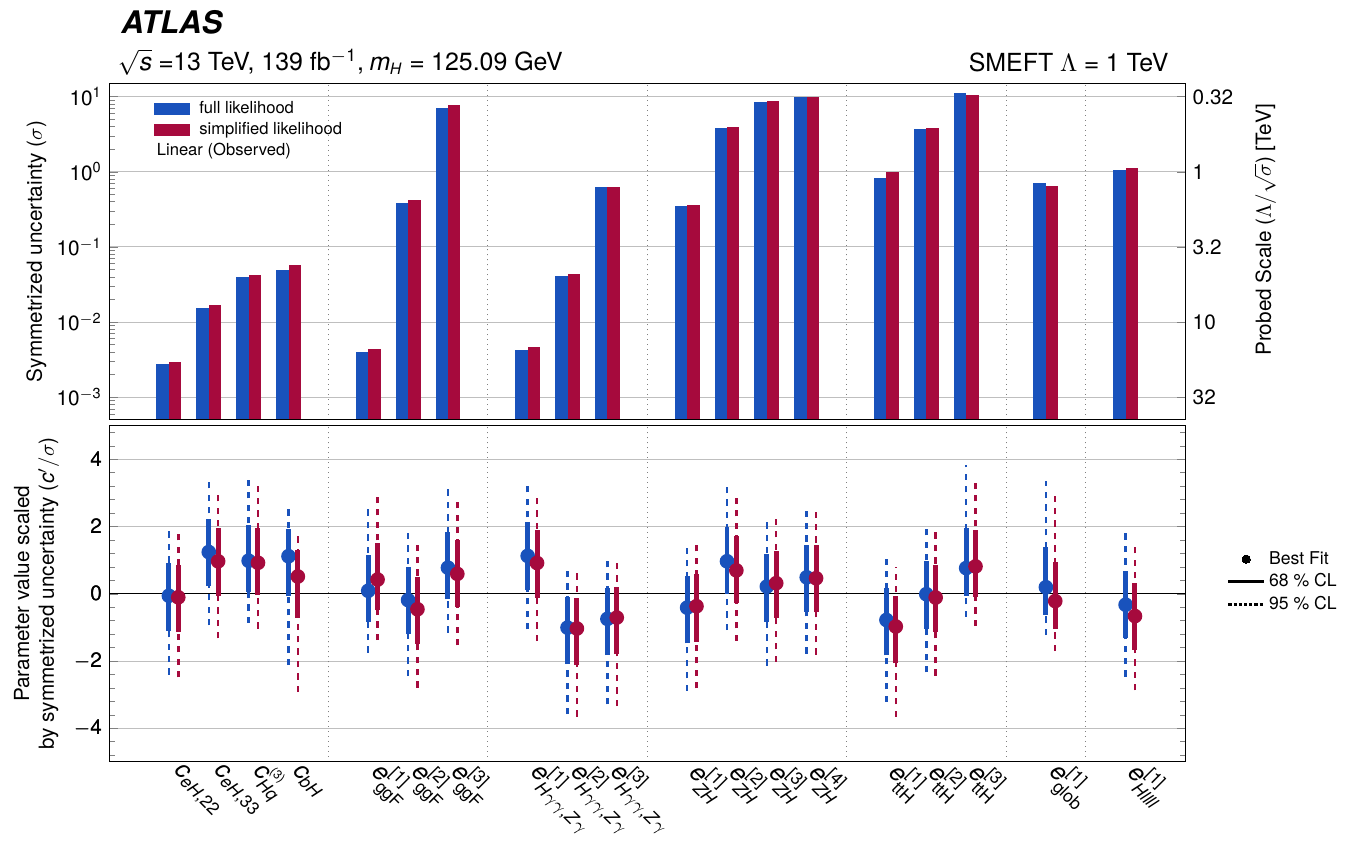}
\caption{Comparison of the observed result from the full likelihood fit
in the SMEFT fit basis with only linear terms (blue) to a fit based on a
simplified likelihood constructed from the covariance matrix of the
STXS measurements rotated to the same SMEFT fit basis (purple).
}
\label{fig:EFT_mvg_lin}
\end{figure}

\begin{figure}[p]
\centering
\includegraphics[width=0.9\textwidth]{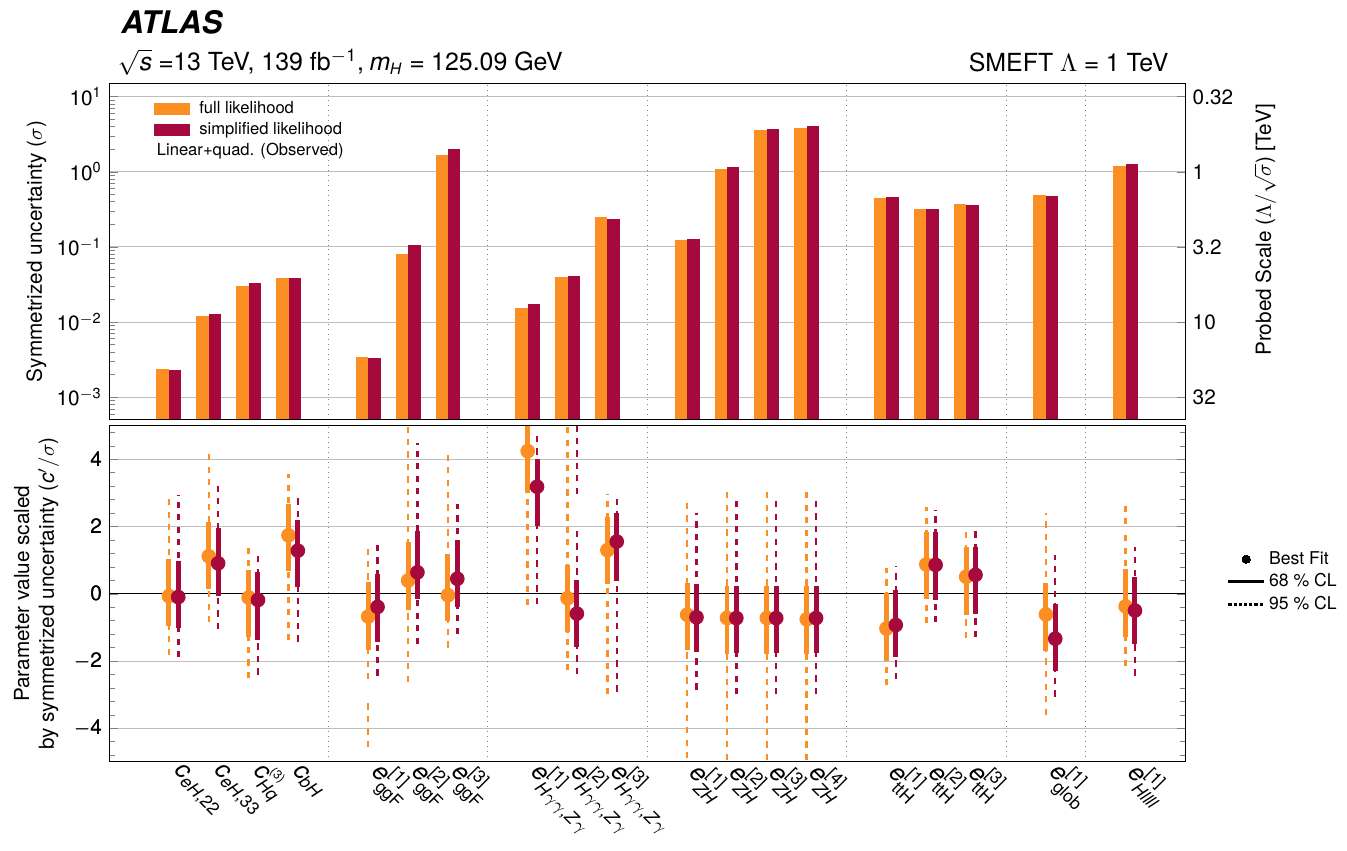}
\caption{Comparison of the observed result from the full likelihood fit
in the SMEFT fit basis with linear and quadratic terms (orange) to a
fit based on a simplified likelihood constructed from the covariance
matrix of the STXS measurements rotated to the same SMEFT fit basis
(dark red).
}
\label{fig:EFT_mvg_quad}
\end{figure}

\FloatBarrier


\FloatBarrier

\printbibliography
\clearpage
 
\begin{flushleft}
\hypersetup{urlcolor=black}
{\Large The ATLAS Collaboration}

\bigskip

\AtlasOrcid[0000-0002-6665-4934]{G.~Aad}$^\textrm{\scriptsize 102}$,
\AtlasOrcid[0000-0002-5888-2734]{B.~Abbott}$^\textrm{\scriptsize 120}$,
\AtlasOrcid[0000-0002-1002-1652]{K.~Abeling}$^\textrm{\scriptsize 55}$,
\AtlasOrcid[0000-0001-5763-2760]{N.J.~Abicht}$^\textrm{\scriptsize 49}$,
\AtlasOrcid[0000-0002-8496-9294]{S.H.~Abidi}$^\textrm{\scriptsize 29}$,
\AtlasOrcid[0000-0002-9987-2292]{A.~Aboulhorma}$^\textrm{\scriptsize 35e}$,
\AtlasOrcid[0000-0001-5329-6640]{H.~Abramowicz}$^\textrm{\scriptsize 151}$,
\AtlasOrcid[0000-0002-1599-2896]{H.~Abreu}$^\textrm{\scriptsize 150}$,
\AtlasOrcid[0000-0003-0403-3697]{Y.~Abulaiti}$^\textrm{\scriptsize 117}$,
\AtlasOrcid[0000-0002-8588-9157]{B.S.~Acharya}$^\textrm{\scriptsize 69a,69b,m}$
% ...
\bigskip
\\

% ...
$^{29}$Physics Department, Brookhaven National Laboratory, Upton NY; United States of America.\\
% ...
$^{35}$$^{(a)}$Facult\'e des Sciences Ain Chock, Universit\'e Hassan II de Casablanca;$^{(b)}$Facult\'{e} des Sciences, Universit\'{e} Ibn-Tofail, K\'{e}nitra;$^{(c)}$Facult\'e des Sciences Semlalia, Universit\'e Cadi Ayyad, LPHEA-Marrakech;$^{(d)}$LPMR, Facult\'e des Sciences, Universit\'e Mohamed Premier, Oujda;$^{(e)}$Facult\'e des sciences, Universit\'e Mohammed V, Rabat;$^{(f)}$Institute of Applied Physics, Mohammed VI Polytechnic University, Ben Guerir; Morocco.\\
% ...
$^{49}$Fakult\"{a}t Physik , Technische Universit{\"a}t Dortmund, Dortmund; Germany.\\
% ...
$^{55}$II. Physikalisches Institut, Georg-August-Universit\"{a}t G\"ottingen, G\"ottingen; Germany.\\
% ...
$^{69}$$^{(a)}$INFN Gruppo Collegato di Udine, Sezione di Trieste, Udine;$^{(b)}$ICTP, Trieste;$^{(c)}$Dipartimento Politecnico di Ingegneria e Architettura, Universit\`a di Udine, Udine; Italy.\\
% ...
$^{102}$CPPM, Aix-Marseille Universit\'e, CNRS/IN2P3, Marseille; France.\\
% ...
$^{117}$Department of Physics, New York University, New York NY; United States of America.\\
% ...
$^{120}$Homer L. Dodge Department of Physics and Astronomy, University of Oklahoma, Norman OK; United States of America.\\
% ...
$^{150}$Department of Physics, Technion, Israel Institute of Technology, Haifa; Israel.\\
$^{151}$Raymond and Beverly Sackler School of Physics and Astronomy, Tel Aviv University, Tel Aviv; Israel.\\
% ...
$^{a}$ Also Affiliated with an institute covered by a cooperation agreement with CERN.\\
$^{b}$ Also at An-Najah National University, Nablus; Palestine.\\
$^{c}$ Also at Borough of Manhattan Community College, City University of New York, New York NY; United States of America.\\
$^{d}$ Also at Center for High Energy Physics, Peking University; China.\\
$^{e}$ Also at Center for Interdisciplinary Research and Innovation (CIRI-AUTH), Thessaloniki; Greece.\\
$^{f}$ Also at Centro Studi e Ricerche Enrico Fermi; Italy.\\
% ...
$^{m}$ Also at Department of Physics, King's College London, London; United Kingdom.\\
% ...

\end{flushleft}


%

%
%

%

%
%
%
%

\end{document}